\documentclass[10pt]{article} 
 
\usepackage{bm}
\usepackage[latin1]{inputenc}
\usepackage[T1]{fontenc}

\usepackage{graphicx}
\usepackage{color}
\usepackage{rotating}
 
\usepackage{float}
\usepackage{url}

\usepackage{lscape}
\usepackage{rotating}
 
\usepackage{amsfonts}
\usepackage{amsthm}
\usepackage{amscd}
\usepackage{algorithm}
\usepackage{algorithmic}

\usepackage{graphicx}
\usepackage{color}
\usepackage{graphics}
\usepackage{amsmath}
\usepackage{amssymb}
\usepackage{amsthm}

\usepackage{hyperref}

\hypersetup{
    colorlinks,
    linkcolor= {blue!90!black}, %{red!80!black},
    citecolor={blue!60!black},
    urlcolor={blue!80!black}
}

\usepackage{float}
 
\usepackage{subfigure}
\usepackage{latexsym}
\usepackage{epsfig}
 
\usepackage{subfigure}

\usepackage{verbatim}
\usepackage{accents}

\usepackage{natbib} 

\usepackage{enumerate}
\usepackage{enumitem}

\usepackage{soul} %% PARA TACHAR CON \st

\usepackage{enumerate}
\usepackage{enumitem}

\usepackage{array,arydshln} 

 \theoremstyle{plain}

\newtheorem{theorem}{Theorem}[section]
\newtheorem{lemma}[theorem]{Lemma}

\newtheorem{proposition}{Proposition}[section]
\newtheorem{remark}{Remark}[section]

 \usepackage{xcolor}

\newcommand{\cancelo}[1]{}

\newcommand\ba {\mathbf a}

\newcommand\bA {\mathbf A}

\newcommand\bZ {\mathbf Z}

\newcommand\indica {\mathbb{I}}

\newcommand\wa {\widehat{{a}}}
\newcommand\wba {\widehat{\ba}}

\newcommand\wefe {\widehat{f}}

\newcommand\wm {\widehat{m}}

\newcommand\wA {\widehat{A}}
\newcommand\wbA {\widehat{\bA}}

\newcommand\wD {\widehat{D}}

\newcommand\wF {\widehat{F}}

\newcommand\wL {\widehat{L}}

\newcommand\wtD {\widetilde{D}}

\newcommand\wtS {\widetilde{S}}

\newcommand\itI {{\mathcal{I}}}

\newcommand\itK {{\mathcal{K}}}

\newcommand\itR {{\mathcal{R}}}
\newcommand\itS {{\mathcal{S}}}
\newcommand\itT {{\mathcal{T}}}

\newcommand\bSi {\mbox{\boldmath $\Sigma$}}

\newcommand\weps {\widehat{\epsilon}}

\newcommand\wvarphi {\widehat{\varphi}}

\newcommand\wgama {\widehat{\gamma}}

\newcommand\wlam {\widehat{\lambda}}

\newcommand\wmu {\widehat{\mu}}

\newcommand\wsigma {\widehat{\sigma}}

\newcommand\wGamma {\widehat{\Gamma}}

\newcommand\wDelta {\widehat{\Delta}}

\newcommand\wUps  {\widehat{\Upsilon}}
\newcommand\wbSi {\widehat{\bSi}}

\newcommand\wtsigma {\widetilde{\sigma}}

\def\real{\mathbb{R}}

\newcommand{\esp}{\mathbb{E}}
\newcommand{\prob}{\mathbb{P}}
\newcommand{\cov}{\mbox{\sc Cov}}
\newcommand{\var}{\mbox{\sc Var}}

\newcommand{\diag}{\mbox{\sc diag}}

\newcommand{\convpp}{ \buildrel{a.s.}\over\longrightarrow}

\newcommand{\convprob  }{ \buildrel{p}\over\longrightarrow}

\newcommand{\convdist}{ \buildrel{D}\over\longrightarrow}

\newcommand{\trasp}{^{\mbox{\footnotesize \sc t}}}

\newcommand\bcero {{\bf{0}}}

\def\median{\mathop{\mbox{median}}}

\newcommand\noi{\noindent}

\def\square{\ifmmode\sqr\else{$\sqr$}\fi}
\def\sqr{\vcenter{
         \hrule height.1mm
         \hbox{\vrule width.1mm height2.2mm\kern2.18mm
\vrule width.1mm}
         \hrule height.1mm}}

\newcommand\cl {\mbox{\footnotesize  \sc cl}}

\newcommand{\hub}{\mbox{\scriptsize \sc h}}
\newcommand{\tuk}{\mbox{\scriptsize \sc t}}

\newcommand{\rob}{\mbox{\footnotesize \sc r}}

\include{colores}

\begin{document}

\title{\bf Robust tests for equality of regression curves \\ based on characteristic functions}
\author{{Graciela Boente}$^{1}$  {and} { Juan Carlos Pardo-Fern\'andez}$^{2}$\\
{\small$^{1}$ \sl Facultad de Ciencias Exactas y Naturales, Universidad de Buenos Aires and
CONICET, Argentina} \\
{\small$^{2}$ \sl Centro de Investigaci\'{o}n e Tecnolox\'{\i}a Matem\'{a}tica de Galicia (CITMAga), Universidade de Vigo, Spain}\\
{\small e--mail: gboente@dm.uba.ar\hskip1truecm   juancp@uvigo.gal}
}
\date{}
\maketitle
 
\begin{abstract}
This paper focuses on the problem of testing the null hypothesis that the regression functions of several populations are equal under a general nonparametric homoscedastic regression model. It is well known that linear kernel regression estimators are sensitive to atypical responses. These distorted estimates will influence the test statistic constructed from them so the conclusions obtained when testing equality of several regression functions may also be affected.   In recent years, the use of testing procedures based on empirical characteristic functions has shown good practical properties. For that reason, to provide more reliable inferences, we construct a test statistic that combines characteristic functions and residuals obtained from a robust smoother under the null hypothesis. 
The asymptotic distribution of the test statistic is studied under the null hypothesis and under root$-n$ contiguous alternatives. A Monte Carlo study is performed to compare the finite sample behaviour of the proposed test  with the classical one obtained using local averages. The reported numerical experiments show the advantage of the proposed methodology over the one based on Nadaraya--Watson estimators for finite samples. An illustration to a real data set is also provided and enables to investigate the sensitivity of the $p-$value to the bandwidth selection.
\end{abstract}

\noindent {\bf Key Words}: Hypothesis testing, Nonparametric regression models, Robust estimation, Smoothing techniques.

%\keywords{Hypothesis testing\and Nonparametric regression models\and Robust estimation\and Smoothing techniques}

%\begin{keyword}   
%Hypothesis testing \sep  Nonparametric regression models \sep  Robust estimation\sep Smoothing techniques
%\end{keyword}
\date{}
\maketitle

\section{Introduction} \label{introduction}

Let us assume that the random vectors $(X_j, Y_j)\trasp$, $j=1,\dots,k$, follow the homoscedastic nonparametric regression models  
\begin{equation}
Y_j = m_j (X_j) + U_j = m_j (X_j) + \sigma_j \varepsilon_j,
\label{modelo}
\end{equation}
 where $m_j:\real \to \real$ is a nonparametric smooth function and  the error $\varepsilon_j$ is independent of the covariate $X_j$. The
nonparametric nature of model \eqref{modelo} offers more flexibility than the standard linear model when modelling a complicated relationship between the response variable and the covariate.   As is usual in a robust framework,  we will avoid first moment conditions and we will require that the errors distribution  $G_j(\cdot)$ has scale $1$. Furthermore, to identify $m_j$ we will impose an identifiability assumption depending on the score function (see assumption \ref{ass:A4} below) which holds whenever   the errors $\varepsilon_j$  have a symmetric distribution. For instance, if the target, that is, the quantity of interest, is the conditional median, the loss function to be used should be the absolute value. In such  a  situation, to identify $m_j$, the requirement is that the error $\varepsilon_j$ has median $0$.  When  second moments exist, as it is the case of the classical approach, the usual assumption is that $\esp(\varepsilon_j)=0$ and $\var(\varepsilon_j)=1$, which means that $m_j$ represents the conditional mean, while $\sigma_j^2$   equals the residuals variance, i.e., $\sigma_j^2=\var(Y_j - m_j (X_j))$.  Henceforth, we assume that the covariates    $X_j$ have the same support $\itR$,  even when they may have different densities. 
 
 In many situations, it is of interest to  compare the regression functions $m_j$, $j=1,\dots, k$, to decide  
if the same functional form appears in all populations. In particular, in this paper we focus on testing the null hypothesis of equality of the regression curves at least in some region $\itR_0$ of the common support $\itR$, versus a general alternative.  The null hypothesis to be considered is
\begin{equation}
H_0 : m_1(x) = m_2(x)=\dots=m_k(x) \text{ for all } x \in \itR_0, \label{eq:H0}
\end{equation}
while  the alternative hypothesis is $H_1: H_0\mbox{ is not true}$.  

When second moments exist, the problem of testing   equality of two regression curves   has been considered   by several authors such as \citet{dette:munk:1998} and \citet{Neumeyer:dette:2003}, among others. The first paper considered   almost uniform design points and construct an $L^2$ statistic for which the asymptotic distribution is derived under the null hypothesis and under fixed alternatives, while the second one proposed  and studied  a procedure  based on the comparison of marked empirical processes of the residuals. Some possible extensions to the situation of $k>2$ were  already mentioned therein. As mentioned in \citet{Pardo:etal:2007}, the extension of the test statistics used when   comparing two regression curves to the situation of $k>2$ regression functions may not be  straightforward, since some loss of power may arise when performing comparisons pairwise. To solve this issue,  \citet{Pardo:etal:2007} proposed Kolmogorov--Smirnov and
Cram\'er--von Mises type statistics and establish their asymptotic distribution
under the null hypothesis and under root-$n$ local alternatives. These statistics were constructed using the empirical distribution functions of the residuals obtained  from non-parametric kernel estimators.     \citet{Pardo:etal:2015} introduced a statistic based on the residuals characteristic functions which can detect local alternatives converging to the null hypothesis at the rate $\sqrt{n}$ and whose $p-$values do not rely  on bootstrap. In this paper, we will provide   a robust alternative to this procedure.

The main reason to provide a robust counterpart is that the  test statistic  based on characteristic functions  mentioned above is based on  linear kernel regression estimators  which  locally average the responses resulting in estimators  sensitive to atypical observations. More precisely, when estimating the regression function at a value $x$, the effect of an outlier in the responses will be larger as the distance between the related covariate  and the point $x$ is smaller. In this sense, atypical data in the responses in nonparametric regression may lead to a complete distorted estimation which will clearly influence the test statistic and the conclusions of the testing procedure. Hence, robust estimates  are needed   to provide more reliable estimations and inferences. Beyond the importance of developing robust estimators, the problem of obtaining   robust hypothesis testing procedures also deserves attention. In the nonparametric setting, robust testing procedures are   scarce. For instance,  a robust test for homoscedasticity in nonparametric regression was defined in \citet{dette:marchlewski:2010}, while \citet{Bianco:Boente:Martinez:2006} proposed a procedure to test if the nonparametric component equals a fixed given function   in the framework of a partly linear regression model.   On the other hand, 
\citet{sun:2006} proposed a test based on an orthogonal moment condition of
residuals which converges at non--parametric rate, while  \citet{dette:etal:2011, dette:etal:2013} provided a test based on the $L^2-$distance between non-crossing non-parametric estimates of the quantile curves,  the first one  converges at the non--parametric rate $\sqrt{nh}$,  where $h$ is the bandwidth parameter, while  the latter one detects alternatives at rate root-$n$. Finally, the proposal in  \citet{Kuruwita:etal:2014} is based on  a marked empirical process of the residuals detecting also  root-$n$ alternatives. A robust approach to compare two regression functions versus a one-sided alternative, using local $M-$estimators, was studied in \citet{boente:pardo:2016}. Their proposal is based on a test statistic that uses a bounded score function and the residuals obtained from a robust estimate for the regression function under the null hypothesis.  When the errors in both populations have the same distribution and the design points have  equal   densities, \citet{koul:schick:1997} defined a family of covariate--matched statistics  allowing to detect  root$-n$ one--sided local alternatives.   It is worth mentioning that this   family includes   a covariate--matched  Wilcoxon--Mann--Whitney test based on the sign of all response differences, for which the asymptotic properties are derived without requiring second moments to the errors.  To extend their proposal to the situation of different errors distribution and possible different error densities,  \citet{koul:schick:2003}  developed a modified version of one of the covariate--matched statistics introduced in  \citet{koul:schick:1997}, but this statistic assumes the existence of second moments and may be affected by atypical data arise in the responses. 
Finally, \citet{feng:etal:2015} considered a test for $H_0$ versus $H_1$ using a generalized likelihood ratio test  incorporating a Wilcoxon  likelihood function and kernel smoothers, which allows to detect alternatives with  non--parametric rate.   In order to obtain asymptotic results for their proposal \citet{feng:etal:2015}    assumed  that the errors $\varepsilon_j$ have   symmetric distributions with Lipschitz densities as well as   the existence of second moment of the regression errors.  

The aim of this paper is to propose a class of tests for $H_0$ versus $H_1$ in \eqref{eq:H0} which combines the ideas of robust smoothing with those given in  \citet{Pardo:etal:2015} to obtain a procedure   detecting root$-n$ alternatives without requiring  first moments to the errors.  In Section \ref{sec:prelim}, we remind the definition of the  robust estimators. The test statistics is introduced in Section \ref{sec:test}, where its asymptotic behaviour under the null hypothesis and contiguous alternatives is also studied. We present the results of a Monte Carlo study in Section \ref{sec:monte} and an illustration to a real data set in Section \ref{sec:datos}. Final comments are provided in Section \ref{sec:conclusiones}.  All proofs are relegated to the Appendix.

\section{Preliminaries on robust regression estimation}{\label{sec:prelim}}

As mentioned above, the robust statistic to be defined is based on robust local $M-$smoothers. For that reason, in this section, we briefly review their definition and state the notation to be employed.

Let $(X_{j\ell},Y_{j\ell})\trasp$, $1\le i\le n_j$, be independent and identically distributed observations   with the same distribution as $(X_j,Y_j)\trasp$, $j=1,\ldots, k$.  As it is well known, if  $\esp|Y_j|<\infty$, the regression functions  $m_j$ in \eqref{modelo}  equals  $\esp(Y_j|X_j)$, which may be estimated using   the Nadaraya--Watson estimator \citep[see, for example,][]{hardle:libro:1990}. To remind its definition, let $K$ be a kernel function (usually a symmetric density) and $h=h_n$  a sequence of strictly positive real numbers. Furthermore, let $K_h(u)=h^{-1}K(u/h)$. The linear kernel smoother used to estimate $m_j$ is defined as 
\begin{equation}\label{clasico}
\wm_{j,\cl}(x)=\left\{\sum_{\ell=1}^{n_j} K_h\left(x-X_{j\ell}\right)\right\}^{-1}\sum_{\ell=1}^{n_j} K_h\left(x-X_{j\ell}\right) Y_{j\ell}\,.
\end{equation}
As mentioned in the introduction, this estimator is sensitive to outlying values in the response
variable, also known as ``vertical outliers'' in the literature. Robust estimates in a  nonparametric setting  provide  an alternative to obtain estimators insensitive to atypical data. Among the  proposals considered   in the literature, we can mention the local $M-$smoothers studied in    \citet{hardle:tsybakov:1988} and  \citet{boente:fraiman:1989}, among others. These estimators use a preliminary scale  estimator  to measure the size of the residuals to be downweighted. For heteroscedastic models, the scale function can only be estimated at a nonparametric rate. In contrast,  under an homoscedastic regression model, root$-n$ scale estimators may be constructed. In particular,   scale estimators based on differences are widely used, see, for instance, \citet{rice:1984} and \citet{hall:etal:1990}.  \citet{ghement:etal:2008} proposed a robust version of these difference--based  estimators. For random covariates,   let $X_{j,(1)}\le \dots\le X_{j,(n_j)}$ be the ordered statistics of the explanatory variables of the $j-$th population and denote as $(X_{j,(1)},Y_{j,D_{1,j}})\trasp, \dots, (X_{j,(n_j)},Y_{j,D_{n_j,j}})\trasp$ the  sample of observations ordered according to the values of the explanatory variables, that is, $X_{j,(\ell)}=X_{j,D_{\ell,j}}$. The estimators defined in \citet{ghement:etal:2008}  can be adapted to the present situation by taking the differences $Y_{j,D_{\ell+1,j}} -Y_{j,D_{\ell,j}} $, see   also \citet{dette:munk:1998}. From these differences, one may define the   robust consistent root$-n$ scale estimator of   $\sigma_j$ as 
\begin{equation}
\wsigma_j=\frac{1}{\sqrt{2}\Phi^{-1}(3/4)}\median_{1\le \ell \le n_j-1} \left|Y_{j,D_{\ell+1,j}} -Y_{j,D_{\ell,j}}\right|,
\label{wsigmaj}
\end{equation}
where the coefficient $\sqrt{2}\Phi^{-1}(3/4)$ ensures Fisher-consistency for normal errors ($\Phi^{-1}$ denotes the quantile function of the standard normal law).

Let  $\rho_j :\real \to \real$, $j=1,\ldots, k$, be  a $\rho-$function as defined in \citet{maronna:etal:2019libro}, that is,    a continuous and even function
non-decreasing on $[0,+\infty)$ and such that $\rho_j(0)=0$. Moreover, $\lim_{u\to \infty} \rho_j(u)\ne 0$ and if $0 \leq u< v$ with $\rho_j(v) < \sup_u \rho_j(u)$ then $\rho_j(u)<\rho_j(v)$. When $\rho_j$ is bounded, we assume that $ \sup_u \rho_j(u)=1$. If $\rho_j$ is differentiable, we denote as $\psi_j$ its derivative. It is often required that $\psi_j$ is bounded, as happens in the following examples.   Two widely used families of $\rho-$functions are the   Huber's function and the Tukey's bisquare one. In both cases, $\rho_j(u)=\rho_{0}(u/c_j)$, where $c_j>0$ is a tuning constant to achieve a given efficiency.  The $\rho-$function $\rho_{0} $ related to the proposal in \citet{Huber:1964} was extensively used in regression problems with fixed covariates and corresponds to $\rho_{0}(u)=\rho_{\hub}(u) \, = \,u^2 / 2 $ when $ |u| \le 1 $, while $\rho_{\hub} \left( u \right)=|u| - 1 / 2 $, otherwise. It leads to an unbounded $\rho-$function with bounded derivative $\psi_{\hub}(u)= \min\{1, \max(-1, u)\}$. A  a smooth  approximation of the Huber function defined as  $\rho_{0}(u)=\sqrt{1+u^2}-1$ may also be considered. The Tukey's bisquare function corresponds to a bounded $\rho-$function  and is defined as $\rho_{0}(u)=\rho_{\,\tuk}(u)  =\min\left\{1 - (1-u ^2)^3, 1\right\}$.  It is worth mentioning that the bounded  derivative of the $\rho-$function controls the effect of  ``vertical outliers''. Clearly, different tuning constants or $\rho-$functions may be chosen when defining $\rho_j$ for each $j=1,\dots,k$, even when it is preferable to ensure the same efficiency in the estimation procedure across populations.

Define
\begin{equation}
\lambda_j(x,a,\sigma)=\esp\left[\psi_j\left( \frac{Y_j-a}{\sigma}\right) |  X_j=x \right] \mbox{ and } \gamma_j(x,a,\sigma)=\esp\left[\rho_j\left( \frac{Y_j-a}{\sigma}\right) |  X_j=x \right].
\label{funcionlamda}
\end{equation}
Note that if \eqref{modelo} holds,  $\psi_j$ is an odd function  and the errors have a symmetric distribution, then  $\lambda_j(x,m_j(x),\sigma)=\esp\psi_j(\sigma_j\varepsilon_j/\sigma) =0$,  for any $\sigma>0$.  Moreover, taking into account that the errors are independent of the covariates, we have that
$$\gamma_j(x,a,\sigma)= \esp\left[\rho_j\left( \frac{\sigma_j\,\varepsilon_j + m_j(x)-a}{\sigma}\right)\right]\,.$$ 
Therefore, Lemma 3.1 in \citet{yohai:1985}  \citep[see also ][Theorem 10.2]{maronna:etal:2019libro}  entails that  $m_j(x)$ is the unique minimizer of $\gamma_j(x,a,\sigma)$ when $\rho_j$ is a $\rho-$function, the errors $\varepsilon_j$ have a density function $g_j(t)$  that is  even, non-increasing in $|t|$, and strictly decreasing for $|t|$ in a neighbourhood of $0$ . 

Hence, to obtain robust estimators of $m_j(x)$,  we plug into (\ref{funcionlamda}) an estimator of the conditional distribution of $Y_j|X_j=x$ and a robust estimator of the error's scale $\wsigma_j$, such as the one defined in \eqref{wsigmaj}. Based on the samples $\{ (X_{j\ell},Y_{j\ell})\trasp, \ell=1,\ldots,n_j \}$, the robust nonparametric estimator of $m_j(x)$ is defined as the minimizer $ \wm_j(x) $ of $\wgama_j(x,a,\wsigma_j)$, where 
\begin{equation}
\wgama_j(x,a,\sigma)= \sum_{\ell=1}^{n_j} K_h\left(x-X_{j\ell}\right)  \rho_j\left( \frac{ Y_{j\ell}-a}{\sigma}\right)\,.
\label{funcionwgama}
\end{equation}
Hence,  $\wm_j(x) $ is  the solution   of 
\begin{equation}
 \wlam_j(x,\wm_j(x),\wsigma_j)=0\, ,
 \label{estimadorrobusto}
 \end{equation} 
 with
\begin{equation}
\wlam_j(x,a,\sigma)= \sum_{\ell=1}^{n_j} K_h\left(x-X_{j\ell}\right)  \psi_j\left( \frac{ Y_{j\ell}-a}{\sigma}\right)\,,
\label{funcionwlamda}
\end{equation}
Note that different $\rho-$functions $\rho_j$ can be used in the different samples, in this way, we  provide a more flexible setting. 

\section{A class of test statistics}{\label{sec:test}}
As in \citet{Huskova:Meintanis:2009} and \citet{Pardo:etal:2015}, our test will be based on a weighted $L^2-$distance between characteristic functions. We will compare the characteristic functions of the residuals obtained from a robust fit with those constructed under the null hypothesis. For that purpose, let $m_0$ be the common regression curve under the null hypothesis and define
$$\varepsilon_{0j} = \frac{Y_j - m_0(X_j)}{\sigma_j}\,.$$
It turns out that the null hypothesis $H_0$ is true if and only if, for all $1\le j \le k$, the random variables $\varepsilon_{j}$ and $\varepsilon_{0j}$ have the same distribution  for some function $m_0$, see \citet{Pardo:etal:2007}. 

Let $W_j:\real\to \real$ be a non-negative weight function with compact support $\itS_j \subset \accentset{\circ}{\itR}$, where    $\accentset{\circ}{\itR}$ stands for the interior of the set $\itR$. A possible practical choice for $W_j$ is the indicator function of the set $\itR_0$, in which case $\itS_j=\itR_0$ for all $j=1, \dots, k$.  For a given non-negative real--valued function $w$, such that $\int w(t) dt<\infty$ and $\int t^2\, w(t) dt<\infty$, and for any complex-valued measurable function $g$, we denote $\|g\|_w^2 = \int |g(t)|^2 w(t) \,dt$ the norm
in the Hilbert space $L^2(\real, w)$. Let $f_j$ be the probability density function of $X_j$  and define $f(x)=\sum_{j=1}^k \pi_j\; f_j(x)$, where $\sum_{j=1}^k \pi_j =1$. In practice, when the sample of the $j-$th population has size $n_j$ and  $n=\sum_{i=1}^k n_j$, we have that $\pi_j=\lim n_j/n$.

Given   independent observations $\{ (X_{j\ell},Y_{j\ell})\trasp, \ell=1,\ldots,n_j \}$, $j=1,\dots, k$,    such that $(X_{j\ell},Y_{j\ell})\trasp\sim (X_j,Y_j)\trasp$   and let $\wm_j(x)$ be the robust estimator of $m_j(x)$ given in \eqref{estimadorrobusto} and $\wsigma_j$ a robust estimator of the error's scale $\sigma_j$, such as the one defined in \eqref{wsigmaj}.  For a given $x \in \itR$, define 
\begin{equation}
\mu_0(x)=\sum_{j=1}^k \pi_j \, \frac{f_j(x)}{f(x)} m_j(x)\,,
\label{eq:mu0}
\end{equation}
and its estimate as 
\begin{equation}
 \wmu_0(x)=  \sum_{j=1}^k \frac{n_j}{n} \, \frac{\wefe_j(x)}{\wefe(x)} \wm_j(x)\,,
\label{eq:wmu0}
\end{equation}
where $\wefe_j(x)$  is the kernel estimator of $f_j$, i.e.,
$$\wefe_j(x)=\frac{1}{n_j} \sum_{\ell=1}^{n_j} K_h\left(x-X_{j\ell}\right)\,,$$
and
$$\wefe(x)=\sum_{j=1}^k \frac{n_j}{n}\; \wefe_j(x)\,.$$
Under the null hypothesis, $\mu_0\equiv m_0$, hence, for a given $x \in \itR$,  an estimator of   the common regression function under the null hypothesis   is $\wmu_0(x)$.

On the basis of these estimators, for each population $j$, we construct two samples
of residuals 
$$\weps_{j\ell}= \frac{Y_{j\ell}- \wm_j(X_{j\ell})}{\wsigma_j}\quad \mbox{and}\quad \weps_{0j\ell}= \frac{Y_{j\ell}- \wmu_0(X_{j\ell})}{\wsigma_j}\,,$$
and the weighted empirical characteristic functions
$$\wvarphi_j(t)=\frac{1}{n_j} \sum_{\ell=1}^{n_j} W_j(X_{j\ell})\,\exp(i\,t\, \weps_{j\ell})\quad \mbox{and}\quad \wvarphi_{0j}(t)=\frac{1}{n_j} \sum_{\ell=1}^{n_j}  W_j(X_{j\ell})\,\exp(i\,t\, \weps_{0j\ell})\,.$$
The test statistic is defined as
\begin{equation}  \label{teststatistic}
 T = \sum_{j=1}^k \frac{n_j}{n} \,  \|\wvarphi_j-\wvarphi_{0j}\|_w\;.
\end{equation}
The null hypothesis will be rejected for large positive values of the test statistic $T$.  As mentioned already in \citet{Pardo:etal:2015} the  weight function $w$   is necessary in
order to ensure the  finiteness of the norms involved in the definition of $T$. A possible choice for $w$ is the  density corresponding to a $N(0,\sigma_w^2)$, which corresponds to the choice made in our numerical study for $\sigma_w=1$. For further discussion on the choice of $w$, we refer to Section 4.2 in \citet{Pardo:etal:2015}.  

\subsection{Asymptotic behaviour of the test statistic}{\label{sec:consistH0}}

To perform the test for a given significance level, critical values obtained from the (asymptotic) null distribution of $T$ are needed. For that reason, in the sequel, we will analyse the asymptotic distribution of the test statistic under the following   assumptions:

\begin{enumerate}[label=\textbf{A\arabic*}]
\item\label{ass:A2} For $j=1,\dots, k$, $\psi_j:\real \to\real$ are odd, bounded and twice continuously differentiable functions, with bounded derivatives. Besides, the first and second derivatives, $\psi_j^{\prime}$ and $\psi_j^{\prime\prime}$, are such that  $\nu_{j}=\esp[\psi_j^{\prime}(\varepsilon_j)]\ne 0$, and $\zeta_{1,j}(u)=u\psi_j^{\prime}(u)$ and $\zeta_{2,j}(u)=u\psi_j^{\prime\prime}(u)$ are bounded. Denote as  $\tau_{j}=\esp[\psi_j^{2}(\varepsilon_j)]$ and $e_j=\tau_j/\nu_j^2$.

\item\label{ass:A3}   For $j=1,\dots, k$, $W_j:\real \to\real$ are bounded non-negative continuous weight functions with compact support $\itS_j \subset \accentset{\circ}{\itR}$, where $\itR$ stands for the support of $X_j$. Without loss of generality we assume that $\|W_j\|_{\infty}=1$.
 
\item\label{ass:A4}   For $j=1,\dots, k$, $\esp \psi_j(a\;\varepsilon_j)=0$, for any $a>0$.

\item\label{ass:A5}  For $j=1,\dots, k$, the regression function  $m_j$ is twice continuously differentiable in a neighbourhood of the support, $\itR$, of the density of $X_j$. 

\item\label{ass:A6}  For $j=1,\dots, k$, the random variable $X_j$ has a density   $f_j$  twice continuously differentiable in a neighbourhood of the support  $\itS_j$  of   $W_j$  and such that $i(f_j)=\inf_{x\in \itS_j} f_j(x)>0$.

\item\label{ass:A7}  The kernel $K:\real \to \real$ is an even, bounded and Lipschitz continuous function with bounded support, say $[-1,1]$ and such that $\int K(u)du=1$.

\item\label{ass:A8}  
\begin{enumerate}  
\item\label{ass:A8a}  The sample sizes are such that $n_j/n\to \pi_j$ and  $n^{1/4}\left(n_j/n-\pi_j\right)\to 0$ where $0<\pi_j<1$ and $n=\sum_{j=1}^k n_j \rightarrow \infty$. 
\item\label{ass:A8b}   Furthermore, $n^{1/2}\left(n_j/n-\pi_j\right)\to 0$.
\end{enumerate}

\item\label{ass:A9}  The bandwidth sequence is such that $h_n\to 0$, $nh_n/\log n\to \infty$, $\sqrt{n h_n^2}/\log n\to \infty$, $n h_n^4\to 0$  as $n \rightarrow \infty$.
\item\label{ass:A10}  For some  $1/4<\gamma_0\le 1/2$, $n_j^{\gamma_0}(\wsigma_j-\sigma_j)=O_{\prob}(1)$.  
\item\label{ass:A11}  $\esp |\varepsilon_j|^{\theta_0}<\infty$,  with $0<\theta_0=1/ (3/4+\gamma_0)<1$ and $\gamma_0$ given in assumption \ref{ass:A10}. 
\end{enumerate}

\begin{remark}
Assumptions \ref{ass:A3} and \ref{ass:A5} to \ref{ass:A7} are standard conditions in the nonparametric literature,  while \ref{ass:A8}  and \ref{ass:A9} are usually a requirement  when dealing with testing problems.  As mentioned in \citet{Pardo:etal:2007}, from a theoretical point of view, assumption  \ref{ass:A8} excludes  the optimal bandwidth used for estimating the regression function which has  order $n^{-1/5}$. This comment regarding the bandwidth rate is also valid for the proposal considered in   \citet{dette:etal:2013} who required the same convergence rate  stated in \ref{ass:A9}  for the bandwidth used to estimate the conditional distribution function. We also refer to \citet{zhang:2003} who provides an interesting insight on the problem of bandwidth selection  in testing problems.  On the other hand,  \ref{ass:A2} and \ref{ass:A4} are usual requirements in a robust setting. In particular, \ref{ass:A4} holds if, for  $j=1,\dots, k$, the distribution $G_j$ of $\varepsilon_j$ is symmetric around $0$ and $\psi_j $ is an odd function. Furthermore, the condition $\nu_j \ne 0 $ in assumption \ref{ass:A2} ensures that  $\wvarphi_j(t)-\wvarphi_{0j}(t)$ has convergence order $n^{1/2}$ allowing   the test statistic to detect root-$n$ alternatives. It is worth mentioning that assumption \ref{ass:A11} is fulfilled when the errors have a Cauchy distribution, meaning that our procedure may be applied when the practitioner suspects that the errors may  be heavy tailed.   A discussion on robust  scale estimators satisfying   \ref{ass:A10} is given in Section \ref{sec:A10}.
\end{remark}

 For the sake of simplicity, in the sequel, we will assume that the same bandwidth is used when estimating the regression functions $m_j$, $j=1,\dots, k$. Similar results can be obtained when different bandwidths are considered as far as they satisfy \ref{ass:A9}.  

From now on, denote as $\omega_{j}=\esp   W_j (X_{j} )$, 
\begin{align*}
\beta_{j} ^{(s)} =\esp\left\{  W_j (X_{s} ) \frac{f_j(X_{s } )}{f( X_{s} ) } \right\} \,,   \qquad   \beta_{j}  =\esp\left\{  W_j (X_{j} )   \frac{f_j(X_{j } )}{f( X_{j} ) } \right\}\,,\\
\alpha_{j,\ell}^{(s)} =  \esp\left(    \frac{ W_\ell(X_{s} )   f_\ell(X_{s} )\,  W_j(X_{s} )  f_j(X_{s} ) }{f^2( X_{s} ) }\right)\,,   \qquad  
\alpha_{j}^{(s)}   =\esp\left\{  W_j^2(X_{s} )   \frac{f_j^2(X_{s } )}{f^2( X_{s} ) } \right\}    \,,
\end{align*}
and note that $\beta_{j} ^{(j)}=  \beta_{j}$, $\alpha_{j,\ell}^{(s)}=\alpha_{\ell,j}^{(s)}$  and $\alpha_{j,j}^{(s)}=\alpha_{j}^{(s)}$.

The next theorem gives the asymptotic distribution of the test statistic under the null hypothesis, while Theorem \ref{teo:theorem2} analyses its behaviour  under local alternatives.

\vskip0.1in

\begin{theorem}{\label{teo:theorem1}} 
Assume that \eqref{modelo} and \ref{ass:A2} to \ref{ass:A7}, \ref{ass:A8a}), \ref{ass:A9}  and \ref{ass:A11} hold. Let $\wsigma_j$ be a consistent estimator of $\sigma_j$, $j=1,\ldots,k$ satisfying \ref{ass:A10}. 
Then,  
\begin{enumerate}
\item[a)] Under $H_0: m_1=m_2=\dots=m_k$, we have that 
 $$\sqrt{n_j}\left(\wvarphi_j(t)-\wvarphi_{0j}(t)\right)= i\, t \varphi_{j}(t) Z_j+ t\, R_{1,n_j}(t)+t^2 R_{2,n_j}(t)= i\, t \varphi_{j}(t) Z_j+  R_{1,n_j}^{\star}(t)+  R_{2,n_j}^{\star}(t)\,,$$
with $\|R_{s,n_j}^{\star}\|_w=o_{\prob}(1)$, $s=1,2$,   and $\bZ=(Z_1,\dots, Z_k)\trasp\sim N(\bcero, \bSi)$, where 
\begin{align*}
\sigma_{jj} & = \sum_{s=1}^k \pi_j\,\pi_s\, e_s\, \alpha_{j}^{(s)} \,\frac{\sigma_s^2}{\sigma_j ^2} +  e_j  \left\{ \omega_{j} ^2 -2 \pi_j \omega_{j} \beta_j \right\} \,,\\
\sigma_{j\ell} & =  \frac{\pi_\ell^{1/2}\, \pi_j^{1/2}}{\sigma_\ell \, \sigma_j} \,  \sum_{s=1}^k     e_s\;\pi_s \, \sigma_s^2  \alpha_{j,\ell}^{(s)} 
      - \frac{\sigma_\ell}{\sigma_j }    \pi_j^{1/2}\,\pi_\ell ^{1/2} \, e_\ell \, \omega_\ell  \;  \beta_j^{(\ell)}
    - \frac{\sigma_j}{\sigma_\ell }    \pi_j^{1/2}\,\pi_\ell ^{1/2} \, e_j  \omega_j  \;  \beta_\ell^{(j)}  \,.
  \end{align*}
\item[b)] Hence,  $n\,T\convdist \bZ\trasp \bA\bZ $, where $\bA=\diag(a_1,\dots, a_k)$ with $a_j=\|g_j\|_w^2$ and $g_j(t)=t\, \varphi_j(t)$.
\end{enumerate}
\end{theorem}

\vskip0.1in
\begin{theorem}{\label{teo:theorem2}} 
Assume that \eqref{modelo} and \ref{ass:A2} to \ref{ass:A9}  and \ref{ass:A11} hold. Let $\wsigma_j$ be a consistent estimator of $\sigma_j$, $j=1,\ldots,k$ satisfying \ref{ass:A10}. Let $\Delta_j:\real \to \real$ be such that $\esp W_j(X_j) \Delta_j^2(X_{j}) <\infty$.
Then, 
 under $H_{1,n}: m_j=m_0+n^{-1/2} \Delta_j$,  we have that 
\begin{align*}
\sqrt{n_j}\left(\wvarphi_j(t)-\wvarphi_{0j}(t)\right)
& = i\, t \varphi_{j}(t) \left(Z_j + \frac{\pi_j^{1/2}}{\sigma_j}  \esp \left\{W_j(X_{j} )\, \left[\Delta_0(X_{j})-\Delta_j(X_{j})\right]\right\}\right)+  R_{1,n_j}^{\star}(t) \,,
\end{align*}
with $\| R_{1,n_j}^{\star}\|_w=o_{\prob}(1)$ and $\bZ=(Z_1,\dots, Z_k)\trasp \sim N(\bcero, \bSi)$ where $\bSi$ is as in Theorem \ref{teo:theorem1} and
$\Delta_0(x)=  \sum_{j=1}^k \pi_j \,  \Delta_j(x) {f_j(x)}/{f(x)}\,.$ 
\end{theorem}

\begin{remark}
 Note that  Theorem \ref{teo:theorem1} implies that the asymptotic distribution of $nT$ under the null hypothesis is a finite linear combination of independent chi-squared variables of the form $\sum_{j=1}^k \gamma_j \chi_{1,j}^2$, where $\gamma_j$ are the eigenvalues of the matrix $\bA \bSi$    and $\chi_{1,j}^2$, $j=1,\ldots,k$,  are independent chi-squared random variables with 1 degree of freedom. It is worth noticing that \citet{bodenham:adams:2016} provides an account for different methods to calculate the law of linear combinations of chi--squared distributions, some of them are implemented  in the \texttt{R} package \texttt{CompQuadForm}. However, in   the numerical study reported in Section \ref{sec:monte} and in the analysis of the real data set described in Section \ref{sec:datos}, we used the same strategy described in \citet{Pardo:etal:2015} to obtain an estimator of the asymptotic null distribution of $nT$. First, empirical and kernel estimators are used to estimate the elements of $\bA$ and $\bSi$ to obtain estimators of these matrices, say $\wbA$ and $\wbSi$. Then, the eigenvalues of $\wbA \wbSi$   are calculated and, finally, a Monte-Carlo procedure is employed to simulate values of the weighted combination of chi-squares, so quantiles and probabilities can be immediately approximated. For the sake of brevity, we do not give all the details here, as they follow the same reasoning as in the above mentioned paper. 
\end{remark}

\subsection{Regarding assumption \ref{ass:A10}}{\label{sec:A10}}
As mentioned in Section \ref{sec:prelim}, for fixed designs robust  scale estimators based on differences  were considered in  \citet{ghement:etal:2008} where its is shown that the considered proposal is asymptotically normally distributed. For random covariates, the estimator given \eqref{wsigmaj} provides a possible choice, while a more general family can be obtained  by choosing a bounded $\rho-$function $\rho$ and adapting the robust scale estimators in \citet{ghement:etal:2008}  using the differences $Y_{j,D_{\ell+1,j}} -Y_{j,D_{\ell,j}} $, $1\le \ell\le n_j$. For fixed designs, \citet{ghement:etal:2008} have shown that $n_j^{1/2}(\wsigma_j-\sigma_j)=O_{\prob}(1)$, we conjecture that the same holds for random designs when the function $\rho$ is a continuous, twice continuously and even function, strictly increasing on $(0, c)$, $\rho(x)=1$ for $|x|\ge c$ and $\rho(x)<1$ when $|x|< c$, as it is the case when  $\rho(u)=\rho_{\,\tuk}(u/c)$.

Another family of scale estimators was studied in Section S.3.2 of the supplementary file of \citet{boente:martinez:2017}. More precisely, these authors suggest to consider the residuals $r_{j,\ell}= Y_{j\ell} - \wm_j(X_{j\ell})$, where $ \wm_j$ is a preliminary regression estimator such as the local median. Denote as $\wF_{n,j}$ the empirical distribution of the residuals  $r_{j,\ell}$. From Proposition S.3.2 in the above mentioned paper, we have that if $\sup_{x\in \itK} | \wm_j(x)-m_j(x)| \convpp 0$, for any compact set $\itK\subset  \itR$ and $\sigma_{\rob}$ is  a robust scale functional, then $\wsigma_j= \sigma_{\rob}( \wF_{n,j})$ is strongly consistent to $\sigma_j$. This family of estimators include the $M-$scale estimators defined as 
\begin{equation}
 \frac{1}{n_j}\sum_{\ell=1}^{n_j} \rho\left(\frac{ Y_{j\ell} - \wm_j(X_{j\ell})}{\wsigma_j }\right)=b\,,
\label{eq:Mscale}
\end{equation}
where $b<1$ and $\esp \rho(\varepsilon_{j})=b$.  For instance, when  $\rho(u)=\rho_{\,\tuk}(u/c)$, the choice $c=1.54764$ and $b = 1/2$ yield a scale estimator  that is Fisher-consistent when the errors have a normal distribution.  Up to our knowledge, rates of convergence for the estimators defined through \eqref{eq:Mscale}  have not been derived yet. Proposition \ref{prop:tasascale}  states that if the preliminary regression estimator satisfies certain assumptions then \ref{ass:A10} holds  taking   $\gamma_0$ given in assumption \ref{ass:C2} below. Note that for this choice  $1/4<\gamma_0<1/3$. 

\begin{enumerate}[label=\textbf{C\arabic*}]
\item\label{ass:C1} $\rho$ is a   continuous, bounded and even function
non-decreasing on $[0,+\infty)$ and such that $\rho(0)=0$. Moreover, $\lim_{u\to \infty} \rho(u)\ne 0$ and if $0 \leq u< v$ with $\rho(v) < \sup_u \rho(u)$ then $\rho(u)<\rho(v)$.  Besides, $\rho$  is  twice continuously differentiable, with bounded derivatives. Let $\psi=\rho^{\prime}$  and $\eta(u)=u \psi(u)$, then  $\eta$ is a bounded function,  $\esp \psi(\varepsilon_j)=0$ and   $A_j=\esp \eta(\varepsilon_j)\ne 0$.
\item\label{ass:C2} For some $1/4<\gamma_0<1/3$, one of the following hold
\begin{enumerate}
\item[a)] $ n_j^{\gamma_0}\sup_{x\in [0,1]}|\wm_j(x)- m_j(x)|=O_{\prob}(1)$.
\item[b)] $ (1/n_j)\sum_{\ell=1}^{n_j} \left\{\wm_j(X_{j\ell})- m_j(X_{j\ell})\right\}^2=O_{\prob}(n_j^{-2\,\gamma_0})$.
\end{enumerate}
\end{enumerate}

\vskip0.1in
\begin{proposition}{\label{prop:tasascale}}
Let $\wsigma_j$ be defined as in \eqref{eq:Mscale}, where  $\rho$ satisfies assumption \ref{ass:C1} and   the preliminary regression estimator satisfies \ref{ass:C2}. Assume that $ \wsigma_j\convprob \sigma_j$.
Then, we have that $n_j^{\gamma_0}(\wsigma_j-\sigma_j)=O_{\prob}(1)$.
\end{proposition}

\begin{remark}
 Assumption \ref{ass:C1} is a usual requirement when considering robust scale estimators either in location or linear regression models. The smoothness and boundedness conditions on the function $\rho$ and its derivatives  stated in assumption \ref{ass:C1} are fulfilled when considering  $\rho(u)=\rho_{\,\tuk}(u/c)$, since for this choice $\psi(u)=0$ for $|u|\ge c$, so $\eta$ is bounded. If the errors have a symmetric distribution, then from the fact that  $\psi$ is an odd function, we obtain that $\esp \psi(\varepsilon_j)=0$. Note that $\eta$ is an even function and the requirement that $A_j=\esp \eta(\varepsilon_j)\ne 0$ is the counterpart when estimating scale to the assumption that    $\esp[\psi_j^{\prime}(\varepsilon_j)]\ne 0$ given in \ref{ass:A2} for the regression function estimators. 

We now discuss whether assumption \ref{ass:C2}  holds for some preliminary robust estimators. In the sequel we assume that assumptions \ref{ass:A5} and \ref{ass:A6} hold.

If cubic splines are used to estimate   $m_j$, the preliminary estimator $\wm_j(x)$ can be defined as $\wm_j(x)=\sum_{s=1}^{k_{n_j}} \wa_s B_s(x)$ where $\{B_s\}_{1\le s \le k_{n_j}}$ is the $B-$spline basis of order $r=4$ and $\wba= (\wa_1, \dots, \wa_{k_{n_j}} )\trasp$ is the minimizer of
$L_{n_j}(\ba)=\sum_{\ell=1}^{n_j} \left|Y_{j\ell}- \sum_{s=1}^{k_{n_j}} a_s B_s(X_{j\ell})\right|$. This estimator is the $B-$spline counterpart of the local median.  Theorem 2.1 in \citet{he:shi:1994} entails that if $k_{n_j}=O(n_j^{1/5})$, then 
$$\frac{1}{n_j}\sum_{\ell=1}^{n_j} \left\{\wm_j(X_{j\ell})- m_j(X_{j\ell})\right\}^2=O_{\prob}\left(n_j^{-4/5}\right)\,,$$
so \ref{ass:C2}b) holds, since $2\gamma_0-4/5<-2/15<0$ implying that  $n_j^{2\,\gamma_0} (1/n_j)\sum_{\ell=1}^{n_j} \left\{\wm_j(X_{j\ell})- m_j(X_{j\ell})\right\}^2=o_{\prob}(1)$.

If local medians are considered and the kernel $K$ satisfies \ref{ass:A7} and the bandwidth has order $h_{n_j}=O\left( n_j^{1/3} (\log(n_j))^{1/3}\right)$,  the rates provided in Theorem 1 in \citet{hardle:luckhaus:1984}, see example 5 therein and also Theorem 3 in \citet{truong:1989}, imply that  
$$\sup_{x\in [0,1]}|\wm_j(x)- m_j(x)|=O_{\prob}\left(n_j^{-1/3} \left(\log n_j\right)^{1/3}\right)\,.$$ 
Thus, $ n_j^{\gamma_0}\sup_{x\in [0,1]}|\wm_j(x)- m_j(x)|= O_{\prob}\left(n_j^{\gamma_0-1/3} \left(\log n_j\right)^{1/3}\right)$ and using that $\gamma_0<1/3$, we get that $ n_j^{\gamma_0}\sup_{x\in [0,1]}|\wm_j(x)- m_j(x)|=o_{\prob}(1)$, so \ref{ass:C2}a) holds.

\end{remark}

\section{Monte Carlo study}{\label{sec:monte}}

In this section, we summarize the results of a Monte Carlo study designed to evaluate the finite sample performance of our proposal. For that purpose, we have  considered a two population setting, even when similar results regarding the performance of the proposed test and its classical counterpart can be achieved when  considering  more than two populations.   The considered scenarios aim to  illustrate the lack of resistance of the classical
procedure when atypical observations arise. At the same time, the simulation reveals the stability of our proposal.  
 More precisely, the classical procedure  involves estimating the regression function through the local kernel estimators given in \eqref{clasico} and constructing the test statistic using the empirical characteristic functions as in    \citet{Pardo:etal:2015}. In contrast, the robust procedure   uses the kernel $M-$estimators described in Section \ref{sec:prelim}  combined with  empirical characteristic functions and corresponds to the robust counterpart of the  test  introduced by the latter authors. The robust estimation method involves computing scale estimators to  standardize the  residuals as well as selecting the score functions and the  smoothing parameters to perform the nonparametric estimation of the regression functions. We considered as  scale estimators those given in \eqref{wsigmaj} and to estimate both regression functions we use robust local  $M-$estimators  computed using the    bisquare Tukey's function with tuning constant $c = 4.685$,  that is, we choose $\rho_j(u)=\rho_{\,\tuk}(u/c)$, for $j=1,2$, where $\rho_{\,\tuk}(u)=\min\left\{1 - (1-u ^2)^3, 1\right\}$. This value for the tuning constant ensures that the estimators have a 95\% efficiency with respect to the classical ones.  The bandwidths were selected using cross-validation both for the regression and density functions. In particular, when considering robust local  $M-$estimators robust cross-validation as defined in \citet{Bianco:Boente:2007} was implemented using a $\tau-$scale estimator.   Henceforth, $T_{n,\rob}$ stands for the robust procedure considered in this paper and $T_{n,\cl}$ for the testing procedure defined in \citet{Pardo:etal:2015}.   
 
 Section \ref{sec:fijas} reports the  results obtained under several homocedastic models to evaluate the level performance of the test statistics and also    the power performance for  fixed   alternatives.  
 The results obtained for two families of contiguous alternatives to the null hypothesis are summarized in Section  \ref{sec:contiguas}. 
 
\subsection{Performance under the null hypothesis and fixed alternatives}{\label{sec:fijas}}

We have considered  several homoscedastic regression  models where the functions $m_j$ in \eqref{modelo} have different shapes and different sample sizes including balanced settings $n_1=n_2=100$ or $200$ and unbalanced ones, $n_1=200$ and $n_2=100$.
The number of Monte Carlo replications was always equal to $NR=1000$.  
On the one hand, to measure the stability in level approximations, we chose different regression function under the null hypothesis
\begin{enumerate}[label = \textbf{M\arabic*}]
\item\label{modelo:m-1} $m_1(x)=m_2(x)=1$,             
\item\label{modelo:m-lineal}  $m_1(x)=m_2(x)=x$,              
\item\label{modelo:m-seno} $m_1(x)=m_2(x)=\sin(2 \pi x)$,
\item\label{modelo:m-exp} $m_1(x)=m_2(x)=\exp(x)$.  
\end{enumerate}
On the other hand, to evaluate the power performance, we  considered  fixed   alternatives that were set as
\begin{enumerate}[label = \textbf{MA\arabic*}]
\item \label{alter:m-1} $m_1(x)=1$,               $m_2(x) = 1 + 0.5 x$, 
\item \label{alter:m-lineal}  $m_1(x)=x$,               $m_2(x) = x + 0.5 x$, 
\item \label{alter:m-seno}  $m_1(x)=\sin(2 \pi x)$,   $m_2(x) = \sin(2 \pi x) + 0.5 x$,
\item  \label{alter:m-exp}  $m_1(x)=\exp(x)$,        $m_2(x) = \exp(x) + 0.5 x$.
\item  \label{alter:m-twosided-lineal}  $m_1(x)=x$,        $m_2(x) = 1-x = x+   (1-2\,x)$.
\item  \label{alter:m-twosided-seno}   $m_1(x)=1$,        $m_2(x) = 1 + \sin(2 \pi x)$.
\end{enumerate}
It is worth mentioning that,  under the fixed alternatives \ref{alter:m-1} to \ref{alter:m-exp},   $m_2(x)=m_1(x)+  0.5 x \ge m_1(x)$, that is, we have a one-sided alternative. In contrast, alternatives \ref{alter:m-twosided-lineal} and \ref{alter:m-twosided-seno} correspond to two--sided alternatives, that is, the functions $m_1$ and $m_2$ cross each other. They are included to evaluate the test capability the detect  more general differences than those given by superiority between the two regression curves.  In all situations, the   covariates were generated with uniform distribution on $[0,1]$, the scale parameters were $\sigma_1=\sqrt{0.25}$ and $\sigma_2=\sqrt{0.50}$ and the significance level was fixed to $\alpha=0.05$.  The weight functions $W_j$ were chosen as equal to one, since we aim to compare the regression functions over their support, i.e., $\itR_0=\itR$.  

Taking into account that  the    covariate--matched   Wilcoxon--Mann--Whitney statistic $W_{n,h} $ defined in \citet{koul:schick:1997} detects root-$n$ local ordered  alternatives, that is, alternatives where $m_2\ge m_1$, and does not require moment conditions, we also include here some results regarding its performance. We only considered the situation where the observations are generated, under the null hypothesis, using  the common function  given by  \ref{modelo:m-lineal}, similar results are obtained when considering  the  regression functions described in \ref{modelo:m-1}, \ref{modelo:m-seno} and \ref{modelo:m-exp}. Besides,  since the  test based on $W_{n,h} $  is designed to detect one--sided alternatives,  we include the one--sided fixed alternative \ref{alter:m-lineal} in our comparison and also the two--sided one, \ref{alter:m-twosided-lineal}. It is worth noticing that this statistic depends on the  bandwidth and there is no automatic way to select it, for that reason, we choose different smoothing parameters  $h=0.10, 0.15$ and $0.20$ to compute $W_{n,h}$. 

 To analyse the behaviour of the proposed test, we studied  samples without outliers generated from the  standard  normal distribution, samples
contaminated with  5\% or 10\% outliers and also a situation where the errors distribution has heavy tails.  More precisely, the following scenarios were considered to simulate the regression errors:
\begin{itemize}  
\item  The clean samples scenario, denoted as $C_0$, corresponds to the situation where $\varepsilon_j\sim N(0, 1)$. In this case no outliers will appear in the data.
\item In the second scenario, labelled $C_1$, we include a 5\% of vertical outliers in the sample by defining $\varepsilon_j \sim 0.95 N(0,1)+0.05 N(j\; 5, \sigma^2)$ with $\sigma=0.1$, for $j=1,2$.
\item Contamination $C_2$ corresponds to 5\% of mild vertical outliers in opposite directions in both samples, that is,  $\varepsilon_j\sim 0.95 N(0,1)+0.05 N((-1)^j\; 5, \sigma^2)$, with $\sigma=0.1$, for $j=1,2.$
\item Contamination $C_3$ corresponds to 5\% of gross vertical outliers in opposite directions in both samples which are obtained defining  $\varepsilon_j\sim 0.95 N(0,1)+0.05 N((-1)^j\; 10, \sigma^2)$, with $\sigma=0.1$, for $j=1,2$.
\item Finally, contamination $C_4$ stands for  a 10\% contamination of extreme vertical outliers only in the first sample, that is,  $\varepsilon_1\sim 0.90 N(0,1)+0.10 N(10, \sigma^2)$ with $\sigma=0.1$ and $\varepsilon_2\sim N(0,1)$.
\end{itemize}

For the tests based on $T_{n,\cl}$ and $T_{n,\rob}$,  the results corresponding
to clean and contaminated samples under $H_0$ are reported in Table \ref{tab:H0}, while those corresponding to the fixed alternatives mentioned above are given in Table \ref{tab:H1}.     Finally, Table \ref{tab:KS97} reports the empirical level and power of the covariate--matched   Wilcoxon--Mann--Whitney statistic $W_{n,h} $. 
To evaluate the test performance, we also examine if the empirical size  is significantly different from the nominal level $\alpha=0.05$. More precisely, in  Tables \ref{tab:H0} and \ref{tab:KS97}, we indicate in \textbf{bold} the values falling out the interval $\itI=[0.032,0.068]$, that is $\itI=[L_1(\alpha),L_2(\alpha)]$ where $L_j(\alpha)=\alpha+ (-1)^j 2.58\,\{\alpha(1-\alpha)/NR\}^{1/2}$, $j=1,2$,    which corresponds to the acceptance region of a test to check whether the actual level differs from the nominal one at level $0.01$.
%EL INTERVALO ES 0.03224732 0.06775268

\begin{center}
\begin{table}[ht!]
%\small
\setlength{\tabcolsep}{4pt}
\begin{tabular}{c c ccc c ccc}
 &   & \multicolumn{3}{c }{$T_{n,\cl}$}  & & \multicolumn{3}{c}{$T_{n,\rob}$}  \\
    \cline{3-5}  \cline{7-9}
    &   & \multicolumn{3}{c }{$(n_1,n_2)$}  & & \multicolumn{3}{c}{$(n_1,n_2)$}  \\
    \cline{3-5}  \cline{7-9}
Contamination & Model  & \multicolumn{1}{c}{$(100,100)$} &  \multicolumn{1}{c}{$(200,100)$} &  \multicolumn{1}{c}{$(200,200)$} & &  \multicolumn{1}{c}{$(100,100)$} & \multicolumn{1}{c}{$(200,100)$} &  \multicolumn{1}{c}{$(200,200)$} \\
   \hline
$C_0$ & \ref{modelo:m-1} & 0.043 & 0.054 & 0.044 & & 0.055 & 0.063 & 0.051\\
& \ref{modelo:m-lineal} & 0.044 & 0.052 & 0.042 & & 0.056 &  0.061  &  0.055 \\
& \ref{modelo:m-seno} & 0.055 & 0.061 & 0.049 & & \textbf{0.074} & \textbf{0.078} & 0.060\\
& \ref{modelo:m-exp} & 0.047 & 0.056 & 0.046 & & 0.060 & 0.066 & 0.053\\
 \hline
 
$C_1$ & \ref{modelo:m-1} & \textbf{0.152} & \textbf{0.163} & \textbf{0.375} & & 0.058 & 0.050 & 0.056\\
& \ref{modelo:m-lineal} & \textbf{0.153} & \textbf{0.159} & \textbf{0.376} & & 0.059 & 0.055 & 0.055\\
& \ref{modelo:m-seno} & \textbf{0.158} & \textbf{0.173} & \textbf{0.384} & & \textbf{0.076} & 0.062 & 0.060\\
& \ref{modelo:m-exp} & \textbf{0.150} & \textbf{0.161} & \textbf{0.374} & & 0.065 & 0.055 & 0.059\\
 \hline

$C_2$ & \ref{modelo:m-1} & \textbf{0.629} & \textbf{0.748} & \textbf{0.916 }& & 0.068 & 0.057 & \textbf{0.070}\\
& \ref{modelo:m-lineal} & \textbf{0.627} & \textbf{0.740} & \textbf{0.917} & & \textbf{0.070} & 0.059 & \textbf{0.070}\\
& \ref{modelo:m-seno} & \textbf{0.617} & \textbf{0.734} & \textbf{0.915} & & \textbf{0.089} & \textbf{0.080} & \textbf{0.077}\\
& \ref{modelo:m-exp} & \textbf{0.619} & \textbf{0.736} & \textbf{0.919} & & \textbf{0.075} & 0.066 & \textbf{0.072}\\
 \hline
 
$C_3$ & \ref{modelo:m-1} & \textbf{0.860} & \textbf{0.941} & \textbf{0.996} & & 0.054 & 0.051 & 0.053\\
& \ref{modelo:m-lineal} & \textbf{0.859} & \textbf{0.935} & \textbf{0.996} & & 0.054 & 0.053 & 0.050\\
& \ref{modelo:m-seno} & \textbf{0.848} & \textbf{0.939} & \textbf{0.993} & & \textbf{0.069} & 0.062 & 0.063\\
& \ref{modelo:m-exp} & \textbf{0.861} & \textbf{0.937}& \textbf{0.996} & & 0.058 & 0.057 & 0.054\\
 \hline
 
$C_4$ & \ref{modelo:m-1}  & \textbf{0.827} & \textbf{0.980} & \textbf{0.996} & & 0.054 & 0.055 & 0.059\\
& \ref{modelo:m-lineal} & \textbf{0.825} & \textbf{0.980} & \textbf{0.995} & & 0.059 & 0.053 & 0.061\\
& \ref{modelo:m-seno}& \textbf{0.817} & \textbf{0.977} & \textbf{0.995} & & \textbf{0.076} & 0.067 & 0.065\\
& \ref{modelo:m-exp} & \textbf{0.830} & \textbf{0.977} & \textbf{0.994} & & 0.062 & 0.057 & 0.057\\
 \hline
\end{tabular}
\caption{\label{tab:H0} Empirical level   of the test statistics $T_{n,\cl} $ and $T_{n,\rob} $,  for clean and contaminated samples.}
\end{table}
\end{center}
 
As expected for clean samples, the  classical procedure based on $T_{n,\cl} $ and its robust counterpart  have a similar performance both in level and power. The empirical level of the test based on $T_{n,\rob}$ seems to be more affected when unbalanced sample sizes are considered specially for model \ref{modelo:m-seno}. For contaminated samples, the empirical level of the classical procedure breaks down, the worst effect is observed under $C_3$,  where the frequency of rejection is almost constant  or under $C_4$ where the frequency of rejection decreases under the considered alternatives. The  robust test is more stable in level and power under the considered contaminations. However, when considering  the sine function (model \ref{modelo:m-seno}) the test becomes liberal for $n_1=n_2=100$ under all contamination schemes. Moreover, the empirical level of the robust test is sensitive to contamination $C_2$ where mild vertical outliers in opposite directions are introduced. These outliers are more difficult to detect for the considered models explaining the test performance under the null hypothesis. Note that, under this contamination as well as   under $C_3$, the empirical level of the classical method based on $T_{n,\cl}$ is always larger than   0.8, while its empirical power is almost 1, becoming completely uninformative.  The same behaviour is observed under $C_4$, when considering   the   alternatives \ref{alter:m-twosided-lineal} and \ref{alter:m-twosided-seno}. In contrast, for the alternatives  \ref{alter:m-1} to  \ref{alter:m-exp},   the empirical power of $T_{n,\cl}$ under $C_4$  is smaller than its empirical level. This Hauck--Donner effect is also observed below  for contiguous alternatives.

Regarding the performance of the   covariate--matched  Wilcoxon--Mann--Whitney test, Table \ref{tab:KS97} reveals that for normal errors, the test respects the level and can detect the one--sided alternative \ref{alter:m-lineal} with slightly higher power than $T_{n,\cl}$ and $T_{n,\rob}$, whereas it is unable to detect the two--sided alternative  \ref{alter:m-twosided-lineal}. The results under   $C_1$ are similar to those obtained for normal errors.   
%In contrast, when considering \ref{alter:m-twosided-lineal}, $W_{n,h}$ fails and results uninformative   for normal errors.
 Under scenarios $C_{2}$ to $C_4$, the level of $W_{n,h}$  breaks down. In particular,  under $C_{2}$ and $C_{3}$ and for sample sizes $n_1=n_2=200$, the  empirical level is always larger than $0.5$ and the power equals $1$, while under $C_4$ the empirical level is almost $0$.    We hence conclude that the   covariate--matched  Wilcoxon--Mann--Whitney test is not adequate when outliers appear in the sample.  Besides, as mentioned above,  this test is unable to detect general alternatives as the one considered in \ref{alter:m-twosided-lineal} and this is reflected on the trivial powers obtained for normal errors which are almost equal to the empirical level. This effect  where the power under  \ref{alter:m-twosided-lineal} is similar to the empirical level  is also observed for the different contaminations considered.

\begin{center}
\begin{table}[ht!]
%\small
\setlength{\tabcolsep}{4pt}
\begin{tabular}{cc ccc c ccc}
 &   & \multicolumn{3}{c }{$T_{n,\cl}$}  & & \multicolumn{3}{c}{$T_{n,\rob}$}  \\
    \cline{3-5}  \cline{7-9}
    &   & \multicolumn{3}{c }{$(n_1,n_2)$}  & & \multicolumn{3}{c}{$(n_1,n_2)$}  \\
    \cline{3-5}  \cline{7-9}
    Contamination & Model  &  {$(100,100)$} &  {$(200,100)$} &  {$(200,200)$} & &  {$(100,100)$} &  {$(200,100)$} &  {$(200,200)$} \\
   \hline
$C_0$ & \ref{alter:m-1} &  0.794 & 0.876 & 0.986 & & 0.789 & 0.871 & 0.985 \\
& \ref{alter:m-lineal} &  0.796 & 0.874 & 0.985 & & 0.788 & 0.868 & 0.985\\
& \ref{alter:m-seno} & 0.806 & 0.885 & 0.987 & & 0.807 & 0.880 & 0.987\\
& \ref{alter:m-exp} & 0.796 & 0.873 & 0.985 & & 0.791 & 0.867 & 0.985\\
& \ref{alter:m-twosided-lineal} & 0.548 & 0.588 & 0.878 & & 0.584 & 0.610 & 0.878 \\
& \ref{alter:m-twosided-seno}   & 0.875 & 0.914 & 0.996 & & 0.876 & 0.904 & 0.996 \\
 \hline
 $C_1$ & \ref{alter:m-1} & 0.765 & 0.817 & 0.976 & & 0.749 & 0.851 & 0.958 \\  
& \ref{alter:m-lineal} & 0.768 & 0.819 & 0.978 & & 0.749 & 0.847 & 0.958 \\
& \ref{alter:m-seno} & 0.782 & 0.834 & 0.982 & & 0.757 & 0.850 & 0.961 \\
& \ref{alter:m-exp} & 0.770 & 0.814 & 0.977 & & 0.746 & 0.847 & 0.960 \\
& \ref{alter:m-twosided-lineal} & 0.217 & 0.211 & 0.488 & & 0.532 & 0.551 & 0.837 \\
& \ref{alter:m-twosided-seno}   & 0.285 & 0.263 & 0.654 & & 0.835 & 0.875 & 0.992 \\
 \hline 
$C_2$ & \ref{alter:m-1} & 0.995 & 1.000 & 1.000 & & 0.843 & 0.904 & 0.977 \\
& \ref{alter:m-lineal} & 0.995 & 1.000 & 1.000 & & 0.848 & 0.898 & 0.976 \\
& \ref{alter:m-seno} & 0.994 & 0.998 & 1.000 & & 0.850 & 0.910 & 0.982 \\
& \ref{alter:m-exp} & 0.994 & 0.999 & 1.000 & & 0.840 & 0.903 & 0.977 \\
& \ref{alter:m-twosided-lineal} & 0.701 & 0.802 & 0.974 & & 0.518 & 0.552 & 0.822 \\
& \ref{alter:m-twosided-seno}   & 0.823 & 0.892 & 0.996 & & 0.810 & 0.856 & 0.986 \\
 \hline 
$C_3$ & \ref{alter:m-1} & 1.000 & 1.000 & 1.000 & & 0.806 & 0.880 & 0.971 \\
& \ref{alter:m-lineal} & 1.000 & 1.000 & 1.000 & & 0.805 & 0.875 & 0.971 \\
& \ref{alter:m-seno} & 0.999 & 1.000 & 1.000 & & 0.815 & 0.880 & 0.976 \\
& \ref{alter:m-exp} & 1.000 & 1.000 & 1.000 & & 0.805 & 0.878 & 0.969 \\
& \ref{alter:m-twosided-lineal} & 0.859 & 0.934 & 0.994 & & 0.540 & 0.562 & 0.835 \\
& \ref{alter:m-twosided-seno}   & 0.871 & 0.946 & 0.998 & & 0.839 & 0.872 & 0.990 \\
 \hline 
 
$C_4$ & \ref{alter:m-1}  & 0.226 & 0.410 & 0.515 & & 0.804 & 0.878 & 0.972 \\
& \ref{alter:m-lineal} & 0.230 & 0.419 & 0.518 & & 0.804 & 0.876 & 0.968 \\
& \ref{alter:m-seno}& 0.230 & 0.420 & 0.508 & & 0.813 & 0.888 & 0.974 \\
& \ref{alter:m-exp} & 0.233 & 0.424 & 0.517 & & 0.800 & 0.879 & 0.966 \\
& \ref{alter:m-twosided-lineal} & 0.849 & 0.981 & 0.998 & & 0.536 & 0.586 & 0.830  \\
& \ref{alter:m-twosided-seno}   & 0.882 & 0.989 & 0.998 & & 0.847 & 0.916 & 0.995\\
 \hline  
\end{tabular}
\caption{\label{tab:H1} Frequency of rejection  of the test statistics $T_{n,\cl} $ and $T_{n,\rob} $,  for fixed alternatives for clean and contaminated samples. }
\end{table}
\end{center}

\begin{center}
\begin{table}[ht!]
%\small
\setlength{\tabcolsep}{3.4pt}
\begin{tabular}{c c c ccc c ccc c ccc}
&&& \multicolumn{11}{c }{$W_{n,h} $}\\
    \cline{4-14}   
  &    &   & \multicolumn{3}{c }{$(n_1,n_2)=(100,100)$} & & \multicolumn{3}{c }{$(n_1,n_2)=(200,100)$} & &  \multicolumn{3}{c }{$(n_1,n_2)=(200,200)$} \\  
 \cline{4-6}  \cline{8-10}\cline{12-14}  
Contamination  & Model  &    &  \small $h=0.10$  &  \small  $h=0.15$  &   \small $h=0.20$  &&  
  \small $h=0.10$  &   \small $h=0.15$  &   \small $h=0.20$   &&  
  \small $h=0.10$  &   \small $h=0.15$  &   \small $h=0.20$  \\
   \hline
$N(0,1)$  &  \ref{modelo:m-lineal} &  &  0.048  &  0.055  &  0.060  &&  0.055  &  0.054  &  0.057  &&  0.046  &  0.044  &  0.044  \\
  &  \ref{alter:m-lineal}    &  &  0.817  &  0.840  &  0.847  &&  0.908  &  0.913  &  0.914  &&  0.990  &  0.991  &  0.991  \\
  &  \ref{alter:m-twosided-lineal}  &  &  0.050  &  0.053  &  0.053  &&  0.045  &  0.048  &  0.050  &&  0.045  &  0.050  &  0.049  \\
\hline
$C_1$   &  \ref{modelo:m-lineal}   &  &  0.050  &  0.053  &  0.055  &&  0.053  &  0.053  &  0.054  &&  0.055  &  0.058  &  0.057  \\
  &  \ref{alter:m-lineal}    &  &  0.783  &  0.799  &  0.803  &&  0.881  &  0.881  &  0.890  &&  0.958  &  0.958  &  0.959  \\
  &  \ref{alter:m-twosided-lineal}  &  &  0.056  &  0.054  &  0.053  &&  0.060  &  0.060  &  0.057  &&  0.050  &  0.056  &  0.057  \\
\hline
$C_2$   &  \ref{modelo:m-lineal}   &  &  \textbf{0.293}  &  \textbf{0.294}  &  \textbf{0.301}  &&  \textbf{0.369}  &  \textbf{0.387}  &  \textbf{0.394}  &&  \textbf{0.509}  &  \textbf{0.513}  &  \textbf{0.522}  \\
  &  \ref{alter:m-lineal}    &  &  0.966  &  0.969  &  0.972  &&  0.988  &  0.992  &  0.992  &&  1.000  &  1.000  &  1.000  \\
  &  \ref{alter:m-twosided-lineal}  &  &  0.259  &  0.270  &  0.270  &&  0.312  &  0.322  &  0.323  &&  0.437  &  0.447  &  0.452  \\
\hline
$C_3$   &  \ref{modelo:m-lineal}   &  &  \textbf{0.293}  &  \textbf{0.294}  &  \textbf{0.301}  &&  \textbf{0.369}  &  \textbf{0.387}  &  \textbf{0.394}  &&  \textbf{0.509}  &  \textbf{0.513}  &  \textbf{0.522}  \\
  &  \ref{alter:m-lineal}    &  &  0.966  &  0.969  &  0.972  &&  0.988  &  0.992  &  0.992  &&  1.000  &  1.000  &  1.000  \\
  &  \ref{alter:m-twosided-lineal}  &  &  0.260  &  0.270  &  0.272  &&  0.313  &  0.323  &  0.324  &&  0.437  &  0.449  &  0.453  \\
\hline
$C_4$   &  \ref{modelo:m-lineal}   &  &  \textbf{0.000}  &  \textbf{0.000}  &  \textbf{0.000}  &&  \textbf{0.003}  &  \textbf{0.002}  &  \textbf{0.002}  &&  \textbf{0.001}  &  \textbf{0.000}  &  \textbf{0.001}  \\
  &  \ref{alter:m-lineal}    &  &  0.303  &  0.308  &  0.313  &&  0.395  &  0.407  &  0.399  &&  0.500  &  0.500  &  0.506  \\
  &  \ref{alter:m-twosided-lineal}  &  &  0.003  &  0.003  &  0.003  &&  0.002  &  0.003  &  0.003  &&  0.000  &  0.000  &  0.000  \\
\hline
\end{tabular}
\caption{\label{tab:KS97}  Empirical level under model \ref{modelo:m-lineal} of the test statistic $W_{n,h} $  and  the corresponding frequency of rejection under the fixed alternatives \ref{alter:m-lineal} and \ref{alter:m-twosided-lineal}, for clean and contaminated samples.}
\end{table}
\end{center}

\begin{center}
\begin{table}[ht!]
\begin{tabular}{c c ccc c ccc}
 &   & \multicolumn{3}{c }{$T_{n,\cl}$}  & & \multicolumn{3}{c}{$T_{n,\rob}$}  \\
    \cline{3-5}  \cline{7-9}
        &   & \multicolumn{3}{c }{$(n_1,n_2)$}  & & \multicolumn{3}{c}{$(n_1,n_2)$}  \\
    \cline{3-5}  \cline{7-9}
Error distribution & Model  &  {$(100,100)$} &  {$(200,100)$} &  {$(200,200)$} & &  {$(100,100)$} &  {$(200,100)$} &  {$(200,200)$} \\
\hline
$N(0,1)$ &   & 0.044 & 0.052 & 0.042 & & 0.056 & 0.061 & 0.055\\
$\itT_2$ & \ref{modelo:m-lineal}  & \textbf{0.027} &  \textbf{0.029} &  0.046 & &  0.053 &  0.045 &  0.049\\
$\itT_1$ &   &  \textbf{0.012} &  \textbf{0.016} & \textbf{0.011} & &  0.051 &  0.059 &  0.051\\
\hline
$N(0,1)$ &  & 0.796 & 0.874 & 0.985 & & 0.788 & 0.868 & 0.880\\
$\itT_2$ & \ref{alter:m-lineal}  &  0.199 &  0.259 & 0.333 & & 0.534 & 0.611  & 0.832 \\
$\itT_1$ &  &  0.022 &  0.027 &  0.012 & & 0.319  &  0.408 &  0.572 \\
\hline
$N(0,1)$ &  & 0.548 & 0.588 & 0.878 & & 0.584 & 0.610 & 0.878 \\
$\itT_2$ & \ref{alter:m-twosided-lineal}  & 0.051   & 0.043    &  0.061 & & 0.216  & 0.226   &   0.391  \\
$\itT_1$ &   &  0.012  &  0.017   & 0.011 & & 0.092  & 0.092  &  0.125  \\
\hline
\end{tabular}
\caption{\label{tab:H0T}  Empirical level  of the test statistics $T_{n,\cl} $ and $T_{n,\rob} $,  under model \ref{modelo:m-lineal}  and  the corresponding frequency of rejection under the fixed alternatives \ref{alter:m-lineal} and \ref{alter:m-twosided-lineal}, for samples with normal and heavy tailed errors.}
\end{table}
\end{center}

To illustrate the level performance when no moments exist, for model \ref{modelo:m-lineal}, we generate errors with Cauchy distribution, labelled $\itT_1$, and with a Student's distribution with two degrees of freedom, labelled $\itT_2$.  Alternatives corresponding to models \ref{alter:m-lineal} and \ref{alter:m-twosided-lineal}  for the same errors distribution  were  considered to study the power behaviour.  We report the results obtained only under this model for the sake of brevity. The results for $T_{n,\cl}$ and $T_{n,\rob}$ under Cauchy and $\itT_2$ errors are summarized in Table  \ref{tab:H0T}, where we repeat the results obtained for normal errors to facilitate comparisons. For errors with heavy tails, the classical test becomes conservative, except when $n_1=n_2=200$ and the errors are $\itT_2$. Moreover, under $\itT_1$, the  test based on $T_{n,\cl}$ shows no power not only for the fixed alternative reported in Table \ref{tab:H0T} but also under contiguous ones,  see Figures  \ref{fig:m-linealT} and {fig:m-two-sided-linealT} below. In contrast, the robust test based on $T_{n,\rob}$  shows a stable empirical level and achieves a reasonable power under \ref{alter:m-lineal}, even when no moments exist.   Under \ref{alter:m-twosided-lineal}, $T_{n,\rob}$ looses power for heavy tailed errors with respect to the one obtained for normal ones, especially under $\itT_1$, where the power is at least five times smaller than the one obtained under normality. 

 \subsection{Performance under contiguous alternatives}{\label{sec:contiguas}}
  In this section we will  study the tests performance for contiguous alternatives.
We consider two families of contiguous alternatives. The first family corresponds to  one-sided contiguous alternatives having the form $H_{\Delta,n}^{(1)}: m_2(x)=m_1(x)+\Delta \, x/\sqrt{n}$, with $n=n_1+n_2$. The second one  has the form $H_{\Delta,n}^{(2)}: m_2(x)=m_1(x)+\Delta \, (3-6x)/\sqrt{n}$, with $n=n_1+n_2$. In both cases, we chose  $\Delta=0,2,4,6$ and $8$. Note that when $\Delta  /\sqrt{n}=1/3$, $H_{\Delta,n}^{(2)}$ equals the fixed alternative \ref{alter:m-twosided-lineal}, while if $\Delta  /\sqrt{n}=0.5$, $H_{\Delta,n}^{(1)}$ corresponds to \ref{alter:m-lineal}. 

\begin{figure}[ht!]
	\begin{center}
		\small
\renewcommand{\arraystretch}{0.1}
\newcolumntype{G}{>{\centering\arraybackslash}m{\dimexpr.33\linewidth-1\tabcolsep}}
  		\begin{tabular}{GGG}
	 $C_0$ & $C_3$ & $C_4$\\
			\includegraphics[scale=0.3]{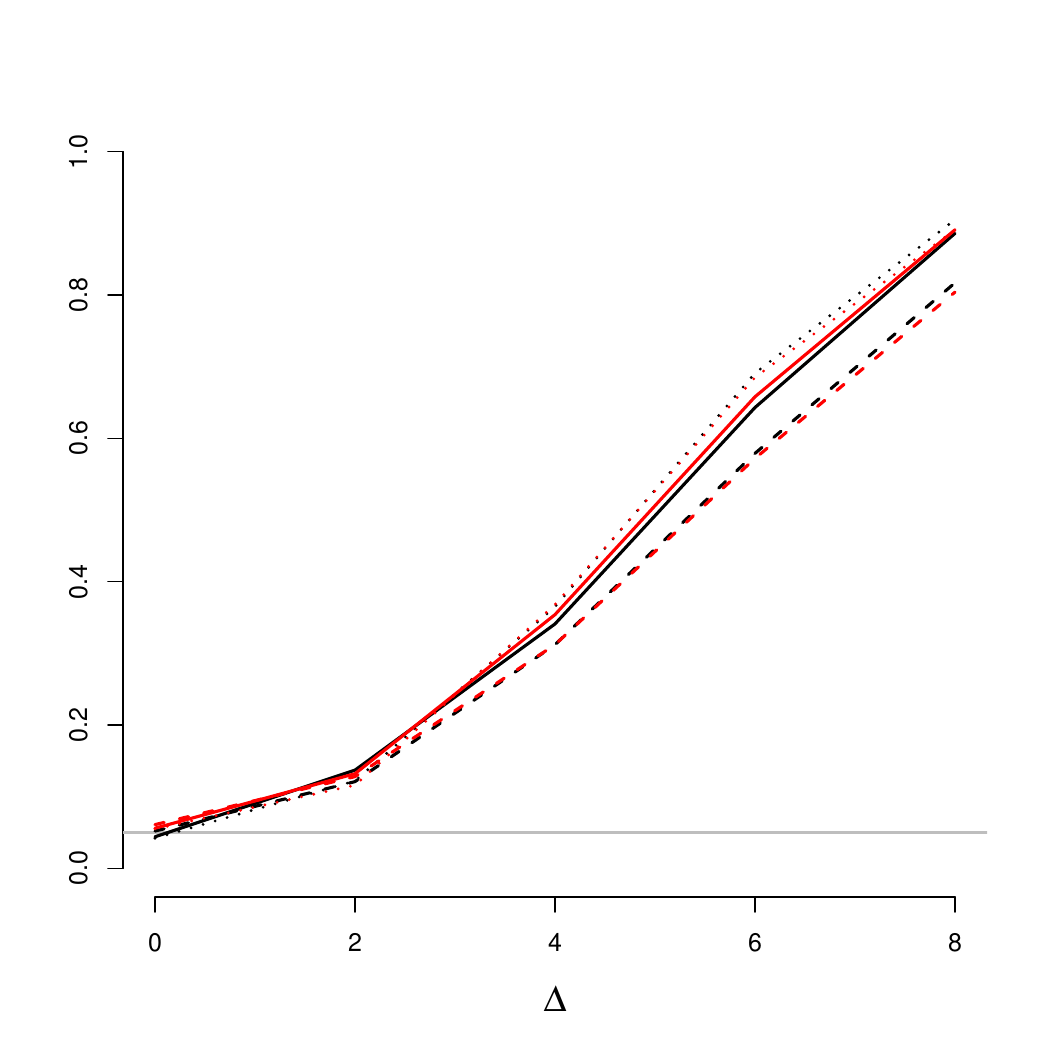} &  
			\includegraphics[scale=0.3]{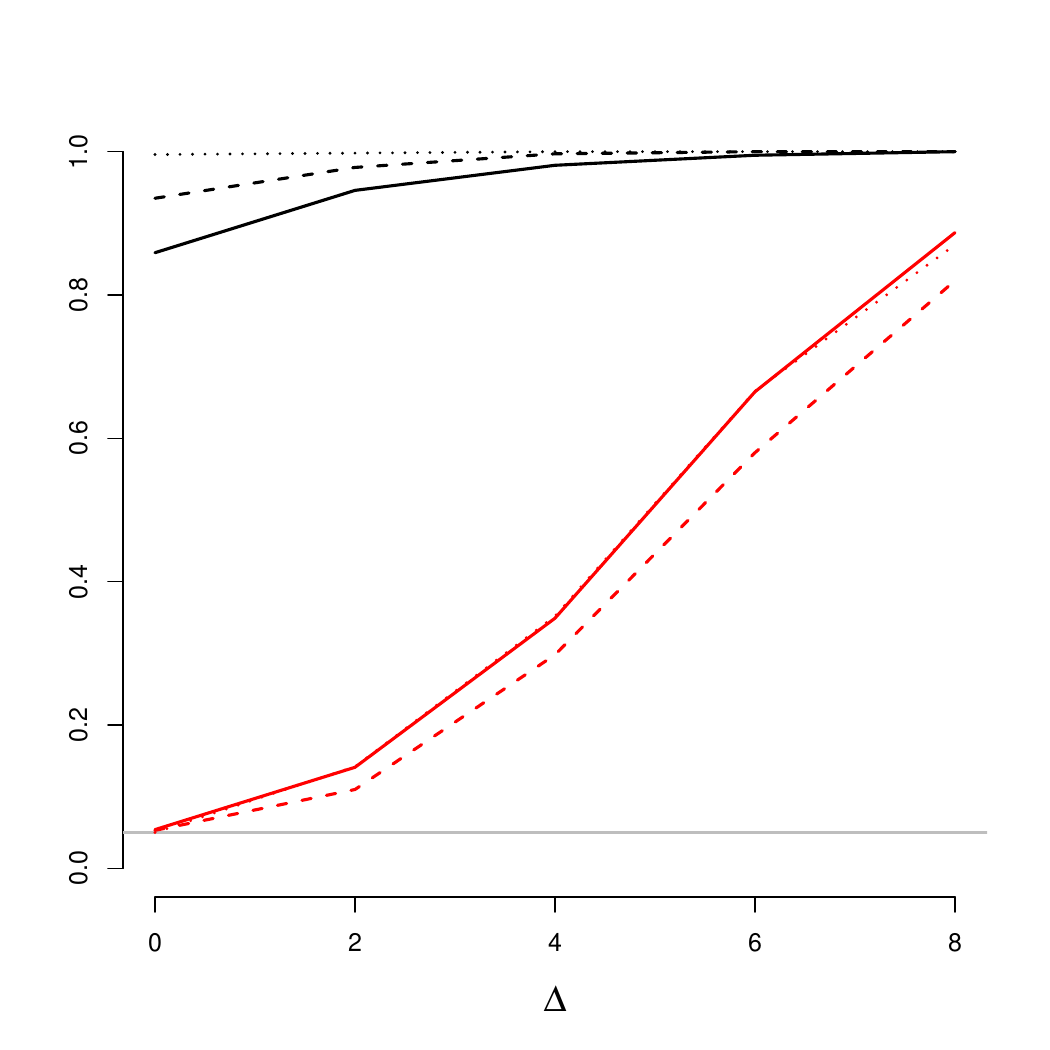} &
			\includegraphics[scale=0.3]{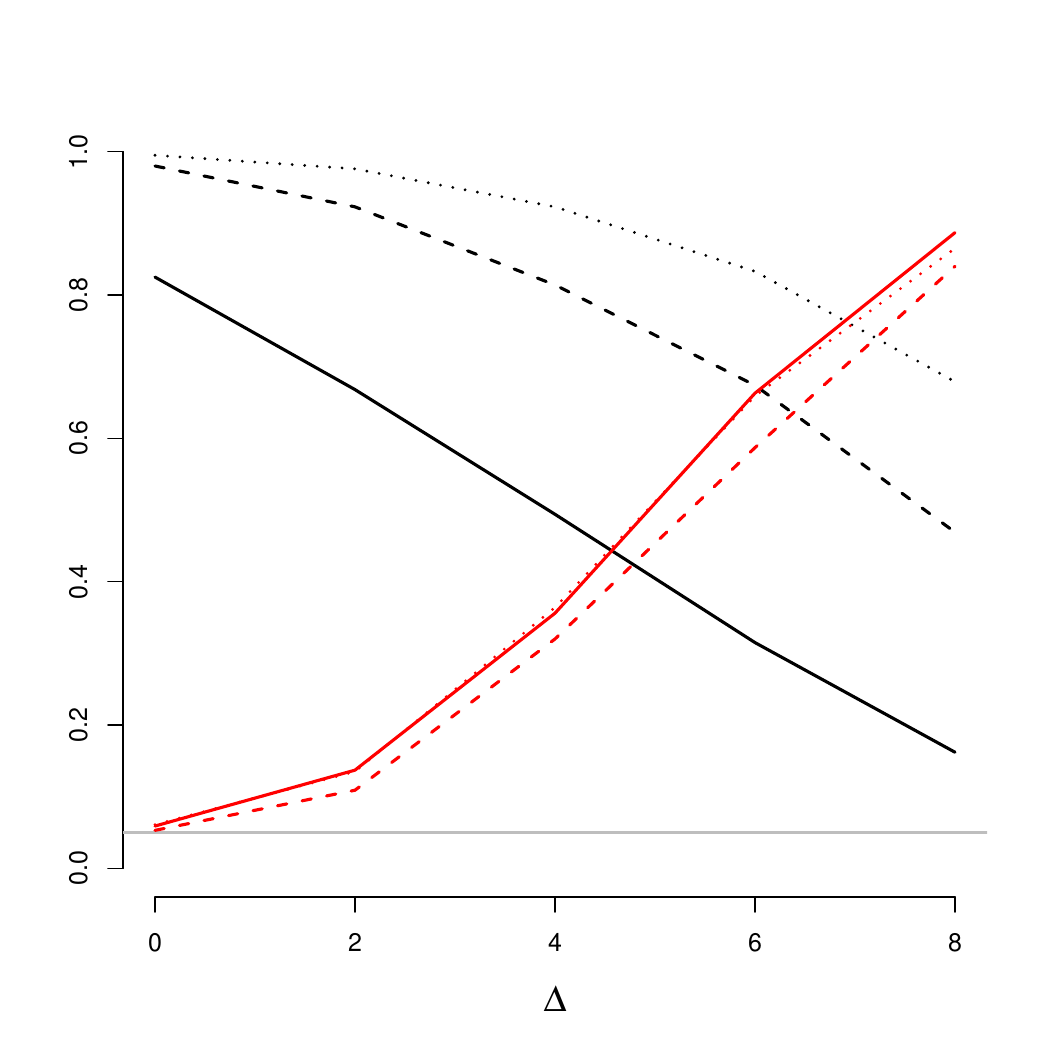} 
		\end{tabular}
		\vskip-0.1in \caption{\label{fig:m-lineal} \small Observed frequencies of rejection for clean and contaminated samples under model \ref{modelo:m-lineal} and the contiguous alternatives $H_{\Delta,n}^{(1)}$.  The solid  and dotted lines correspond to $n_1=n_2=100$ and $200$, respectively, while the dashed line to  $n_1=200$ and $n_2=100$. The frequencies of rejection of $T_{n,\cl}$ and  $T_{n,\rob}$ are given in black and red, respectively.}
	\end{center} 
\end{figure}

\begin{figure}[ht!]
	\begin{center}
		\small
\renewcommand{\arraystretch}{0.1}
\newcolumntype{G}{>{\centering\arraybackslash}m{\dimexpr.33\linewidth-1\tabcolsep}}
  		\begin{tabular}{GGG}
	 $C_0$ & $C_3$ & $C_4$\\[2ex]
	\includegraphics[scale=0.3]{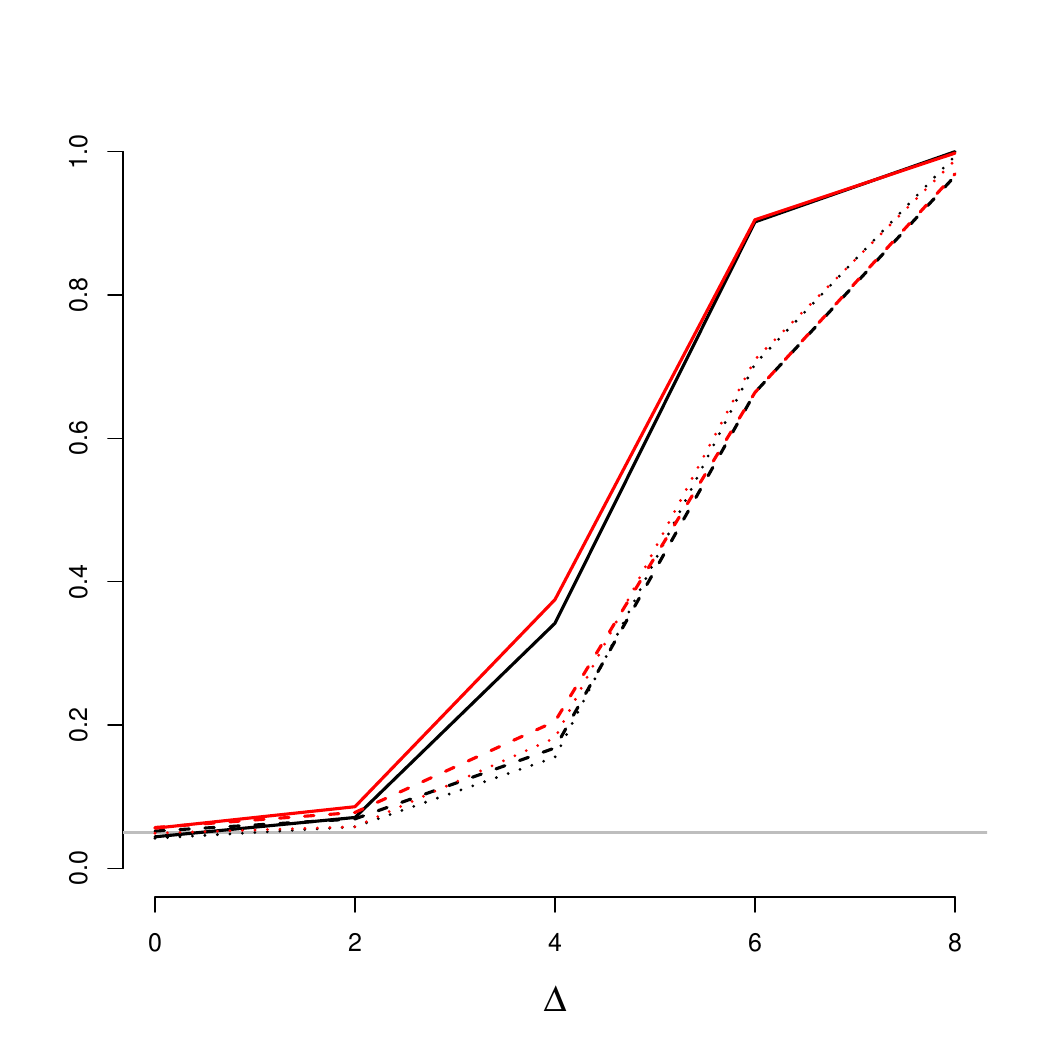} & 
	\includegraphics[scale=0.3]{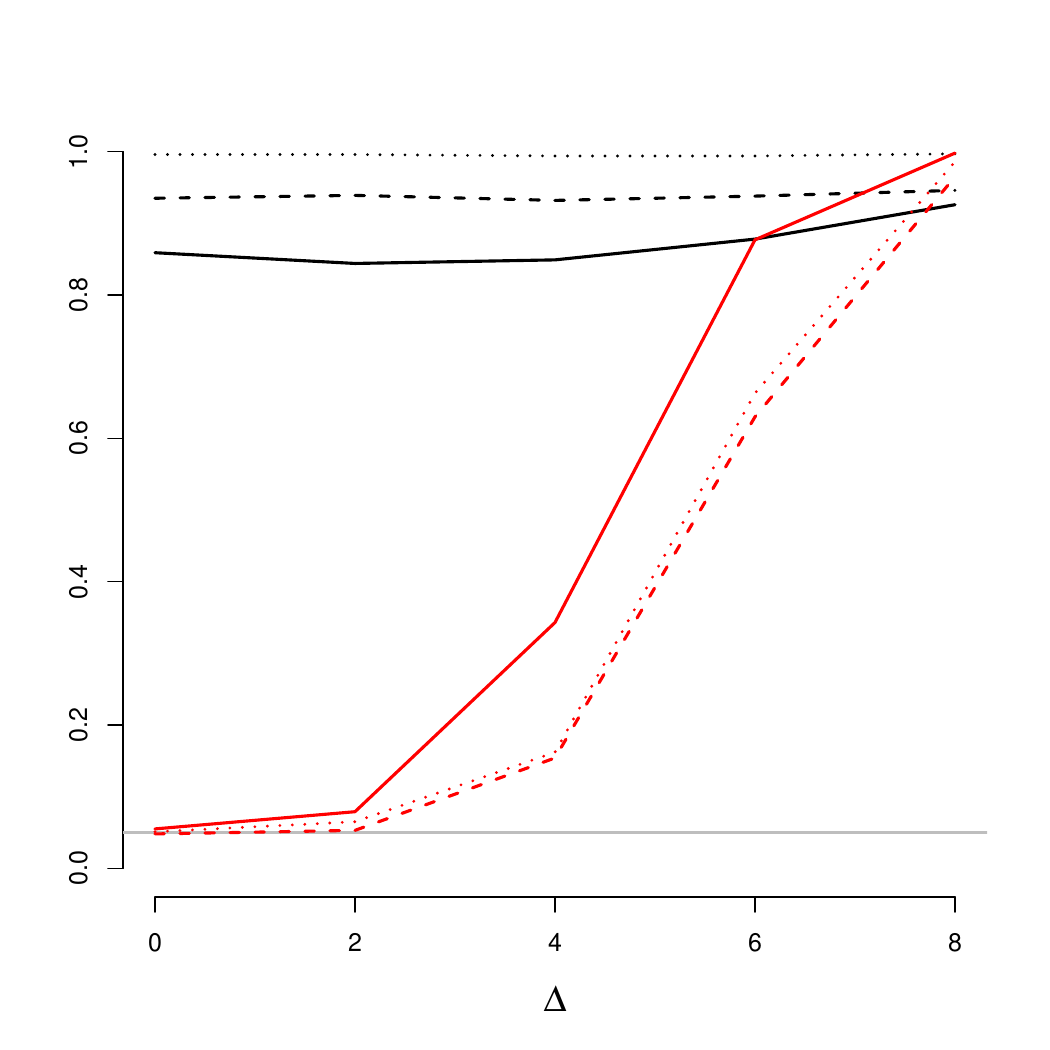} &
	\includegraphics[scale=0.3]{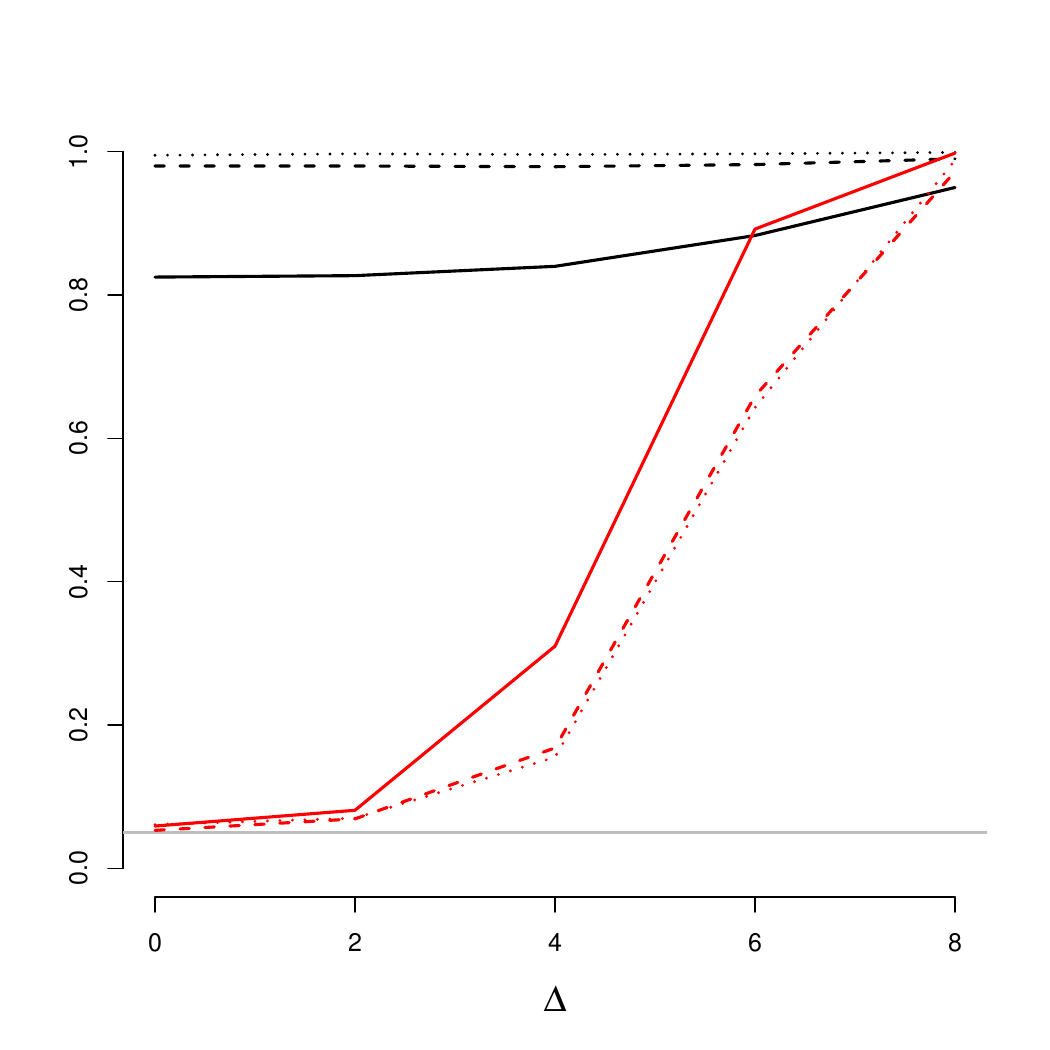} 
		\end{tabular}
		\vskip-0.1in \caption{\label{fig:m-two-sided-lineal} \small Observed frequencies of rejection for clean and contaminated samples under model \ref{modelo:m-lineal} and the contiguous alternatives $H_{\Delta,n}^{(2)}$.  The solid  and dotted lines correspond to $n_1=n_2=100$ and $200$, respectively, while the dashed line to  $n_1=200$ and $n_2=100$. The frequencies of rejection of $T_{n,\cl}$ and  $T_{n,\rob}$ are given in black and red, respectively.}
	\end{center} 
\end{figure}

\begin{figure}[ht!]
	\begin{center}
		\small
\renewcommand{\arraystretch}{0.1}
\newcolumntype{G}{>{\centering\arraybackslash}m{\dimexpr.33\linewidth-1\tabcolsep}}
  		\begin{tabular}{GGG}
	 $C_0$ & $C_3$ & $C_4$\\
			\includegraphics[scale=0.33]{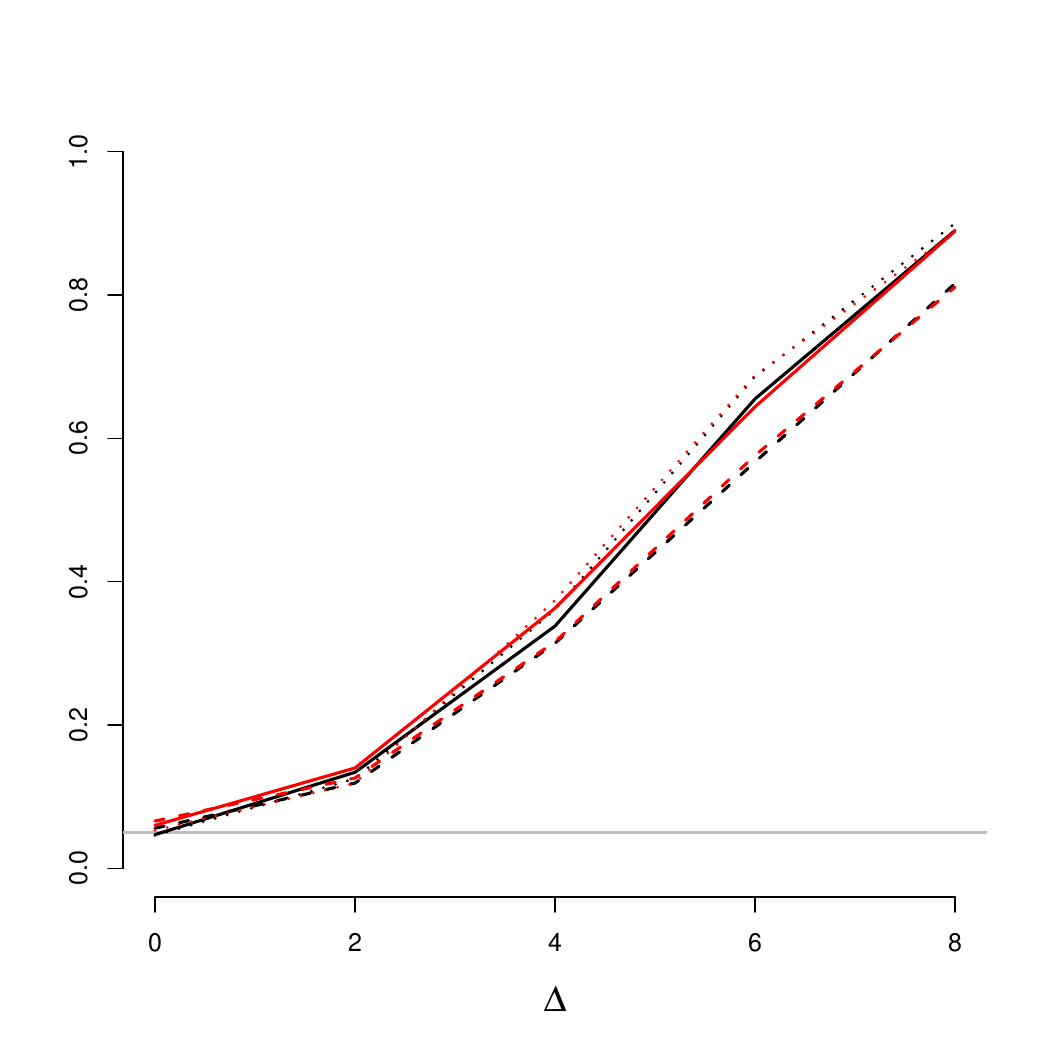} & 
			\includegraphics[scale=0.33]{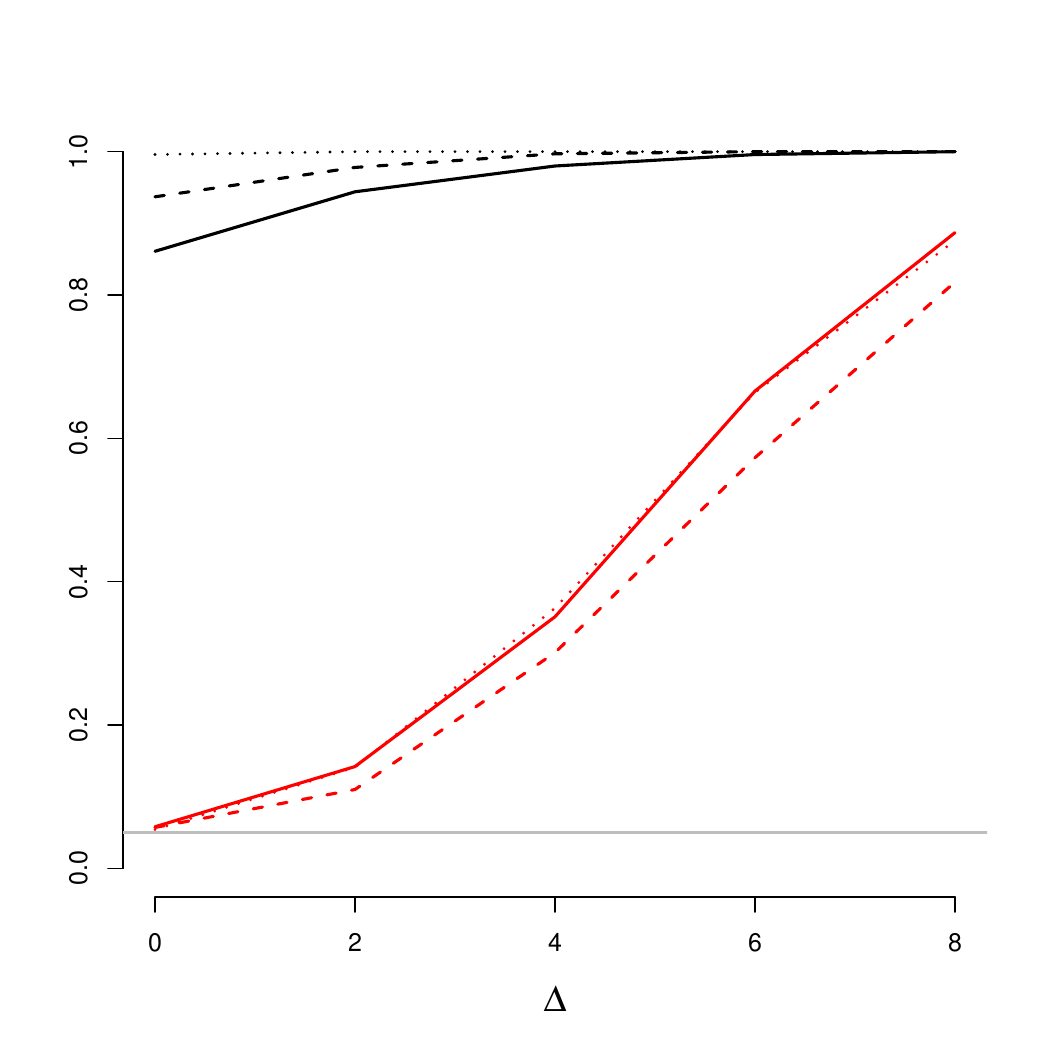} &
			\includegraphics[scale=0.33]{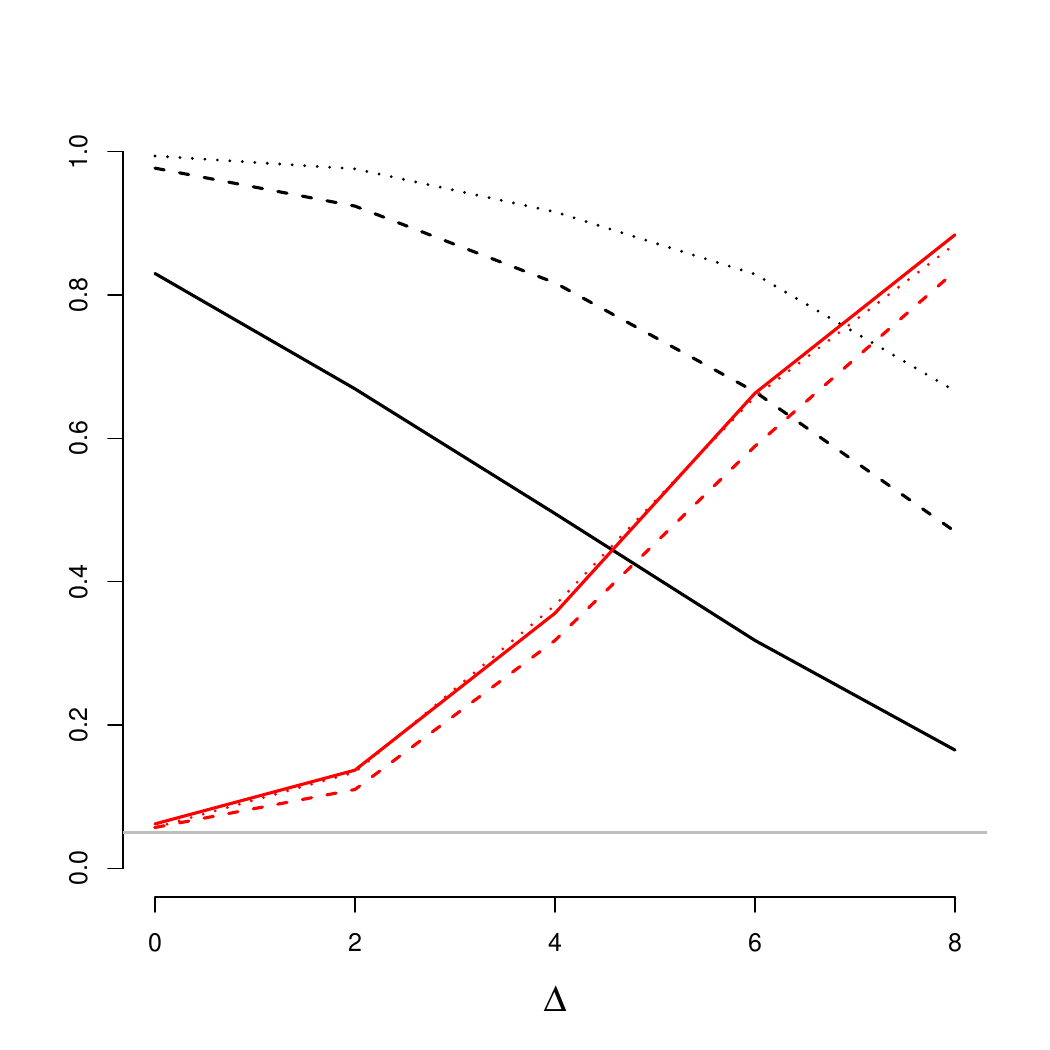} 
		\end{tabular}
		\vskip-0.1in \caption{\label{fig:m-exp} \small Observed frequencies of rejection for clean and contaminated samples under model \ref{modelo:m-exp} and the contiguous alternatives $H_{\Delta,n}^{(1)}$. The solid  and dotted lines correspond to $n_1=n_2=100$ and $200$, respectively, while the dashed line to  $n_1=200$ and $n_2=100$. The frequencies of rejection of $T_{n,\cl}$ and  $T_{n,\rob}$ are given in black and red, respectively.}
	\end{center} 
\end{figure}

 The results for all models are quite similar and for that reason, we only report here the power performance   under model  \ref{modelo:m-lineal}, while for model \ref{modelo:m-exp} we only report the results under  the set of  alternatives $H_{\Delta,n}^{(1)}$.
When considering model  \ref{modelo:m-lineal}, the observed frequencies of rejection  for clean samples and for samples generated under the contamination schemes $C_3$ and $C_4$, are displayed in Figures  \ref{fig:m-lineal} and \ref{fig:m-two-sided-lineal} for the families of   contiguous alternatives $H_{\Delta,n}^{(1)}$ and   $H_{\Delta,n}^{(2)}$, respectively, while for model  \ref{modelo:m-exp} the results under the set of alternative $H_{\Delta,n}^{(1)}$ are given in Figure \ref{fig:m-exp}.
 The solid  and dotted lines correspond to $n_1=n_2=100$ and $200$, respectively, while the dashed line to the unbalanced setting $n_1=200$ and $n_2=100$. Besides, we display in black the frequency curves corresponding to  $T_{n,\cl}$ and in red those obtained with  $T_{n,\rob}$.  

The left panels of Figures  \ref{fig:m-lineal}  to \ref{fig:m-exp} illustrate  that both procedures have a similar performance under $C_0$ with a small loss of power when unbalanced designs are considered. The two contaminations considered do not affect the robust test introduced in this paper that still provides reliable results.   Regarding the  performance  of the test based on $T_{n,\cl}$  under contamination, different behaviours can be described. When gross vertical outliers are introduced in both populations, the test becomes non--informative under the family of alternatives $H_{\Delta,n}^{(1)}$ with an almost constant frequency of rejection. The same effect on $T_{n,\cl}$  is observed under $C_4$ when considering the two--sided local alternatives $H_{\Delta,n}^{(2)}$. When considering the one--sided alternatives $H_{\Delta,n}^{(1)}$ and the contamination scheme $C_4$, a Hauck--Donner effect may be observed, since its  power  decreases almost to the level of significance as the alternative  moves away  from the null hypothesis, when $n_1=n_2=100$. We guess that the same effect would be observed for the other sample sizes when larger values of $\Delta$ are considered. In contrast, the two contaminations  $C_3$ and $C_4$  considered do not affect the robust test introduced in this paper that still provides reliable results.

\begin{figure}[ht!]
	\begin{center}
		\small
\renewcommand{\arraystretch}{0.1}
\newcolumntype{G}{>{\centering\arraybackslash}m{\dimexpr.33\linewidth-1\tabcolsep}}
  		\begin{tabular}{GGG}
	 $N(0,1)$ & $\itT_2$ & $\itT_1$\\
			\includegraphics[scale=0.3]{grafico_contiguas_C0_modelo-M2.pdf} & 
			\includegraphics[scale=0.3]{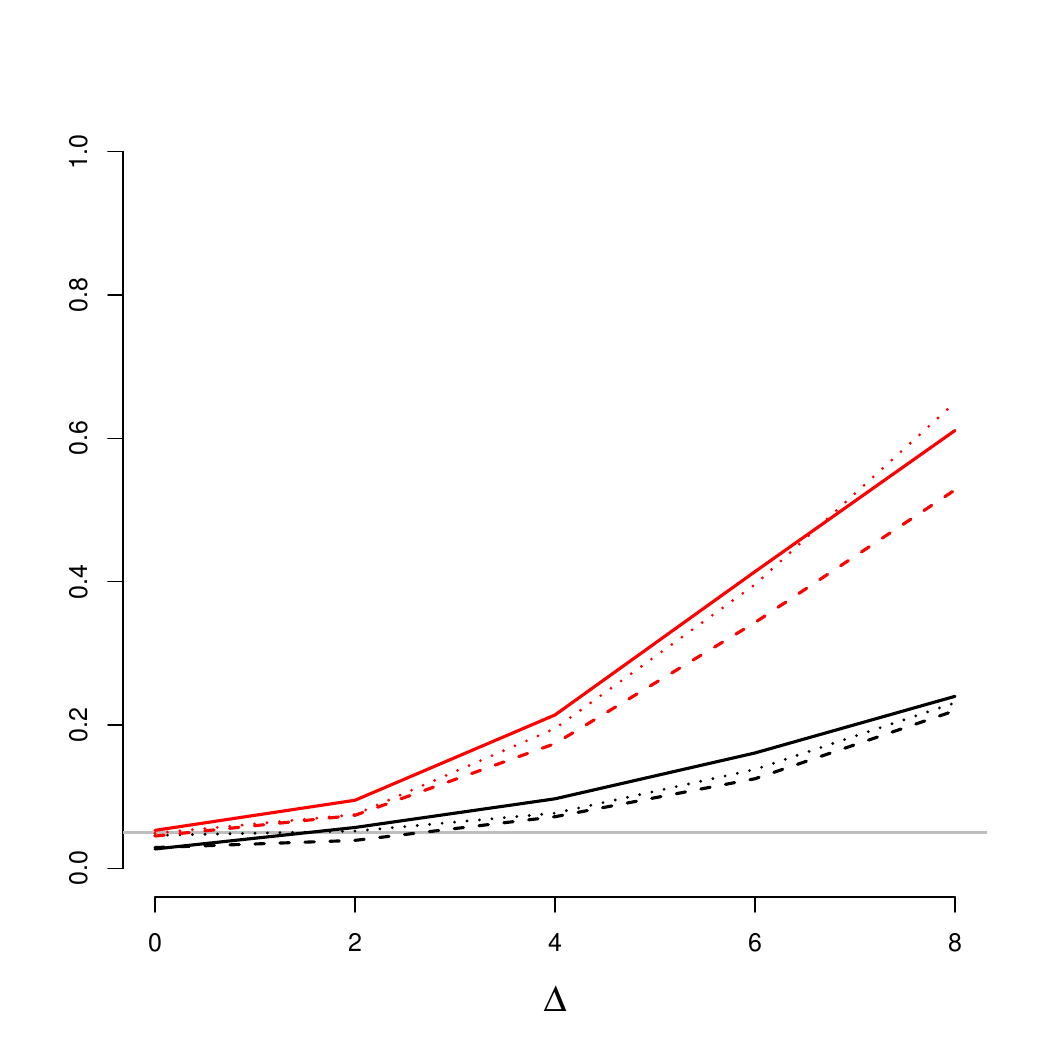} &
			\includegraphics[scale=0.3]{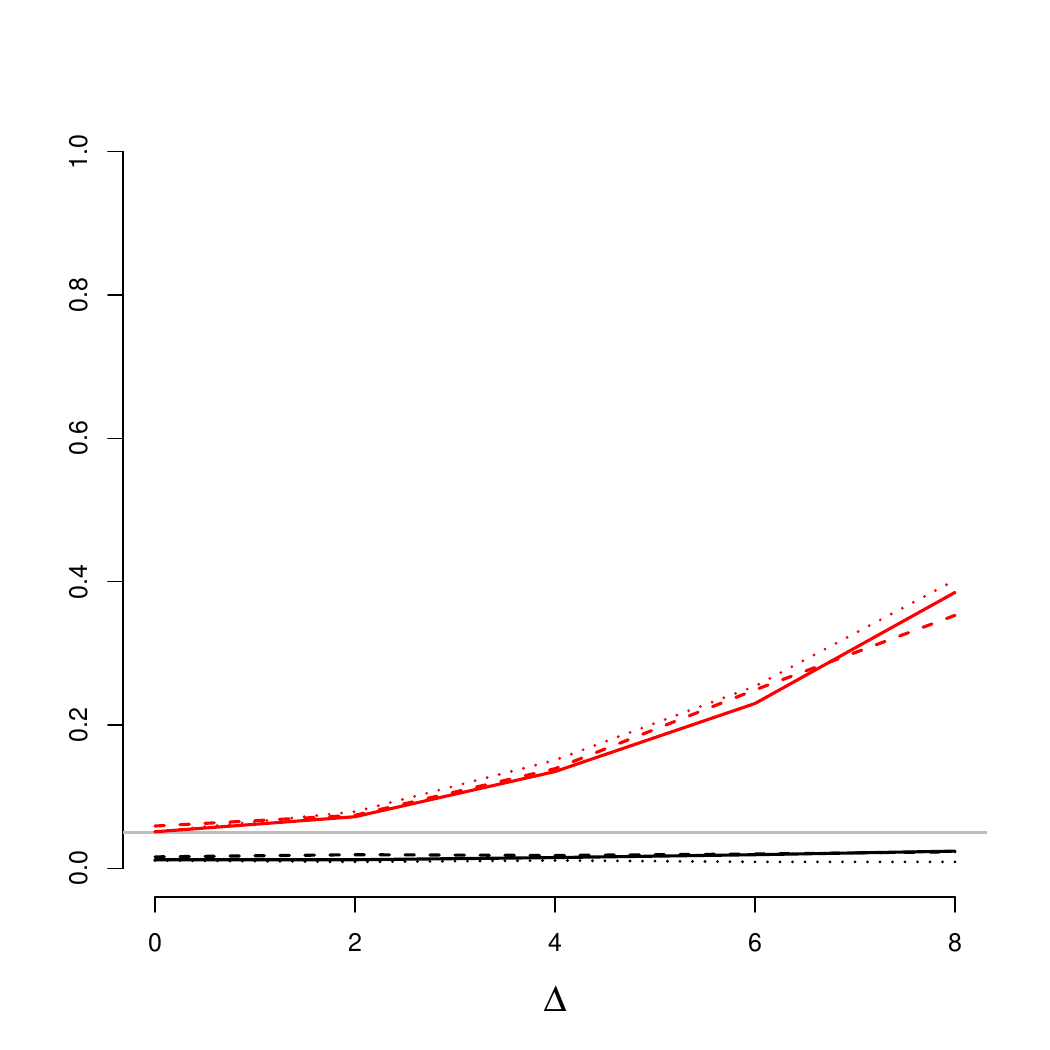} 
		\end{tabular}
		\vskip-0.1in \caption{\label{fig:m-linealT} \small Observed frequencies of rejection for samples with normal and heavy tailed errors under model \ref{modelo:m-lineal}  and the contiguous alternatives $H_{\Delta,n}^{(1)}$.  The solid  and dotted lines correspond to $n_1=n_2=100$ and $200$, respectively, while the dashed line to  $n_1=200$ and $n_2=100$. The frequencies of rejection of $T_{n,\cl}$ and  $T_{n,\rob}$ are given in black and red, respectively.}
	\end{center} 
\end{figure}

\begin{figure}[ht!]
	\begin{center}
		\small
\renewcommand{\arraystretch}{0.1}
\newcolumntype{G}{>{\centering\arraybackslash}m{\dimexpr.33\linewidth-1\tabcolsep}}
  		\begin{tabular}{GGG}
	 $N(0,1)$ & $\itT_2$ & $\itT_1$\\[2ex]
	\includegraphics[scale=0.3]{grafico_contiguas_C0_modelo-MA5.pdf} &  
	 	\includegraphics[scale=0.3]{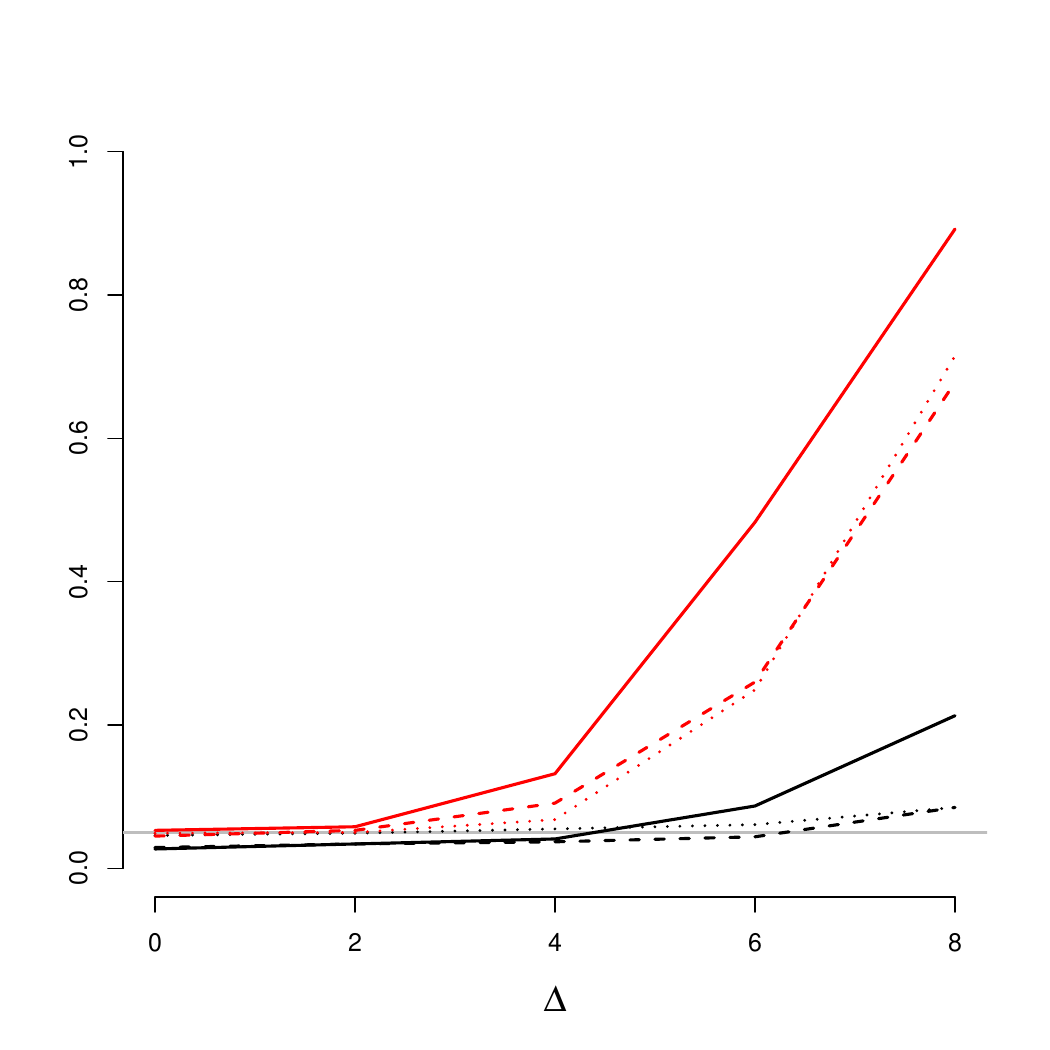} & 
	 	\includegraphics[scale=0.3]{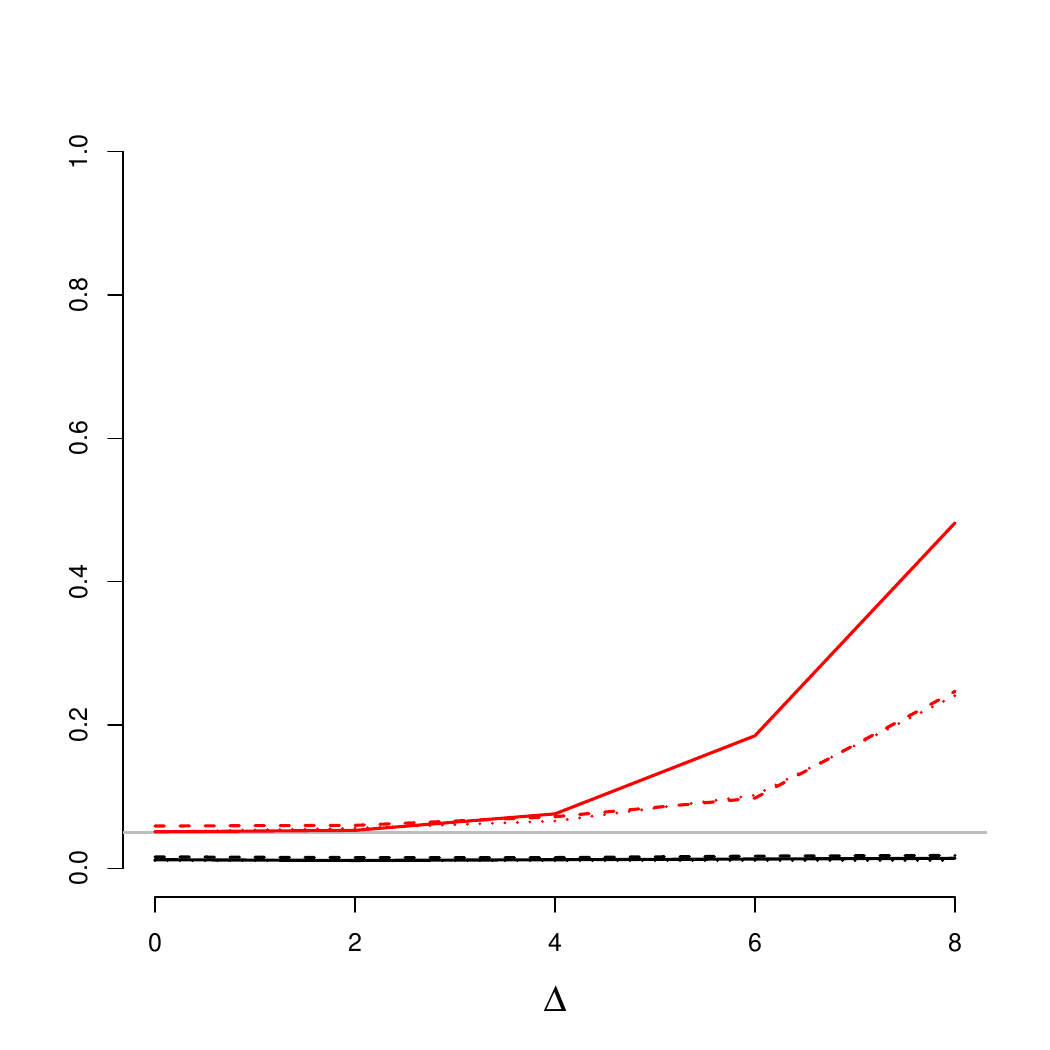} 
		\end{tabular}
		\vskip-0.1in \caption{\label{fig:m-two-sided-linealT} \small Observed frequencies of rejection for samples with normal and heavy tailed errors under model \ref{modelo:m-lineal}  and the contiguous alternatives $H_{\Delta,n}^{(2)}$.  The solid  and dotted lines correspond to $n_1=n_2=100$ and $200$, respectively, while the dashed line to  $n_1=200$ and $n_2=100$. The frequencies of rejection of $T_{n,\cl}$ and  $T_{n,\rob}$ are given in black and red, respectively.}
	\end{center} 
\end{figure}

 Figures \ref{fig:m-linealT} and  \ref{fig:m-two-sided-linealT} display   the corresponding frequencies of rejection when the errors are heavy tailed.
 To facilitate comparisons the left panel in both  Figures repeats the plot for normal errors already displayed in Figures  \ref{fig:m-lineal}  and \ref{fig:m-two-sided-lineal}.  When the errors have a $\itT_2$ distribution, the classical test shows a clear lack of power
underperforming the robust method. For Cauchy errors, the classical method shows no power, as already described for fixed alternatives in Table \ref{tab:H0T} making the test unreliable. With respect to the behaviour of the robust test, even though some loss of power is observed, specially when the errors have a Cauchy distribution, the  test  still provides reliable results, since the empirical level is not affected (see Table \ref{tab:H0T}) and the power increases with $\Delta$.

\section{Real data analysis}{\label{sec:datos}}

In environmental studies the relation between rainfall and acid rain has been studied to decide the pollution impact. In this section, we consider a data set that was previously studied in \citet{hall:hart:1990} and \citet{Neumeyer:dette:2003} which contains, the week, the amount of rainfall and the logarithm of the sulfate concentration  along   a five-year period 1979-1983  in two locations of North Carolina, Coweeta and  Lewiston.  For some weeks, data are not available, so we only have information   on 
215 weeks  in Lewiston and  on  220 weeks  in Coweeta.
As mentioned in \citet{hall:hart:1990}  the data were part of the National Atmospheric Deposition Program.  Both \citet{hall:hart:1990} and \citet{Neumeyer:dette:2003} used the data to   compare the logarithm of acidity, i.e., the logarithm of the sulfate concentration  previously adjusted  for the
amount of rainfall  as a function of time in the two locations.  In our analysis we are instead interested in the relation between the logarithm of the sulfate concentration  and the rainfall, that is, the response variable is the logarithm of the sulfate concentration which was modelled nonparametrically as a function of the rainfall. From now on the observations corresponding to Coweeta are identified as $(X_{i1}, Y_{i1})\trasp$ and those of Lewiston as $(X_{i2}, Y_{i2})\trasp$, so that we deal with the regression model \eqref{modelo}.

\begin{figure}[ht!]
\begin{center}
\begin{tabular}{cc}
	(a) Coweeta & (b) Lewiston\\[-0.2in]
			\includegraphics[scale=0.36]{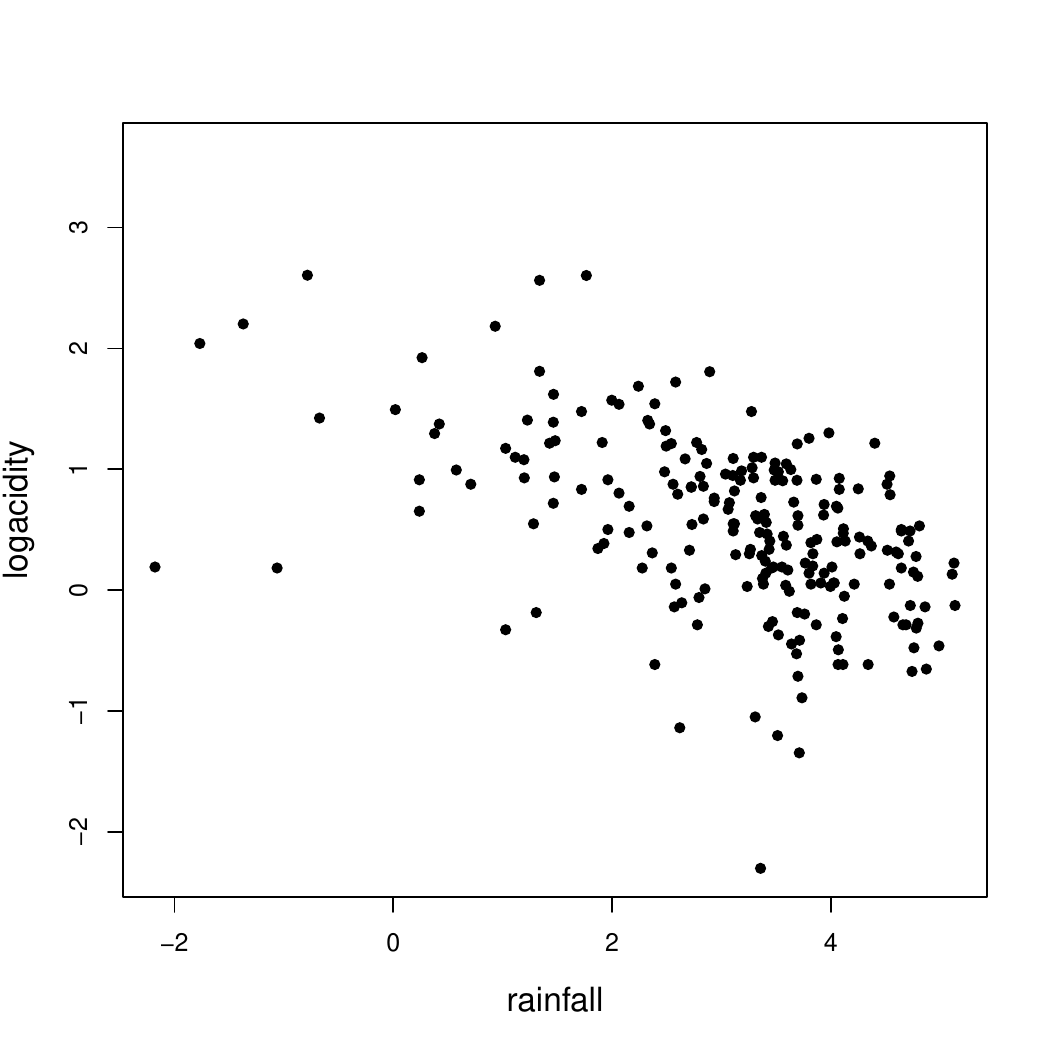} &
			\includegraphics[scale=0.36]{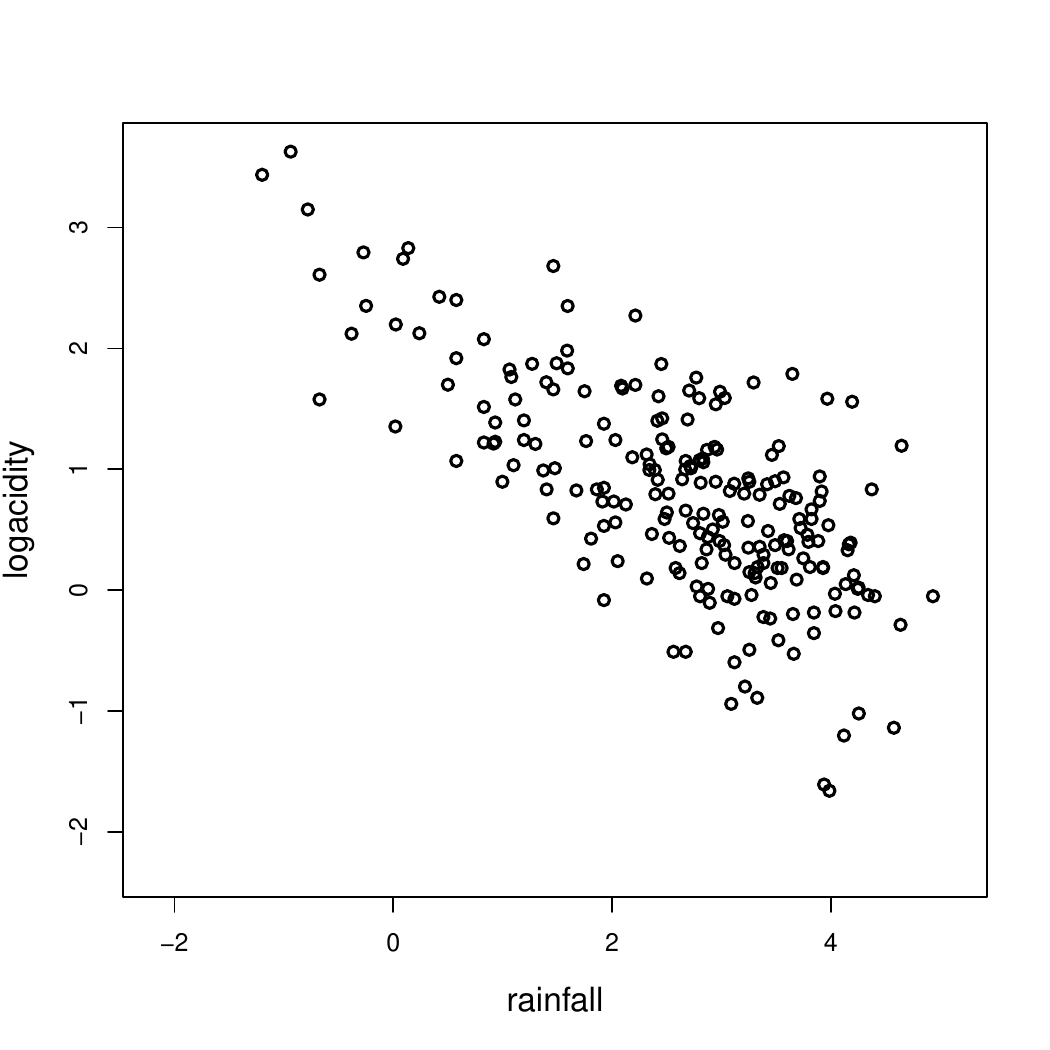} \\[-0.1in]			
			\multicolumn{2}{c}{	(c)}\\[-0.2in]
			\multicolumn{2}{c}{			\includegraphics[scale=0.36]{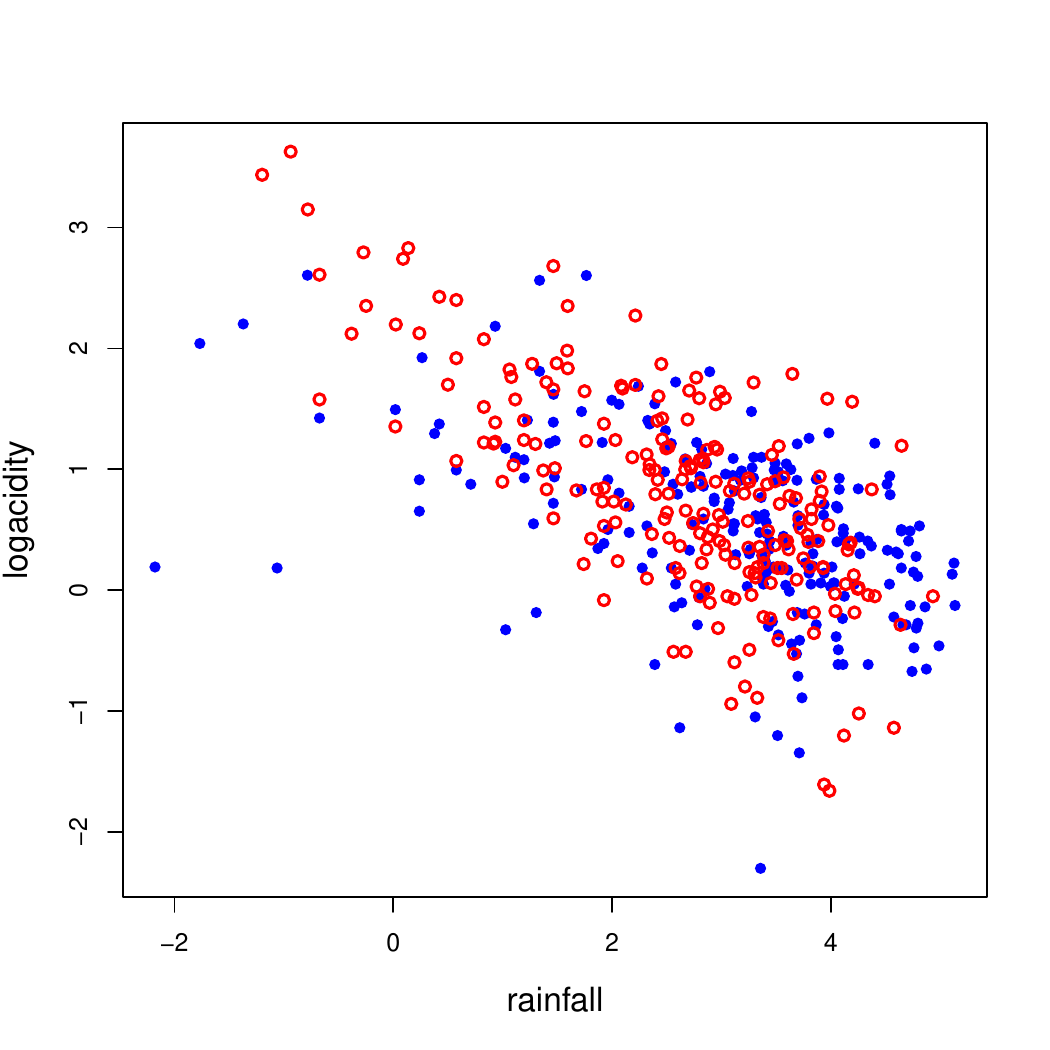} }
\end{tabular}
		\vskip-0.1in \caption{\label{fig:datos} \small Scatterplot of the logarithm of the sulfate concentration  (\lq \lq logacidity\rq \rq) versus rainfall. Panel (a) corresponds to the data recorded at Coweeta, while panel (b) to those in  Lewiston. The lower panel (c) the observations at both locations are plotted jointly, the blue filled points and the red circles  correspond to the measurements in  Coweeta and   Lewiston, respectively. }
	\end{center} 
\end{figure}

Figure \ref{fig:datos} displays the observations corresponding to Coweeta and Lewiston. The upper plot presents the data in separate panels, while in the lower one the observations corresponding to Coweeta are shown in blue filled points  and those related to Lewiston as red circles.

\begin{figure}[ht!]
\begin{center}
\begin{tabular}{cc}
	(a) Coweeta & (b) Lewiston\\[-0.2in]
			\includegraphics[scale=0.36]{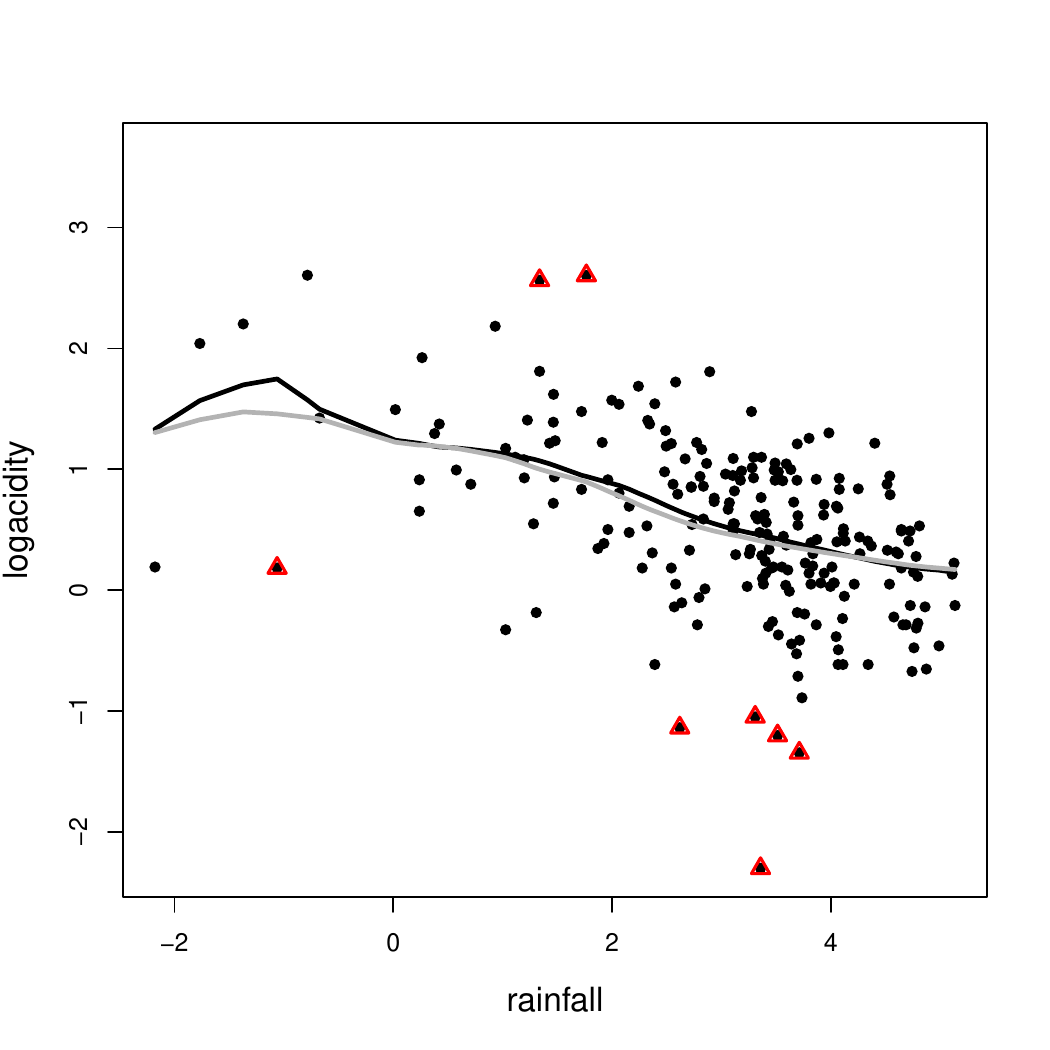} &
			\includegraphics[scale=0.36]{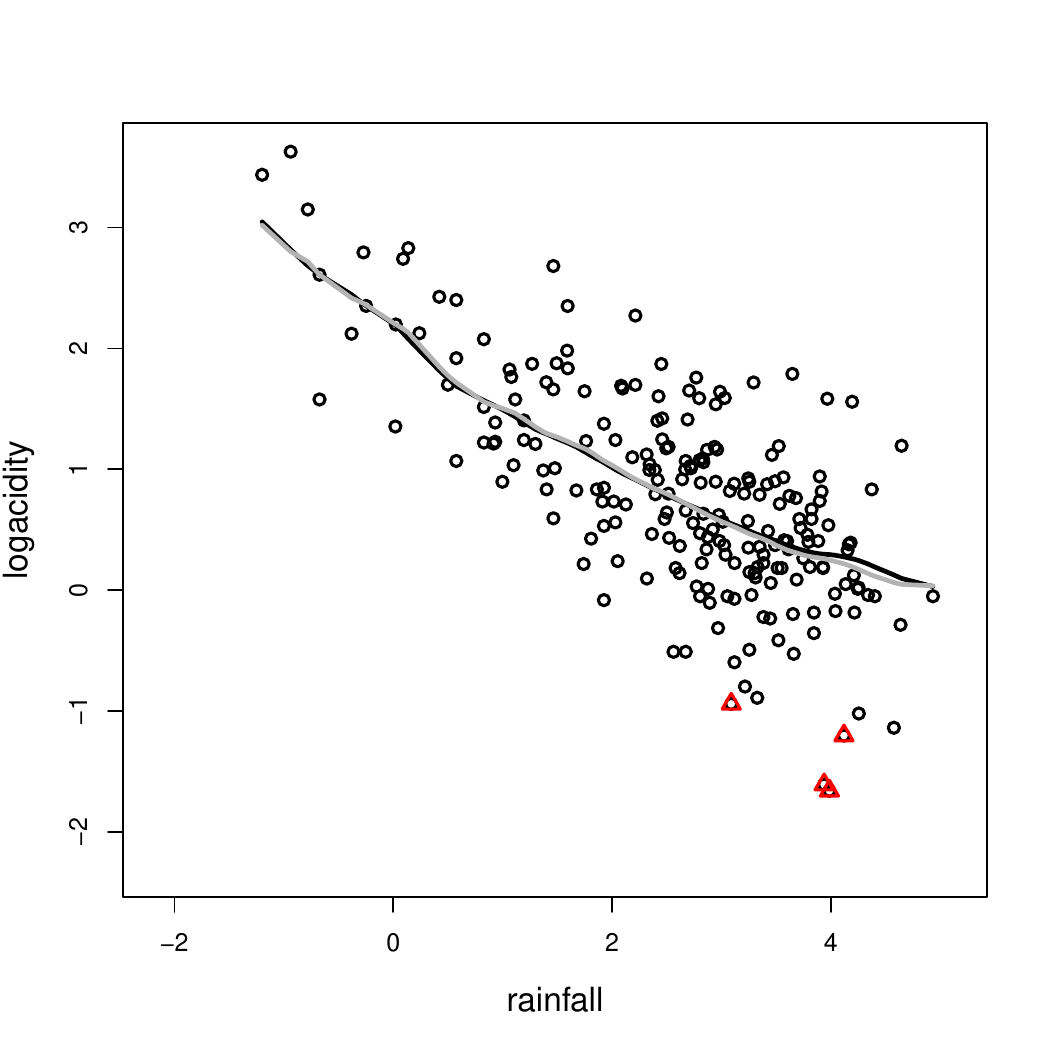} 
\end{tabular}
		\vskip-0.1in \caption{\label{fig:datos-ajuste} \small Scatterplot of the logarithm of the sulfate concentration  (\lq \lq logacidity\rq \rq) versus rainfall together with the Nadaraya--Watson and robust smoothers in gray and black lines, respectively. The red triangles highlight the observations detected as atypical by the robust fit. Panel (a) corresponds to the data recorded at Coweeta, while panel (b) to those in  Lewiston.}
	\end{center} 
\end{figure}

The fits obtained for each city using the classical and robust smoothers are given in Figure  \ref{fig:datos-ajuste} together with the observations detected as atypical (in red triangles) using the boxplot of the residuals from the robust fit. The main differences between the two fits are observed in Coweeta for low values of rainfall. 
For the Nadaraya--Watson estimator the cross--validation bandwidths equal $h_1=1.6$ and $h_2=0.8$, while when using a local $M-$smoother and a robust cross-validation criterion, we obtain $h_1=1.3$ and $0.9$. The classical test statistic proposed in \citet{Pardo:etal:2015} rejects at level 0.05 the null hypothesis with a $p-$value equal to $0.0496$, while the robust procedure does not detect differences between both locations ($p-$value=  0.1117). 

\begin{figure}[ht!]
\begin{center}
\begin{tabular}{cc}
(a)   & (b)  \\[-0.1in]
			\includegraphics[scale=0.4]{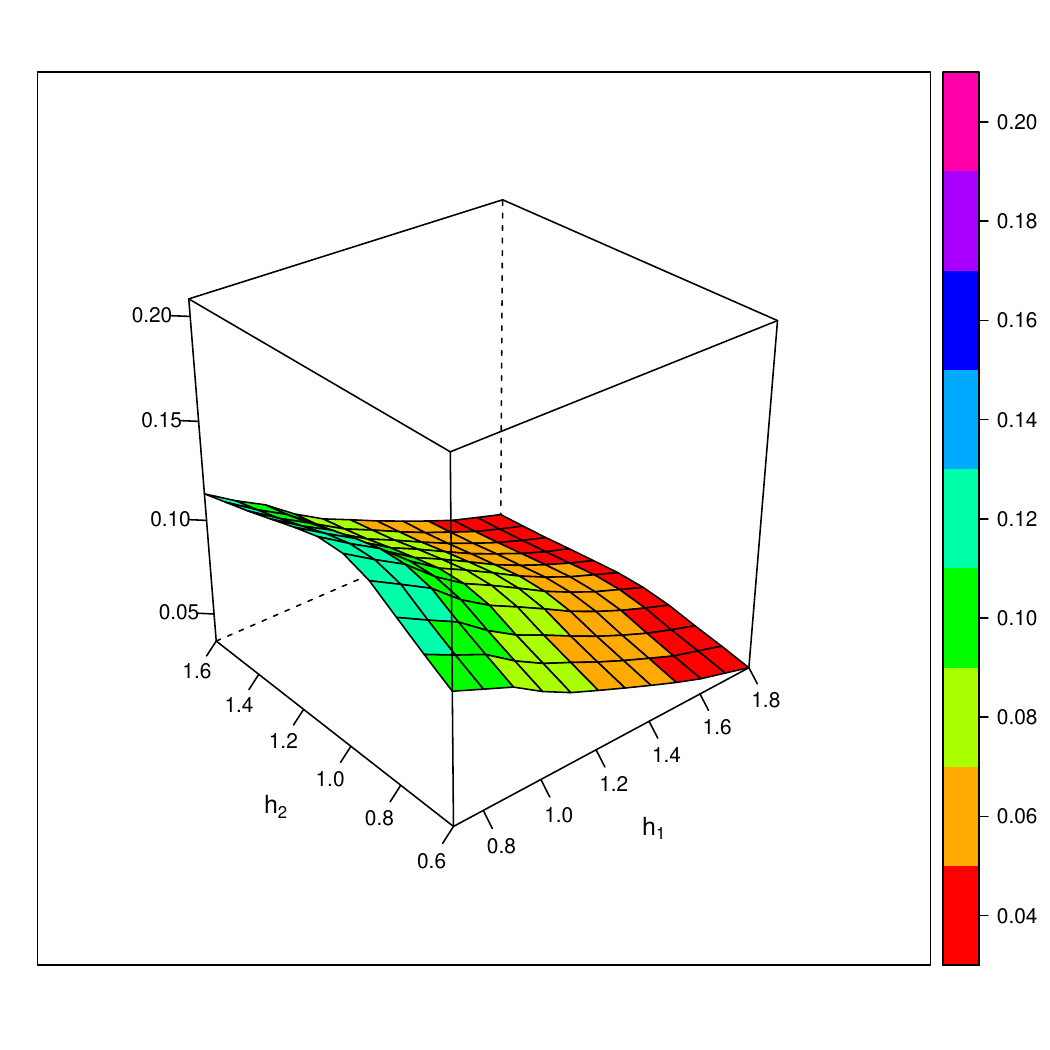} &
			\includegraphics[scale=0.4]{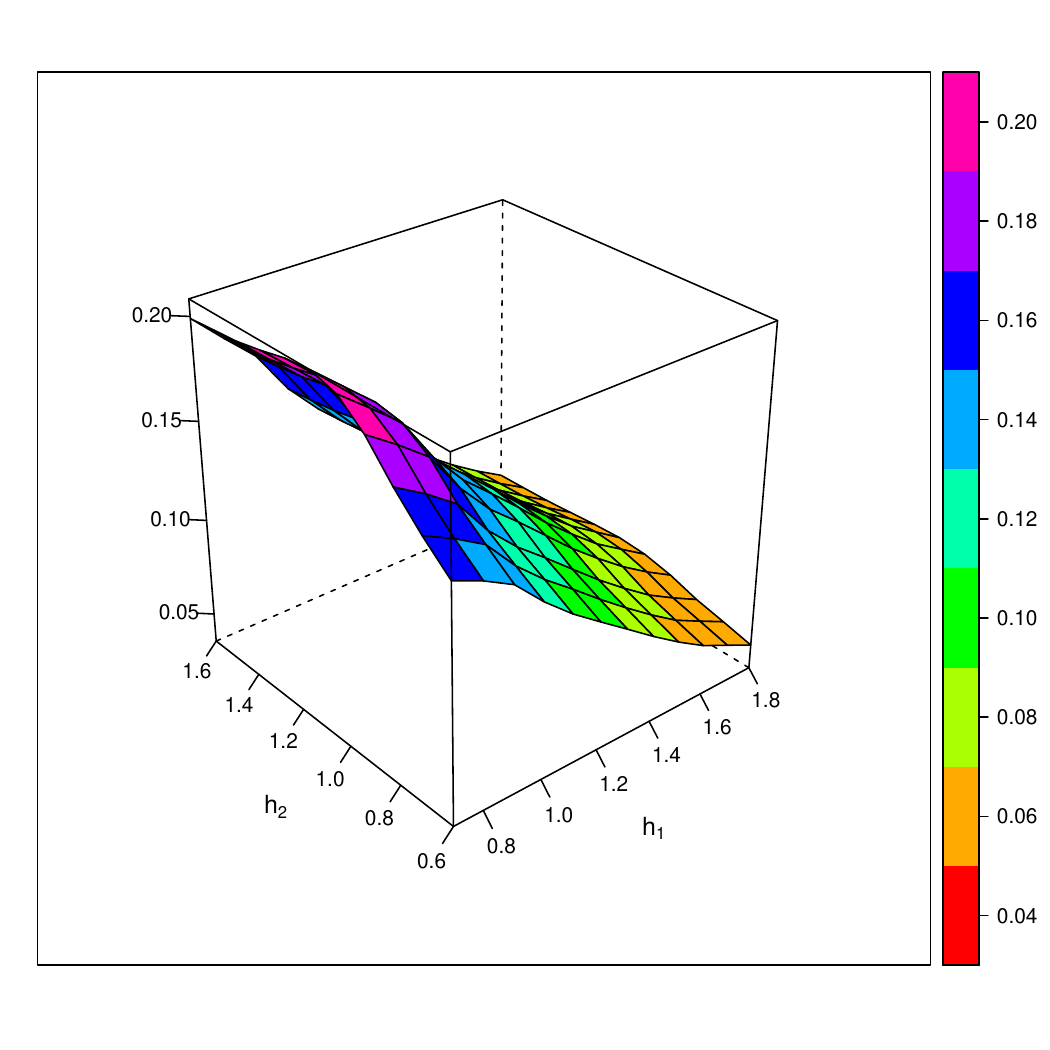} 
\end{tabular}
		\vskip-0.1in 
		\caption{\label{fig:datos-pvalores} \small $p-$values as a function of the bandwidths used to estimate the regression functions. Panel (a) corresponds to the  $p-$values obtained when using the procedure proposed in \citet{Pardo:etal:2015} and panel (b) to  those of the robust testing method defined in this paper.}
	\end{center} 
\end{figure}

To detect the possible influence of the bandwidth choice on the resulting $p-$value, we choose a grid of values for $h_1$ and $h_2$ ranging in the range 0.7 to 1.8 and 0.6 to 1.6, respectively, with a step of 0.1. Figure \ref{fig:datos-pvalores} displays the surface of the obtained $p-$values. The left panel corresponds to the  classical  procedure   which is based on empirical characteristic functions using the residuals from  the Nadaraya--Watson  smoother, while the right one to the method proposed in this paper. The obtained surfaces show  that the decision taken by the test based on the statistic $T_{n,\rob}$ is less dependent to the bandwidth choice, while the classical one leads to $p-$values varying from  $ 0.035$ to $0.132$ changing the decision at 5\% level. This effect can be explained by the effect that the observations, whose residuals from the robust fit are detected as outliers by the boxplot, have on the classical procedure.  

\section{Final Comments}\label{sec:conclusiones}

In this paper we proposed and studied a new robust procedure to test  equality of several regression curves in a nonparametric setup, which detects alternatives converging to the null hypothesis at the parametric rate $n^{-1/2}$. Our proposal adapts the ideas in  \citet{Pardo:etal:2015} by considering the empirical characteristic functions of the residuals obtained from a robust fit.  In this way,   first moment conditions for the errors distribution  are avoided.  The robust procedure introduced does not assume that the design points have the same density. Simulations have shown a good practical behaviour of the new test under different regression models and contamination settings. If no outliers are present in the sample, the behaviour of the new test is almost equal to that of the procedure given in  \citet{Pardo:etal:2015}, but when outliers appear in the samples, the robust test clearly outperforms the latter. The influence of the smoothing parameter on the test $p-$values is also studied on a real data set, revealing that the robust testing procedure is more stable with respect to the bandwidth choice.

\vskip0.1in
 {\small
\noi\textbf{Acknowledgements.}  This research was partially supported by Grants 20020170100022BA from the Universidad de Buenos Aires and \textsc{pict}  2021-I-A-00260  from \textsc{anpcyt} at Buenos Aires, Argentina (G. Boente) and also by the Spanish Projects from the Ministry of Science and Innovation (MCIN/ AEI /10.13039/501100011033) {PID2020-116587GB-I00. CoDyNP} (G. Boente) and PID2020-118101GB-I00 (J. C. Pardo-Fern\'andez). The research was begun while G. Boente was visiting the Universidade de Vigo supported by this last project.}
 
 \setcounter{section}{0}
\renewcommand{\thesection}{\Alph{section}}

\setcounter{equation}{0}
\renewcommand{\theequation}{A.\arabic{equation}}

\section{Appendix: Proofs}{\label{sec:appen}}

The following Lemma states an asymptotic distribution result that will be useful in the proof of Theorem \ref{teo:theorem1}.

\begin{lemma}{\label{lemma:lema1}}
Assume that \eqref{modelo} and \ref{ass:A2} to \ref{ass:A4}, \ref{ass:A6} and \ref{ass:A8a}) hold. Define 
 $$  
 Z_{n,j}  =  \frac{1}{\sigma_j}  \pi_j^{1/2} \sum_{s=1}^k  \frac{\sigma_s}{\nu_{s}}   \pi_s ^{1/2}   \, \frac{1}{\sqrt{n_s}} \sum_{r=1}^{n_s}  W_j(X_{sr} )   \frac{f_j(X_{sr} )}{f( X_{sr} ) }  \psi_s\left( \varepsilon_{sr}\right)    - \frac{1}{\nu_j }  \esp\left(W_j(X_{j})\right) \frac{1}{ \sqrt{n_j}} \sum_{s=1}^{n_j}      \psi_j\left( \varepsilon_{js}\right)\,.$$
 Then,    $\bZ_n=(Z_{n,1}, \dots,Z_{n,k})\trasp \convdist  N(\bcero, \bSi)$  where $\bSi$ is defined in Theorem \ref{teo:theorem1}.
 \end{lemma}
  
  \begin{proof}
  Note that
  \begin{align*}
 Z_{n,j}& =    \frac{1}{\sigma_j}  \pi_j^{1/2} \sum_{s\ne j}  \frac{\sigma_s}{\nu_{s}}   \pi_s ^{1/2}   \, \frac{1}{\sqrt{n_s}} \sum_{r=1}^{n_s}  W_j(X_{sr} )   \frac{f_j(X_{sr} )}{f( X_{sr} ) }  \psi_s\left( \varepsilon_{sr}\right)  \\
 & +      \frac{\pi_j }{\nu_{j}}   \, \frac{1}{\sqrt{n_j}} \sum_{r=1}^{n_j}  W_j(X_{jr} )   \frac{f_j(X_{jr} )}{f( X_{jr} ) }  \psi_j\left( \varepsilon_{jr}\right)    - \frac{1}{\nu_j }  \esp\left(W_j(X_{j})\right) \frac{1}{ \sqrt{n_j}} \sum_{s=1}^{n_j}      \psi_j\left( \varepsilon_{js}\right)\\
 &  =     \sum_{s\ne j}  \frac{\sigma_s}{\sigma_j }   \frac{\pi_j^{1/2}\pi_s ^{1/2}}{\nu_s}   \, \frac{1}{\sqrt{n_s}} \sum_{r=1}^{n_s}  W_j(X_{sr} )   \frac{f_j(X_{sr} )}{f( X_{sr} ) }  \psi_s\left( \varepsilon_{sr}\right)  \\
 & +        \frac{1}{\sqrt{n_j}} \sum_{r=1}^{n_j} \psi_j\left( \varepsilon_{jr}\right)  \left[ \frac{\pi_j }{\nu_{j}}  W_j(X_{jr} )   \frac{f_j(X_{jr} )}{f( X_{jr} ) }     - \frac{\esp\left(W_j(X_{j})\right)}{\nu_j }   \right] \,.
  \end{align*} 
 Recall that $\nu_{j}=\esp[\psi_j^{\prime}(\varepsilon_j)]\ne 0$, $\tau_{j}=\esp[\psi_j^{2}(\varepsilon_j)]$, $e_j=\tau_j/\nu_j^2$ and $\esp\psi_j(\varepsilon_j)=0$. Then, using that the populations are independent we get that
    \begin{align*}
 \var(Z_{n,j})& = 
     \sum_{s\ne j} \pi_j\pi_s\,  \frac{\sigma_s^2}{\sigma_j ^2}   \frac{  \tau_s}{\nu_s^2}    \esp\left\{  W_j^2(X_{s} )   \frac{f_j^2(X_{s } )}{f^2( X_{s} ) } \right\}   +        \tau_j \esp  \left[ \frac{\pi_j }{\nu_{j}}  W_j(X_{j } )   \frac{f_j(X_{j } )}{f( X_{j } ) }     - \frac{\esp\left(W_j(X_{j})\right)}{\nu_j }   \right]^2\\
     & = 
     \sum_{s\ne j} \pi_j\pi_s\,  \frac{\sigma_s^2}{\sigma_j ^2}   \frac{  \tau_s}{\nu_s^2}    \esp\left\{  W_j^2(X_{s} )   \frac{f_j^2(X_{s } )}{f^2( X_{s} ) } \right\}   +        \frac{\tau_j}{\nu_{j}^2} \esp  \left[ \pi_j    W_j(X_{j } )   \frac{f_j(X_{j } )}{f( X_{j } ) }     -  \esp\left(W_j(X_{j})\right)   \right]^2\\
     & = 
     \sum_{s\ne j} \pi_j\pi_s\, e_s\,  \frac{\sigma_s^2}{\sigma_j ^2}       \esp\left\{  W_j^2(X_{s} )   \frac{f_j^2(X_{s } )}{f^2( X_{s} ) } \right\}   +        e_j \esp  \left[ \pi_j    W_j(X_{j } )   \frac{f_j(X_{j } )}{f( X_{j } ) }     -  \esp\left(W_j(X_{j})\right)   \right]^2 \\
     & = 
     \sum_{s=1}^k \pi_j\pi_s\, e_s\,  \frac{\sigma_s^2}{\sigma_j ^2}       \esp\left\{  W_j^2(X_{s} )   \frac{f_j^2(X_{s } )}{f^2( X_{s} ) } \right\}\\
  &   -  \pi_j^2\, e_j\,         \esp\left\{  W_j^2(X_{j} )   \frac{f_j^2(X_{j } )}{f^2( X_{j} ) } \right\}    +        e_j \esp  \left[ \pi_j    W_j(X_{j } )   \frac{f_j(X_{j } )}{f( X_{j } ) }     -  \esp\left(W_j(X_{j})\right)   \right]^2\\
   & = 
     \sum_{s=1}^k \pi_j\pi_s\, e_s\,  \frac{\sigma_s^2}{\sigma_j ^2}       \esp\left\{  W_j^2(X_{s} )   \frac{f_j^2(X_{s } )}{f^2( X_{s} ) } \right\}     +        e_j   \left\{\left[    \esp W_j(X_{j})   \right]^2 -2   \pi_j \esp\left(W_j(X_{j})\right)  \esp  \left[     W_j(X_{j } )   \frac{f_j(X_{j } )}{f( X_{j } ) } \right]  \right\}\,.
  \end{align*} 
Noting that
$$  \alpha_{j}^{(s)}=\esp\left\{  W_j^2(X_{s} )   \frac{f_j^2(X_{s } )}{f^2( X_{s} ) } \right\}\qquad   \beta_{j}=\esp\left\{  W_j (X_{j} )   \frac{f_j(X_{j } )}{f( X_{j} ) } \right\} \qquad   \omega_{j}=\esp   W_j (X_{j} )\,,$$
we obtain
  \begin{align*}
 \var(Z_{n,j})& = 
     \sum_{s=1}^k \pi_j\,\pi_s\, e_s\,      \alpha_{j}^{(s)} \,\frac{\sigma_s^2}{\sigma_j ^2}       +        e_j    \left\{ \omega_{j}^2 -2   \pi_j \omega_{j} \beta_j
 \right\}\;.
  \end{align*} 
  We now compute $ \cov(Z_{n,j}, Z_{n,\ell})$, for $\ell\ne j$. Recall that
 \begin{align*} 
 Z_{n,j}  & =     \sum_{s\ne j}  \frac{\sigma_s}{\sigma_j }   \frac{\pi_j^{1/2}\pi_s ^{1/2}}{\nu_s}   \, \frac{1}{\sqrt{n_s}} \sum_{r=1}^{n_s}  W_j(X_{sr} )   \frac{f_j(X_{sr} )}{f( X_{sr} ) }  \psi_s\left( \varepsilon_{sr}\right)    +        \frac{1}{\sqrt{n_j}} \sum_{r=1}^{n_j} \psi_j\left( \varepsilon_{jr}\right)  \left[ \frac{\pi_j }{\nu_{j}}  W_j(X_{jr} )   \frac{f_j(X_{jr} )}{f( X_{jr} ) }     - \frac{\omega_j}{\nu_j }   \right]\\
 & =     \sum_{s\ne j, \ell}  \frac{\sigma_s}{\sigma_j }   \frac{\pi_j^{1/2}\pi_s ^{1/2}}{\nu_s}   \, \frac{1}{\sqrt{n_s}} \sum_{r=1}^{n_s}  W_j(X_{sr} )   \frac{f_j(X_{sr} )}{f( X_{sr} ) }  \psi_s\left( \varepsilon_{sr}\right)   \\
 & + \frac{\sigma_\ell}{\sigma_j }   \frac{\pi_j^{1/2}\pi_\ell ^{1/2}}{\nu_\ell}   \, \frac{1}{\sqrt{n_\ell}} \sum_{r=1}^{n_\ell}  W_j(X_{\ell r} )   \frac{f_j(X_{\ell r} )}{f( X_{\ell r} ) }  \psi_\ell \left( \varepsilon_{\ell r}\right)    +        \frac{1}{\sqrt{n_j}} \sum_{r=1}^{n_j} \psi_j\left( \varepsilon_{jr}\right)  \left[ \frac{\pi_j }{\nu_{j}}  W_j(X_{jr} )   \frac{f_j(X_{jr} )}{f( X_{jr} ) }     - \frac{\omega_j}{\nu_j }   \right]\,,
  \end{align*} 
  while
  \begin{align*} 
 Z_{n,\ell}  & =     \sum_{s\ne \ell}  \frac{\sigma_s}{\sigma_\ell }   \frac{\pi_\ell^{1/2}\pi_s ^{1/2}}{\nu_s}   \, \frac{1}{\sqrt{n_s}} \sum_{r=1}^{n_s}  W_\ell(X_{sr} )   \frac{f_\ell(X_{sr} )}{f( X_{sr} ) }  \psi_s\left( \varepsilon_{sr}\right)    +        \frac{1}{\sqrt{n_\ell}} \sum_{r=1}^{n_\ell} \psi_\ell\left( \varepsilon_{\ell r}\right)  \left[ \frac{\pi_\ell }{\nu_{\ell}}  W_\ell(X_{\ell r} )   \frac{f_\ell(X_{\ell r} )}{f( X_{\ell r} ) }     - \frac{\omega_\ell}{\nu_\ell}   \right]\\
 & =    \sum_{s\ne \ell, j}  \frac{\sigma_s}{\sigma_\ell }   \frac{\pi_\ell^{1/2}\pi_s ^{1/2}}{\nu_s}   \, \frac{1}{\sqrt{n_s}} \sum_{r=1}^{n_s}  W_\ell(X_{sr} )   \frac{f_\ell(X_{sr} )}{f( X_{sr} ) }  \psi_s\left( \varepsilon_{sr}\right) \\
 & + \frac{\sigma_j}{\sigma_\ell }   \frac{\pi_j^{1/2}\pi_\ell ^{1/2}}{\nu_j}   \, \frac{1}{\sqrt{n_j}} \sum_{r=1}^{n_j}  W_\ell(X_{jr} )   \frac{f_\ell(X_{jr} )}{f( X_{jr} ) }  \psi_j\left( \varepsilon_{jr}\right)    +        \frac{1}{\sqrt{n_\ell}} \sum_{r=1}^{n_\ell} \psi_\ell\left( \varepsilon_{\ell r}\right)  \left[ \frac{\pi_\ell }{\nu_{\ell}}  W_\ell(X_{\ell r} )   \frac{f_\ell(X_{\ell r} )}{f( X_{\ell r} ) }     - \frac{\omega_\ell}{\nu_\ell}   \right]\,.
  \end{align*} 
For simplicity, denote
 \begin{align*}
A_{s,\ell} & = \frac{\sigma_s}{\sigma_\ell }   \frac{\pi_\ell^{1/2}\pi_s ^{1/2}}{\nu_s}   \, \frac{1}{\sqrt{n_s}} \sum_{r=1}^{n_s}  W_\ell(X_{sr} )   \frac{f_\ell(X_{sr} )}{f( X_{sr} ) }  \psi_s\left( \varepsilon_{sr}\right)\,,\\
 c_{j,\ell} & =\cov\left(\frac{\sigma_\ell}{\sigma_j }   \frac{\pi_j^{1/2}\pi_\ell ^{1/2}}{\nu_\ell}   \, \frac{1}{\sqrt{n_\ell}} \sum_{r=1}^{n_\ell}  W_j(X_{\ell r} )   \frac{f_j(X_{\ell r} )}{f( X_{\ell r} ) }  \psi_\ell \left( \varepsilon_{\ell r}\right)   ,   \frac{1}{\sqrt{n_\ell}} \sum_{r=1}^{n_\ell} \psi_\ell\left( \varepsilon_{\ell r}\right)  \left[ \frac{\pi_\ell }{\nu_{\ell}}  W_\ell(X_{\ell r} )   \frac{f_\ell(X_{\ell r} )}{f( X_{\ell r} ) }     - \frac{\omega_\ell}{\nu_\ell}   \right] \right)\,.
 \end{align*}
 Then, we have that 
   \begin{align} 
  \cov(Z_{n,j}, Z_{n,\ell}) &   = \cov\left(
   \sum_{s\ne \ell, j} A_{s,\ell} , \sum_{r\ne j, \ell} A_{r,j}  \right) +  c_{j,\ell} +  c_{\ell,j} \,.
 \label{eq:covZnellj}
  \end{align} 
 We begin by computing $\cov\left(
   \sum_{s\ne \ell, j} A_{s,\ell} , \sum_{r\ne j, \ell} A_{r,j}  \right)  $. Note that the samples independence entail that
  \begin{align*}
 & \cov\left(
   \sum_{s\ne \ell, j} A_{s,\ell} , \sum_{r\ne j, \ell} A_{r,j}  \right)  = \sum_{s \ne \ell, j} \cov\left(A_{s,\ell}, A_{s,j} \right ) \\
   & = \sum_{s \ne \ell, j} \cov\left(  \frac{\sigma_s}{\sigma_\ell} \frac{ \pi_s^{1/2} \pi_\ell^{1/2}}{\nu_s}  \, \frac{1}{\sqrt{n_s}} \sum_{r=1}^{n_s}  W_\ell(X_{sr} )   \frac{f_\ell(X_{sr} )}{f( X_{sr} ) }  \psi_s\left( \varepsilon_{sr}\right) ,  \frac{\sigma_s}{\sigma_j} \frac{\pi_s^{1/2} \pi_j^{1/2}}{\nu_s} \, \frac{1}{\sqrt{n_s}} \sum_{r=1}^{n_s}  W_j(X_{sr} )   \frac{f_j(X_{sr} )}{f( X_{sr} ) }  \psi_s\left( \varepsilon_{sr}\right)\right)\\
   & = \frac{\pi_\ell^{1/2}\, \pi_j^{1/2}}{\sigma_\ell \, \sigma_j} \,  \sum_{s \ne \ell, j}     \frac{\pi_s \, \sigma_s^2}{\nu_s^2}   \cov\left(  W_\ell(X_{s} )   \frac{f_\ell(X_{s} )}{f( X_{s} ) }  \psi_s\left( \varepsilon_{s}\right),  W_j(X_{s} )   \frac{f_j(X_{s} )}{f( X_{s} ) }  \psi_s\left( \varepsilon_{s}\right)\right) \,.
  \end{align*}
  Using again that $\esp\psi_j(\varepsilon_j)=0$ and that $\esp\psi_s^2\left( \varepsilon_{s}\right) /\nu_s^2=e_s$ and that
  $$ \alpha_{j,\ell}^{(s)}=  \esp\left(    \frac{ W_\ell(X_{s} )   f_\ell(X_{s} )\,  W_j(X_{s} )  f_j(X_{s} ) }{f^2( X_{s} ) }\right)  \,,$$
  we get that
   \begin{align}
   \cov\left(
   \sum_{s\ne \ell, j} A_{s,\ell} \right. & \left., \sum_{r\ne j, \ell} A_{r,j}  \right)  
      = \frac{\pi_\ell^{1/2}\, \pi_j^{1/2}}{\sigma_\ell \, \sigma_j} \,  \sum_{s \ne \ell, j}     \frac{\pi_s \, \sigma_s^2}{\nu_s^2}   \esp\left(  W_\ell(X_{s} )   W_j(X_{s} )    \frac{f_\ell(X_{s} )\ f_j(X_{s} ) }{f^2( X_{s} ) }  \psi_s^2\left( \varepsilon_{s}\right) \right)  
   \nonumber \\
 & = \frac{\pi_\ell^{1/2}\, \pi_j^{1/2}}{\sigma_\ell \, \sigma_j} \,  \sum_{s \ne \ell, j}     e_s\;\pi_s \, \sigma_s^2  \esp\left(    \frac{ W_\ell(X_{s} )   f_\ell(X_{s} )\,  W_j(X_{s} )  f_j(X_{s} ) }{f^2( X_{s} ) }\right)  
 \nonumber \\
 & = \frac{\pi_\ell^{1/2}\, \pi_j^{1/2}}{\sigma_\ell \, \sigma_j} \,  \sum_{s \ne \ell, j}     e_s\;\pi_s \, \sigma_s^2  \alpha_{j,\ell}^{(s)}
 \nonumber \\
 & =  \frac{\pi_\ell^{1/2}\, \pi_j^{1/2}}{\sigma_\ell \, \sigma_j} \,  \sum_{s=1}^k     e_s\;\pi_s \, \sigma_s^2  \alpha_{j,\ell}^{(s)} 
 -  \pi_\ell^{1/2}\, \pi_j^{1/2}\, \frac{ \sigma_\ell}{ \sigma_j} \,       e_\ell \;\pi_\ell  \,  \alpha_{j,\ell}^{(\ell)}
 -    \pi_\ell^{1/2}\, \pi_j^{1/2}\, \frac{ \sigma_j}{ \sigma_\ell}\,      e_j\;\pi_j    \alpha_{j,\ell}^{(j)} \,.
 \label{eq:covA}
  \end{align}
  Let us compute $ c_{j,\ell}$. Taking into account that $\esp  \psi_\ell\left( \varepsilon_{\ell}\right) =0$ and the independence between the errors and the covariates, we obtain
   \begin{align*}
 c_{j,\ell} & =\cov\left(\frac{\sigma_\ell}{\sigma_j }   \frac{\pi_j^{1/2}\pi_\ell ^{1/2}}{\nu_\ell}   \, \frac{1}{\sqrt{n_\ell}} \sum_{r=1}^{n_\ell}  W_j(X_{\ell r} )   \frac{f_j(X_{\ell r} )}{f( X_{\ell r} ) }  \psi_\ell \left( \varepsilon_{\ell r}\right)   ,   \frac{1}{\sqrt{n_\ell}} \sum_{r=1}^{n_\ell} \psi_\ell\left( \varepsilon_{\ell r}\right)  \left[ \frac{\pi_\ell }{\nu_{\ell}}  W_\ell(X_{\ell r} )   \frac{f_\ell(X_{\ell r} )}{f( X_{\ell r} ) }     - \frac{\omega_\ell}{\nu_\ell}   \right] \right) \\
 & =\cov\left(\frac{\sigma_\ell}{\sigma_j }   \frac{\pi_j^{1/2}\pi_\ell ^{1/2}}{\nu_\ell}   \, \frac{1}{\sqrt{n_\ell}} \sum_{r=1}^{n_\ell}  W_j(X_{\ell r} )   \frac{f_j(X_{\ell r} )}{f( X_{\ell r} ) }  \psi_\ell \left( \varepsilon_{\ell r}\right)   ,   \frac{1}{\nu_\ell}\frac{1}{\sqrt{n_\ell}} \sum_{r=1}^{n_\ell} \psi_\ell\left( \varepsilon_{\ell r}\right)  \left[  \pi_\ell    W_\ell(X_{\ell r} )   \frac{f_\ell(X_{\ell r} )}{f( X_{\ell r} ) }     -  {\omega_\ell}  \right] \right) \\
 & =\frac{\sigma_\ell}{\sigma_j }   \frac{\pi_j^{1/2}\pi_\ell ^{1/2}}{\nu_\ell^2} \, \cov\left(    W_j(X_{\ell} )   \frac{f_j(X_{\ell} )}{f( X_{\ell} ) }  \psi_\ell \left( \varepsilon_{\ell }\right)   ,    \psi_\ell\left( \varepsilon_{\ell}\right)  \left[ \pi_\ell   W_\ell(X_{\ell } )   \frac{f_\ell(X_{\ell} )}{f( X_{\ell} ) }     - \omega_\ell    \right] \right) 
 \\
 & =\frac{\sigma_\ell}{\sigma_j }   \frac{\pi_j^{1/2}\pi_\ell ^{1/2}}{\nu_\ell^2} \, \esp\left(\psi_\ell^2\left( \varepsilon_{\ell}\right)  \right)\, \esp\left(    W_j(X_{\ell} )   \frac{f_j(X_{\ell} )}{f( X_{\ell} ) }         \left[ \pi_\ell   W_\ell(X_{\ell } )   \frac{f_\ell(X_{\ell} )}{f( X_{\ell} ) }     - \omega_\ell    \right]  \right) \\
  & =\frac{\sigma_\ell}{\sigma_j }    \pi_j^{1/2}\,\pi_\ell ^{1/2} \, e_\ell \, \left\{\pi_\ell \;\esp\left(      \frac{ W_j(X_{\ell} )f_j(X_{\ell} )\;   W_\ell(X_{\ell } )f_\ell(X_{\ell} )  }{f^2( X_{\ell} ) }   \right)           - \omega_\ell  \; \esp\left(    W_j(X_{\ell} )   \frac{f_j(X_{\ell} )}{f( X_{\ell} ) }  \right)  \right\}
  \\
  & =\frac{\sigma_\ell}{\sigma_j }    \pi_j^{1/2}\,\pi_\ell ^{1/2} \, e_\ell \, \left\{\pi_\ell \;
  \alpha_{j,\ell}^{(\ell)}  - \omega_\ell  \;  \beta_j^{(\ell)}\right\} \,,
 \end{align*}
 where we have used that
% $$\beta_{j} ^{(s)} =\esp\left\{  W_j (X_{s} ) \frac{f_j(X_{s } )}{f( X_{s} ) } \right\} \qquad \alpha_{j,\ell}^{(s)} =  \esp\left(    \frac{ W_\ell(X_{s} )   f_\ell(X_{s} )\,  W_j(X_{s} )  f_j(X_{s} ) }{f^2( X_{s} ) }\right) \,.$$
  $$\beta_{j} ^{(\ell)} =\esp\left\{  W_j (X_{\ell} ) \frac{f_j(X_{\ell } )}{f( X_{\ell} ) } \right\} \qquad \alpha_{j,\ell}^{(\ell)} =  \esp\left(    \frac{ W_\ell(X_{\ell} )   f_\ell(X_{\ell} )\,  W_j(X_{\ell} )  f_j(X_{\ell} ) }{f^2( X_{\ell} ) }\right) \,.$$
 Summarizing we have that
    \begin{align*}
 c_{j,\ell} & =\frac{\sigma_\ell}{\sigma_j }    \pi_j^{1/2}\,\pi_\ell ^{1/2} \, e_\ell \, \left\{\pi_\ell \;
  \alpha_{j,\ell}^{(\ell)}  - \omega_\ell  \;  \beta_j^{(\ell)}\right\} 
  =\frac{\sigma_\ell}{\sigma_j }    \pi_j^{1/2}\,\pi_\ell ^{1/2} \,\pi_\ell \; e_\ell \,  
  \alpha_{j,\ell}^{(\ell)}  - \frac{\sigma_\ell}{\sigma_j }    \pi_j^{1/2}\,\pi_\ell ^{1/2} \, e_\ell \, \omega_\ell  \;  \beta_j^{(\ell)} \,, \\
   c_{\ell,j} & =\frac{\sigma_j}{\sigma_\ell }    \pi_j^{1/2}\,\pi_\ell ^{1/2} \, e_j \, \left\{\pi_j\;
  \alpha_{j,\ell}^{(j)}  - \omega_j  \;  \beta_\ell^{(j)}\right\} 
  =\frac{\sigma_j}{\sigma_\ell }    \pi_j^{1/2}\,\pi_\ell ^{1/2} \, \pi_j\; e_j \, 
  \alpha_{j,\ell}^{(j)}  - \frac{\sigma_j}{\sigma_\ell }    \pi_j^{1/2}\,\pi_\ell ^{1/2} \, e_j  \omega_j  \;  \beta_\ell^{(j)} \,,
 \end{align*}
 which together with \eqref{eq:covA} and \eqref{eq:covZnellj} leads to
   \begin{align*} 
  \cov(Z_{n,j}, Z_{n,\ell}) &   = \cov\left(
   \sum_{s\ne \ell, j} A_{s,\ell} , \sum_{r\ne j, \ell} A_{r,j}  \right) +  c_{j,\ell} +  c_{\ell,j} 
\\
& =  \frac{\pi_\ell^{1/2}\, \pi_j^{1/2}}{\sigma_\ell \, \sigma_j} \,  \sum_{s=1}^k     e_s\;\pi_s \, \sigma_s^2  \alpha_{j,\ell}^{(s)} 
    -  \pi_\ell^{1/2}\, \pi_j^{1/2}\, \frac{ \sigma_\ell}{ \sigma_j} \,       e_\ell \;\pi_\ell  \,  \alpha_{j,\ell}^{(\ell)}
 -    \pi_\ell^{1/2}\, \pi_j^{1/2}\, \frac{ \sigma_j}{ \sigma_\ell}\,      e_j\;\pi_j    \alpha_{j,\ell}^{(j)} \\
 &\hskip-0.4in  +\frac{\sigma_\ell}{\sigma_j }    \pi_j^{1/2}\,\pi_\ell ^{1/2} \,\pi_\ell \; e_\ell \,  
  \alpha_{j,\ell}^{(\ell)}  - \frac{\sigma_\ell}{\sigma_j }    \pi_j^{1/2}\,\pi_\ell ^{1/2} \, e_\ell \, \omega_\ell  \;  \beta_j^{(\ell)}
  + \frac{\sigma_j}{\sigma_\ell }    \pi_j^{1/2}\,\pi_\ell ^{1/2} \, \pi_j\; e_j \, 
  \alpha_{j,\ell}^{(j)}  - \frac{\sigma_j}{\sigma_\ell }    \pi_j^{1/2}\,\pi_\ell ^{1/2} \, e_j  \omega_j  \;  \beta_\ell^{(j)}  \\
  & =  \frac{\pi_\ell^{1/2}\, \pi_j^{1/2}}{\sigma_\ell \, \sigma_j} \,  \sum_{s=1}^k     e_s\;\pi_s \, \sigma_s^2  \alpha_{j,\ell}^{(s)} 
      - \frac{\sigma_\ell}{\sigma_j }    \pi_j^{1/2}\,\pi_\ell ^{1/2} \, e_\ell \, \omega_\ell  \;  \beta_j^{(\ell)}
    - \frac{\sigma_j}{\sigma_\ell }    \pi_j^{1/2}\,\pi_\ell ^{1/2} \, e_j  \omega_j  \;  \beta_\ell^{(j)} \,, 
  \end{align*} 
  and   the proof follows now from the multivariate central limit theorem. 
    \end{proof}

  In the sequel we will use the   consistency rates stated in Lemma \ref{lemma:lema2} which we include without proof. In the case of the robust regression estimator, the proof is a direct consequence of the results  in \citet{boente:pardo:2016} combined with the bandwidth rate given in  \ref{ass:A9}, while for the density estimator the result can be found in \citet{Pardo:etal:2015}. 
  
 From now on,  given a sequence $\{a_n\}_{n\ge 1}$ a sequence of positive numbers and $\{V_n\}$ a sequence of random variables, $V_n=O_{a.co.}(a_n)$ means that for some positive constant $C_0$, $\sum_{n\ge 1} \prob(V_n>C_0 a_n) <\infty$. 
  
   \vskip0.1in
 \begin{lemma}{\label{lemma:lema2}}
Define 
  \begin{eqnarray}
 \wL_j(x,\sigma)&=&\frac{1}{n_j}\sum_{\ell=1}^{n_j}\omega_{j\ell}(x) \psi_j\left( \frac{Y_{j\ell}-m_j(x)}{\sigma}\right)\,,
 \label{eq:WLj}
 \end{eqnarray}
$\omega_{j\ell}(x)=K_h\left(x-X_{j\ell}\right)$ and $\theta_{n_j}=\sqrt{\log n_j/(n_j h)}$. Then, under conditions \ref{ass:A2}, \ref{ass:A4}, \ref{ass:A5}, \ref{ass:A7} and \ref{ass:A8a}),   for any compact set $\itK\subset \accentset{\circ}{\itR}$,  we have
\begin{eqnarray}
\sup_{x\in \itK}|\wm_j(x)-m_j(x)| =  O_{a.co.}\left(h^{2} + \theta_{n_j}\right)=o_{\prob}(n_j^{-1/4})\,,
\label{eq:tasamj}\\
\sup_{x\in \itK}\left|\wm_j(x)-m_j(x)-  \frac{\wsigma_j}{f_j(x)\nu_{j}}\, \wL_j(x,\wsigma_j) \right| =  O_{a.co.}\left(h^{2} + \theta_{n_j}^2\right)=o_{\prob}(n_j^{-1/2}) \,,\label{eq:linearmj} \\
\sup_{x\in \itK}|\wefe_j(x)-f_j(x)| = o_{\prob}(n_j^{-1/4})\,.
\label{eq:tasafj}
\end{eqnarray}
\end{lemma}

\noi A direct consequence of  \ref{ass:A8}, \eqref{eq:tasamj} and \eqref{eq:tasafj} is that, under $H_{1,n}$,
\begin{equation}
\sup_{x\in \itK}|\wmu_0(x)-\mu_0(x)| = o_{\prob}(n_j^{-1/4})\,,
\label{eq:tasamu0}
\end{equation}
where $\mu_0(x)$ and $\wmu_0(x)$ are defined in \eqref{eq:mu0} and \eqref{eq:wmu0}, respectively. Note that under $H_0$, $\mu_0=m_0=m_j$, for all $j=1,\dots, k$,  so that \eqref{eq:tasamu0} follows immediately, while under the alternative $H_{1,n}$, 
$$\mu_0(x)=m_0(x)+ n^{-1/2} \sum_{j=1}^k \pi_j \, \frac{f_j(x)}{f(x)} \Delta_j(x)=m_0(x)+ n^{-1/2} \Delta_0(x)\,.$$ 
Hence, we also have that
\begin{equation}
\sup_{x\in \itK}|\wmu_0(x)-m_0(x)| = o_{\prob}(n_j^{-1/4})\,.
\label{eq:tasam0}
\end{equation}

For the sake of simplicity, from now on, we denote $\wDelta_j(x)=(\wmu_0(x)-\wm_j(x))/{\wsigma_j}$,  $\wUps_{0,j}(x)=(\mu_0(x)-\wmu_0(x))/{\wsigma_j}$,  $\wUps_j(x)=(m_j(x)-\wm_j(x))/{\wsigma_j}$ and $\wGamma_j(x)= n^{-1/2}(\Delta_j(x)-\Delta_0(x))/{\wsigma_j}$. Then, 
\begin{equation}
  \wDelta_j(x)= - \wUps_{0,j}(x)+\wUps_j(x)-\wGamma_j(x)\,.
  \label{eq:wDeltaj}
  \end{equation}

 \vskip0.1in
 
\begin{lemma}{\label{lemma:lema3}}
Assume that \eqref{modelo} and \ref{ass:A2} to   \ref{ass:A7}, \ref{ass:A8a}) and \ref{ass:A9}  hold. Let $\wsigma_j$ be a consistent estimator of $\sigma_j$, $j=1,\ldots, k$ satisfying \ref{ass:A10} and let $\Delta_j:\real \to \real$ be such that $\esp W_j(X_j) \Delta_j^2(X_{j}) <\infty$. Assume that $H_{1,n}: m_j=m_0+n^{-1/2} \Delta_j$ holds  and define
  \begin{align*}
 D_{1,n_j}(t) & =
- \;\frac{1}{\sqrt{n_j}} \sum_{\ell=1}^{n_j} W_j(X_{j\ell})\,\exp(i\,t\, \weps_{0j\ell})\,\left\{  \frac{\wmu_0(X_{j\ell})-\wm_j(X_{j\ell})}{\wsigma_j}\right\}^2 \exp(i\,t\, \xi_{j\ell}^{(n)})\,,\\
 D_{2,n_j}(t) & = 
  \frac{1}{\sqrt{n_j}} \sum_{\ell=1}^{n_j} W_j(X_{j\ell})\, \left[\wUps_{0,j}(X_{j\ell})+ \wGamma_j(X_{j\ell})\right]  \wDelta_j(X_{j\ell}) \exp(i\,t\, \xi_{j\ell}^{(n)}) \,, \\
 D_{3,n_j}(t)&=  
 \left(\frac{\wsigma_j-\sigma_j}{  \sigma_j}\right) \frac{1}{\sqrt{n_j}} \sum_{\ell=1}^{n_j} W_j(X_{j\ell})\,\exp\left\{i\,t\, \frac{\sigma_j}{\wsigma_j}\varepsilon_{j\ell}\right\}\,  \left\{ \wUps_j(X_{j\ell})- \wUps_0(X_{j\ell})  - \wGamma_j(X_{j\ell}) \right\}\,,
\end{align*}
where $\xi_{j\ell}^{(n)}$ are random variables that may depend on the sample size. Then, for $s=1,2,3$, we have that $\sup_{t\in \real} |D_{s,n_j}(t)|=o_{\prob}(1)$
\end{lemma}
 
\begin{proof}
Using that  $m_j=m_0+n^{-1/2} \Delta_j$ and that $\mu_0(x)=m_0(x) + n^{-1/2} \Delta_0(x)$, we get that
\begin{align}
 \wmu_0(x)-\wm_j(x)&= \left(\wmu_0(x)-m_0(x)\right)-  \left(\wm_j(x)-m_{j}(x)\right) - n^{-1/2}  \Delta_j(x)\nonumber\\
 &= \left(\wmu_0(x)-\mu_0(x)\right)-  \left(\wm_j(x)-m_{j}(x)\right)+ n^{-1/2}  \left(\Delta_0(x) -\Delta_j(x) \right) \,,
\label{eq:difwm}
\end{align}
which implies
\begin{align*}
\sup_{t\in \real} |D_{1,n_j}(t)|& \le \frac{1}{\sqrt{n_j}} \sum_{\ell=1}^{n_j} W_j(X_{j\ell})\, \frac{\left[\wmu_0(X_{j\ell})-\wm_j(X_{j\ell})\right]^2}{\wsigma_j^2} \\
& \le \frac{4}{\sqrt{n_j}\, \wsigma_j^2} \sum_{\ell=1}^{n_j} W_j(X_{j\ell})\,  \left\{\left[\wmu_0(X_{j\ell})-m_0(X_{j\ell})\right]^2+ \left[\wm_j(X_{j\ell})-m_j(X_{j\ell})\right]^2   + \, n^{-1} \Delta_j^2(X_{j\ell})  \right\}
 \\
& \le \frac{4}{\wsigma_j^2} \left\{\sqrt{n_j} \sup_{x\in \itS_j} \left[\wmu_0(x)-m_0(x)\right]^2+   \sqrt{n_j} \sup_{x\in \itS_j}\left[\wm_j(x)-m_j(x)\right]^2\right\}\\
& + \frac{4}{\wsigma_j^2} \, \frac{n_j}{n}\,\frac{1}{\sqrt{n_j}}\,\left( \frac 1{n_j}  \sum_{\ell=1}^{n_j} W_j(X_{j\ell})\, \Delta_j^2(X_{j\ell})  \right) \,. 
\end{align*}
From \eqref{eq:tasamj}, taking $\itK=\itS_j$, and using that $\wsigma_j\convprob \sigma_j$, $\esp W_j(X_j) \Delta_j^2(X_{j}) <\infty$ and $n_j/n\to \kappa_j$, we easily get that $
\sup_{t\in \real} |D_{1,n_j}(t)|=o_{\prob}(1)$.

Let us show that $\sup_{t\in \real} |D_{2,n_j}(t)|=o_{\prob}(1)$. Using  \eqref{eq:difwm} and denoting $A(x)=|\Delta_0(x)|+|\Delta_j(x)|$, $\wA_0= \sup_{x\in \itS_j} \left|\wmu_0(x)-\mu_0(x)\right|$ and $\wA_j=\sup_{x\in \itS_j} \left|\wm_j(x)-m_j(x)\right|$, we can bound $\sup_{t\in \real} |D_{2,n_j}(t)|$ as 
\begin{align*}
 \sup_{t\in \real} |D_{2,n_j}(t)|& \le \frac{1}{\sqrt{n_j}} \sum_{\ell=1}^{n_j} W_j(X_{j\ell})\, \left|\wUps_{0,j}(X_{j\ell})+ \wGamma_j(X_{j\ell})\right|\; \left|\wDelta_j(X_{j\ell}) \right|\\
 &  \le  \frac{1}{\wsigma_j^2}   \frac{1}{\sqrt{n_j}} \sum_{\ell=1}^{n_j} W_j(X_{j\ell})\left|\mu_0(X_{j\ell})-\wmu_0(X_{j\ell}) +n^{-1/2}\left[\Delta_j(X_{j\ell})-\Delta_0(X_{j\ell})\right] \right|\;
 \left| \wmu_0(X_{j\ell})-\wm_j(X_{j\ell}) \right|\\
 &  \le  \frac{1}{\wsigma_j^2}   \frac{1}{\sqrt{n_j}} \sum_{\ell=1}^{n_j} W_j(X_{j\ell})\left\{\left|\wmu_0(X_{j\ell})-\mu_0(X_{j\ell})\right|\; +n^{-1/2}\; A(X_{j\ell}) \right\}\;\times\\
& \left\{ \left| \wmu_0(X_{j\ell})-\mu_0(X_{j\ell}) \right| + \left|\wm_j(X_{j\ell})-m_{j}(X_{j\ell})\right| + n^{-1/2} A(X_{j\ell})\right\}\\
&  \le  \frac{1}{\wsigma_j^2}  \left\{ \sqrt{n_j} \wA_0^2 +2\, \wA_0  \;   \left(\frac{n_j}{n}\right)^{1/2}\, \frac{1}{n_j}\; \sum_{\ell=1}^{n_j} W_j(X_{j\ell}) A(X_{j\ell})  + \sqrt{n_j}  \wA_0 \wA_j \right.\\
& \left. +\wA_j  \;   \left(\frac{n_j}{n}\right)^{1/2}\, \frac{1}{n_j}\; \sum_{\ell=1}^{n_j} W_j(X_{j\ell}) A(X_{j\ell})+ n^{-1/2} \left(\frac{n_j}{n}\right)^{1/2}\, \frac{1}{n_j}\; \sum_{\ell=1}^{n_j} W_j(X_{j\ell}) A^2(X_{j\ell}) \right\}\,.
  \end{align*}
Hence, using \eqref{eq:tasamj} and \eqref{eq:tasam0}, together with the fact that $\esp W_j(X_j) \Delta_j^2(X_{j}) <\infty$, we obtain that 
$\sup_{t\in \real} |D_{2,n_j}(t)|=o_{\prob}(1)$.

Finally, to prove that $\sup_{t\in \real} |D_{3,n_j}(t)|=o_{\prob}(1)$, note that
 \begin{align} 
  \sup_{t\in \real}   |D_{3,n_j}(t)| & \le \left(\frac{|\wsigma_j-\sigma_j|}{\wsigma_j\, \sigma_j}\right) \frac{1}{\sqrt{n_j}} \sum_{\ell=1}^{n_j} W_j(X_{j\ell})\,\Big\{ \left|\wmu_0(X_{j\ell})-\mu_0(X_{j\ell})\right|+ \left|\wm_j(X_{j\ell})-m_j(X_{j\ell})\right|
\nonumber
\\
 &  + n^{-1/2}\left|\Delta_j(X_{j\ell})-\Delta_0(X_{j\ell})\right|\Big\} \nonumber
 \\
&\le \left(\frac{|\wsigma_j-\sigma_j|}{\wsigma_j\, \sigma_j}\right) \sqrt{n_j} \left(\wA_0  + \wA_j\right)\nonumber \\
& + \left(\frac{|\wsigma_j-\sigma_j|}{\wsigma_j\, \sigma_j}\right) \left(\frac{n_j}{n}\right)^{1/2}\, \frac{1}{n_j}\; \sum_{\ell=1}^{n_j} W_j(X_{j\ell})\left|\Delta_j(X_{j\ell})-\Delta_0(X_{j\ell})\right|
\,.
\label{eq:cotaR3}
 \end{align}
 Using the consistency of $\wsigma_j$ and the fact that $\esp W_j(X_j) \Delta_j^2(X_{j}) <\infty$, we obtain that the second term on the right hand side of \eqref{eq:cotaR3} converges to 0 in probability, while \eqref{eq:tasamj} and \eqref{eq:tasam0} together with the fact that $ {n_j}^{1/4} (\wsigma_j-\sigma_j)=o_{\prob}(1)$ entail  that the first term is $o_{\prob}(1)$, concluding the proof.
\end{proof}
\vskip0.1in

 \begin{lemma}{\label{lemma:lema4}}
 Assume that \eqref{modelo} and \ref{ass:A2} to \ref{ass:A9} hold. Let $\wsigma_j$ be a consistent estimator of $\sigma_j$, $j=1,\ldots, k$ satisfying \ref{ass:A10} and consider functions $\Delta_j:\real \to \real$ such that $\esp W_j(X_j) \Delta_j^2(X_{j}) <\infty$. Assume that $H_{1,n}: m_j=m_0+n^{-1/2} \Delta_j$ holds. Denote $V_s(x)= (n_s/n) (f_s(x)/f(x))$, $M_s(x)=m_s(x)/f(x)$ and $C_s(x)= (n_s/n) M_s(x)$. Furthermore, define
  \begin{align*}
   D_{1,n_j}(\sigma,t) & = \frac{1}{\sigma_j}\, \left(\frac{n_j}{n}\right)^{1/2}\,\frac{1}{n_j} \sum_{\ell=1}^{n_j} W_j(X_{j\ell})\,\exp\left\{i\,t\, \frac{\sigma_j}{\sigma}\varepsilon_{j\ell}\right\}  (\Delta_j(X_{j\ell})-\Delta_0(X_{j\ell})) \,, \\
 D_{2,n_j}(\sigma,t) & = \frac{1}{\sigma_j}\frac{1}{\sqrt{n_j}} \sum_{\ell=1}^{n_j} W_j(X_{j\ell})\,\exp\left\{i\,t\, \frac{\sigma_j}{\sigma}\varepsilon_{j\ell}\right\}\,  \left\{ \wm_j(X_{j\ell})-m_j(X_{j\ell})\right\} \,, \\
   D_{3,n_j}(\sigma,t) & = \sum_{s=1}^k \frac{1}{\sigma_j}\frac{1}{\sqrt{n_j}} \sum_{\ell=1}^{n_j} W_j(X_{j\ell})V_s(X_{j\ell})\,\exp\left\{i\,t\, \frac{\sigma_j}{\sigma}\varepsilon_{j\ell}\right\}\,  \left\{ \wm_s(X_{j\ell})-m_s(X_{j\ell})\right\} \, , \\
  D_{4,n_j}(\sigma,t)  & = \sum_{s=1}^k \frac{1}{\sigma_j}\frac{1}{\sqrt{n_j}} \sum_{\ell=1}^{n_j} W_j(X_{j\ell}) C_s(X_{j\ell})\,\exp\left\{i\,t\, \frac{\sigma_j}{\sigma}\varepsilon_{j\ell}\right\}\,  \left\{ \wefe_s(X_{j\ell})-f_s(X_{j\ell})\right\} \, ,\\
 D_{5,n_j}(\sigma,t)  & = \sum_{s=1}^k \frac{1}{\sigma_j}\frac{1}{\sqrt{n_j}} \sum_{\ell=1}^{n_j} W_j(X_{j\ell}) V_s(X_{j\ell}) M_s(X_{j\ell})\,\exp\left\{i\,t\, \frac{\sigma_j}{\sigma}\varepsilon_{j\ell}\right\}\,  \left\{ \wefe(X_{j\ell})-f(X_{j\ell})\right\}\,.
\end{align*}
\begin{enumerate}
\item[a)] If $\wD(t)=i\,t\,\left(D_{1,n_j} (\wsigma_j,t)-D_{1,n_j} (\sigma_j,t)\right)$, then $\|\wD\|_w=o_{\prob}(1)$. 
\item[b)] Moreover, if  \ref{ass:A11} holds, we have that, for $s=1,\dots, 5$, $D_{s,n_j}(\wsigma_j, t)= D_{s,n_j}(\sigma_j, t) + i\,t\,R_{s,n_j}(t) $, where $\sup_{t\in \real} |R_{s,n_j}(t)|=o_{\prob}(1)$.
\end{enumerate}
\end{lemma}

 \begin{proof}
  We begin by proving b). Using a  Taylor's expansion of order one, we obtain that
  \begin{align*}
   D_{1,n_j} (\wsigma_j,t)-D_{1,n_j} (\sigma_j,t) & = i\,t\, \frac{1}{\sigma_j}\, \left( \frac{\sigma_j}{\wsigma_j}-1\right)\left(\frac{n_j}{n}\right)^{1/2}\,\frac{1}{n_j} \sum_{\ell=1}^{n_j} W_j(X_{j\ell})\,\exp\left\{ i\,t\,\xi_{j\ell} \varepsilon_{j\ell} \right\}  (\Delta_j(X_{j\ell})-\Delta_0(X_{j\ell}))  \varepsilon_{j\ell} \\
   &=  i\,t\, R_{1,n_j}(t) \,,
   \end{align*}
 with  $\xi_{j\ell}$ and intermediate point. Hence, noting that
 \begin{align*}
 \sup_{t\in \real} |R_{1,n_j}(t)| 
  & \le \left(\frac{n_j}{n}\right)^{1/2}\, \frac{1}{\wsigma_j\,\sigma_j}\,  n_j^{1/\theta_0-1}  \left| {\wsigma_j}- {\sigma_j}\right|\,\frac{1}{n_j^{1/\theta_0}} \sum_{\ell=1}^{n_j}|\varepsilon_{j\ell}|\,  W_j(X_{j\ell})\, \left|\Delta_j(X_{j\ell})-\Delta_0(X_{j\ell})\right| \,. 
  \end{align*}
 Using that $\esp|\varepsilon_{j}|^{\theta_0}$, $W_j$ is bounded and  $  \Delta_j $ is bounded in the support of $W_j$,   we get that $\esp|Z_{j}|^{\theta_0}<\infty$, where $Z_j= |\varepsilon_{j }|\,  W_j(X_{j} )\, \left|\Delta_j(X_{j})-\Delta_0(X_{j})\right|$. Thus, from  the Marcinkiewicz--Zygmund strong law of large numbers,  see Appendix A in  \citet{shao:tu:1995}, we get that 
 $$\frac{1}{n_j^{1/\theta_0}} \sum_{\ell=1}^{n_j}|\varepsilon_{j\ell}|\,  W_j(X_{j\ell})\, \left|\Delta_j(X_{j\ell})-\Delta_0(X_{j\ell})\right| \convpp 0\,,$$
which together with the facts that ${1/\theta_0-1}=  \gamma_0-1/4$ and    $n^{\gamma_0}(\wsigma_j- \sigma_j)=O_{\prob}(1)$, imply that $\sup_{t\in \real} |R_{1,n_j}(t)| =o_{\prob}(1)$. 
 
 Note that $ D_{2,n_j}(\wsigma_j,t)-D_{2,n_j}(\sigma_j,t)= i\,  t R_{2,n_j}(t)$, where 
 \begin{align*}
R_{2,n_j}(t)  &= \frac{1}{\sigma_j}\frac{1}{\sqrt{n_j}} \sum_{\ell=1}^{n_j} W_j(X_{j\ell})\, \exp\left\{i\,t\, \xi_{j\ell}\right\} \,\left( \frac{\sigma_j}{\wsigma_j}-1\right)  \varepsilon_{j\ell}  \left\{ \wm_j(X_{j\ell})-m_j(X_{j\ell})\right\} \,,
\end{align*}
with $\xi_{j\ell}$ an intermediate point. We have the following bound for $R_{2,n_j}(t) $
\begin{align*}
\left|R_{2,n_j}(t)\right| &= 
\left|\frac{1}{\sigma_j}\frac{1}{\sqrt{n_j}} \sum_{\ell=1}^{n_j} W_j(X_{j\ell})\, \exp\left\{i\,t\, \xi_{j\ell}\right\} \,\left( \frac{\sigma_j}{\wsigma_j}-1\right)  \varepsilon_{j\ell}  \left\{ \wm_j(X_{j\ell})-m_j(X_{j\ell})\right\}\right|\\
& \le   \frac{1}{\sigma_j} \sup_{x\in \itS_j}\left| \wm_j(x)-m_j(x)\right|\;\left| \frac{\sigma_j}{\wsigma_j}-1\right|\frac{1}{\sqrt{n_j}} \sum_{\ell=1}^{n_j} W_j(X_{j\ell}) \left|\varepsilon_{j\ell}\right|\\
& \le  \frac{1}{\wsigma_j\,\sigma_j} n_j^{1/4}\sup_{x\in \itS_j}\left| \wm_j(x)-m_j(x)\right|\;  n_j^{\gamma_0}\left|\sigma_j- \wsigma_j\right| \frac{1}{n_j^{1/\theta_0}} \sum_{\ell=1}^{n_j}  \left|\varepsilon_{j\ell}\right| \,,
 \end{align*}
 where in the last inequality we have used that $1/\theta_0- \gamma_0-1/4=1/2$.  Therefore, using again  that $\esp|\varepsilon_{j}|^{\theta_0}$,  the Marcinkiewicz--Zygmund strong law of large numbers, that   $n^{\gamma_0} (\wsigma_j- \sigma_j)=O_{\prob}(1)$,  \eqref{eq:tasamj} and \ref{ass:A10}, we obtain that  $\sup_{t\in \real} |R_{2,n_j}(t)|=o_{\prob}(1)$ as desired. Similarly, using  \eqref{eq:tasamj}, \ref{ass:A6}, \ref{ass:A10} and the fact that  $\esp |\varepsilon_j|^{\theta_0}<\infty$, we obtain that $ D_{3,n_j} (\wsigma_j,t)=D_{3,n_j} (\sigma_j,t)  + i\,t \, R_{3,n_j}(t)$, where $\sup_{t\in \real} |R_{3,n_j}(t)|=o_{\prob}(1)$.
 Finally, using  \eqref{eq:tasafj}, similar arguments allow to conclude that, for $s=4,5$, $ D_{s,n_j} (\wsigma_j,t)=D_{s,n_j} (\sigma_j,t)  + i\,t \, R_{s,n_j}(t)$, where $\sup_{t\in \real} |R_{s,n_j}(t)|=o_{\prob}(1)$.

Let us show a). Denote  $D_{1,n_j}^{(1)}(t)=D_{1,n_j} (\wsigma_j,t)-D_{1,n_j} (\sigma_j,t)$. Then, for any $M>0$, 
\begin{equation}
D_{1,n_j}^{(1)}(t)=    \frac{1}{\sigma_j}\, \left(\frac{n_j}{n}\right)^{1/2}\left(D_{1,n_j}^{(2)}(M,t)+ D_{1,n_j}^{(3)}(M,t)\right) \,,
\label{eq:R34}
\end{equation}
  with
\begin{align*}
D_{1,n_j}^{(2)}(M,t)& =   \frac{1}{n_j} \sum_{\ell=1}^{n_j} U(X_{j\ell})\,\left\{\exp\left\{i\,t\, \frac{\sigma_j}{\wsigma_j}\varepsilon_{j\ell}\right\} -\exp\left\{i\,t\, \varepsilon_{j\ell}\right\} \right\}  \indica_{|\varepsilon_{j\ell}|\le M}\,, \\
D_{1,n_j}^{(3)}(M,t)& =   \frac{1}{n_j} \sum_{\ell=1}^{n_j} U(X_{j\ell})\,\left\{\exp\left\{i\,t\, \frac{\sigma_j}{\wsigma_j}\varepsilon_{j\ell}\right\} -\exp\left\{i\,t\, \varepsilon_{j\ell}\right\} \right\}  \indica_{|\varepsilon_{j\ell}|\ge M}\,,
\end{align*}
where for the sake of simplicity we have denoted  $U(x)=W_j(x)\left(\Delta_j(x)-\Delta_0(x)\right)$.
Note that
\begin{align*}
A_{1,n_j}(M)=\| i\,t\,D_{1,n_j}^{(3)}(M,t)  \|_w   & \le 2 \frac{1}{n_j} \sum_{\ell=1}^{n_j} |U(X_{j\ell})|  \indica_{|\varepsilon_{j\ell}|\ge M} \left\{  \int t^2  w(t) \, dt \right\}^{1/2} \,,
 \end{align*}
 while 
$$\sup_{t\in \real} |D_{3,n_j}^{(2)}(M,t)|\le   M  \left| \frac{\sigma_j}{\wsigma_j}-1\right| \,\frac{1}{n_j} \sum_{\ell=1}^{n_j} \,   \left|U(X_{j\ell}) \right|  \,, $$  
so that
\begin{equation}
A_{2,n_j}(M)=\| i\,t\, D_{1,n_j}^{(2)}(M,t)  \|_w \le   M  \left| \frac{\sigma_j}{\wsigma_j}-1\right| \,\frac{1}{n_j} \sum_{\ell=1}^{n_j} \,   \left|U(X_{j\ell}) \right| \left\{  \int  t^2  w(t) \, dt \right\}^{1/2}  \,.
\label{eq:A2}
\end{equation}
 Given $\delta>0$, choose $M$ such that $\left\{  \int   t^2 w(t) \, dt \right\}^{1/2}\, \esp |U (X_{j})|  \prob(|\varepsilon_j|>M)<\delta/(8 a_j)$, where  $a_j= \pi_j^{1/2}/{\sigma_j}$. The  law of large numbers entail that  
 $$A_{1,n_j}(M)\convprob 2 \left\{  \int  t^2  w(t) \, dt \right\}^{1/2}\, \esp |U (X_{j})|  \prob(|\varepsilon_j|>M)<\delta/(4 a_j) \,,$$ 
 so that, given $\eta>0$, there exists $n_{j,0}$ such that for $n_j\ge n_{j,0}$, we have that
\begin{equation}
\prob\left(A_{1,n_j}(M) <\delta/(2 a_j) \right)>1-\eta/2\;.
\label{eq:A1}
\end{equation}
On the other hand, the consistency of $\wsigma_j$ together with the fact that $\esp|U (X_{j})| <\infty$ entail that 
$$ A_{3,n_j}=\left| \frac{\sigma_j}{\wsigma_j}-1\right| \,\frac{1}{n_j} \sum_{\ell=1}^{n_j} \,   \left|U(X_{j\ell}) \right| \left\{  \int  t^2 w(t) \, dt \right\}^{1/2} \convprob 0\,,$$
therefore, we can choose $n_{j,1}$ such that for $n_j\ge n_{j,1}$, we have that $\prob\left(A_{3,n_j}  <\delta/(2\,M\, a_j) \right)>1-\eta/2$, implying that
\begin{equation}
\prob\left(A_{2,n_j}(M)  <\delta/(2 a_j)  \right)>1-\eta/2\;.
\label{eq:A3}
\end{equation} 
Taking into account that $n_j/n\to \pi_j$, we get that
$$a_{j,n_j} =\frac{1}{\sigma_j}\, \left(\frac{n_j}{n}\right)^{1/2} \to \frac{1}{\sigma_j}\, \pi_j^{1/2}=a_j \,,$$
so for $n_j\ge n_{j,2}$,   we have that $a_{j,n_j} \le 2 a_j$. Combining \eqref{eq:R34}, \eqref{eq:A2}, \eqref{eq:A1} and \eqref{eq:A3}, we obtain that for $n_j\ge \max(n_{j,0},n_{j,1},n_{j,2})$, 
 $\prob(\|\wD \|_w < \delta )> 1-\eta$,
which entails that $\|\wD \|_w=o_{\prob}(1)$  concluding the proof.
 \end{proof}

\vskip0.1in
\begin{lemma}{\label{lemma:lema5}}
Assume that \eqref{modelo} and \ref{ass:A2} to \ref{ass:A7}, \ref{ass:A8a}) and \ref{ass:A9}  hold. Define for $s=1,\ldots, k$,
\begin{align*}
A_{1,s,n_j}(t) & = \frac{1}{\sqrt{n_j}} \sum_{\ell=1}^{n_j} W_j(X_{j\ell}) M_s(X_{j\ell})\,\exp\left\{i\,t\,  \varepsilon_{j\ell}\right\}\,  \left\{ \wefe_s(X_{j\ell})-f_s(X_{j\ell})\right\} \,, \\
 A_{2,s,n_j}(t) & =  \frac{1}{\sqrt{n_j}} \sum_{\ell=1}^{n_j} W_j(X_{j\ell}) V_s(X_{j\ell}) M_s(X_{j\ell})\,\exp\left\{i\,t\,  \varepsilon_{j\ell}\right\}\,  \left\{ \wefe(X_{j\ell})-f(X_{j\ell})\right\}\,,
 \end{align*}
where $V_s(x)= (n_s/n) (f_s(x)/f(x))$, $M_s(x)=m_s(x)/f(x)$. Then, we have that $\sup_{t\in \real} |A_{\ell,s,n_j}(t)|=o_{\prob}(1)$, for $\ell=1,2$.
\end{lemma}
 
 \begin{proof}
 We will only show that $\sup_{t\in \real} |A_{1,s,n_j}(t)|=o_{\prob}(1)$, since the proof of $\sup_{t\in \real} |A_{2,s,n_j}(t)|=o_{\prob}(1)$ is analogous. Denote 
$$f_{h,s}(x)= \esp \wefe_s(x)= \frac{1}{h}\int K\left(\frac{x-u}h\right) f_s(u) du=\frac{1}{h}\int K\left(\frac{x-u}h\right) \left\{f_s(u)-f_s(x)\right\} du +f_s(x)= r_{h,s}(x)+f_s(x)\,.$$
Recall that \ref{ass:A6} and \ref{ass:A7} imply that for $x\in \itS_j$, 
\begin{align*}
r_{h,s}(x)&=\int K\left(v\right) \left\{f_s(x-h\,v)-f_s(x)\right\} dv= - h f_s^{\prime} (x)\int v\,K\left(v\right) dv + h^2 \int v^2\,K\left(v\right)f_s^{\prime\prime} (\xi_{v,x})dv\\
&=h^2 \int v^2\,K\left(v\right)f_s^{\prime\prime} (\xi_{v,x})dv\,,
 \end{align*}
where $\xi_{v,x}$ is an intermediate point between $x$ and $x-h\,v$. Using that $f_s^{\prime\prime} $ is a continuous function in a neighbourhood of $\itS_j$, we get that, for $h$ small enough, $\sup_{x\in \itS_j, v} |f_s^{\prime\prime} (\xi_{v,x})|= a_{j,s}<\infty$, so
\begin{equation}
\sup_{x\in \itS_j}|r_{h,s}(x)|\le  h^2 a_{j,s} \int v^2\,K\left(v\right)dv\,.
\label{eq:cotarsh}
\end{equation}
Then, $A_{1,s,n_j}(t)= B_{1,s,n_j}(t)+B_{2,s,n_j}(t)$ where 
\begin{align*}
B_{1,s,n_j}(t) & =   \frac{1}{\sqrt{n_j}} \sum_{\ell=1}^{n_j} W_j(X_{j\ell}) M_s(X_{j\ell})\,\exp\left\{i\,t\,  \varepsilon_{j\ell}\right\}\,  \left\{ \wefe_s(X_{j\ell})-f_{h,s}(X_{j\ell})\right\} \,, \\
B_{2,s,n_j}(t) &= \frac{1}{\sqrt{n_j}} \sum_{\ell=1}^{n_j} W_j(X_{j\ell}) M_s(X_{j\ell})\,\exp\left\{i\,t\,  \varepsilon_{j\ell}\right\}\,  r_{h,s}(X_{j\ell}) \,.
 \end{align*}
Using \eqref{eq:cotarsh} and that $\|W_j\|_{\infty}=1$, \ref{ass:A5} and \ref{ass:A6}, we get that
 $$\sup_{t\in \real} |B_{2,s,n_j}(t)|\le \sqrt{n_j}\, h^2 \, a_{j,s} A_{j,s}\,\int v^2\,K\left(v\right)dv  \,,$$
where $A_{j,s}$ is an upper bound of $|M_s(u)|$  in a neighbourhood of $\itS_j$. Hence,  the fact that $n\, h^4\to 0$  entails that $\sup_{t\in \real} |B_{2,s,n_j}(t)|=o_{\prob}(1)$.

Let us consider the situation $s\ne j$. In this case, 
\begin{align*}
B_{1,s,n_j}(t) & =   \frac{1}{\sqrt{n_j}} \sum_{\ell=1}^{n_j} W_j(X_{j\ell}) M_s(X_{j\ell})\,\exp\left\{i\,t\,  \varepsilon_{j\ell}\right\}\,  \left\{ \wefe_s(X_{j\ell})-f_{h,s}(X_{j\ell})\right\} \\
& = \frac{1}{\sqrt{n_j}} \frac{1}{n_s} \sum_{\ell=1}^{n_j}   \sum_{r=1}^{n_s} W_j(X_{j\ell}) M_s(X_{j\ell})\,\exp\left\{i\,t\,  \varepsilon_{j\ell}\right\} \left\{ K_h\left(X_{j\ell}-X_{sr}\right)-f_{h,s}(X_{j\ell})\right\} \\
& = B_{1,s,n_j}^{(1)}(t)+ i \; B_{1,s,n_j}^{(2)}(t) \,.
 \end{align*}
Using that $\esp \left\{K_h\left(X_{j }-X_{s }\right)-f_{h,s}(X_{j })\right\} =0$, standard arguments allow to show that, for $j=1,\ldots, k$,
 $\sup_{t\in \real} \esp \{(B_{1,s,n_j}^{(j)}(t))^2\} =o(1)\,.$
 Hence, $\|B_{1,s,n_j} \|_w=o_{\prob}(1)$. 
 
 Let us consider the situation $s= j$. In this case, $B_{1,s,n_j}(t) =C_{1,n_j}(t)+C_{2,n_j}(t)$ where
\begin{align*}
C_{1,n_j}(t) & = \frac{1}{n_j\sqrt{n_j} } \sum_{\ell=1}^{n_j}     W_j(X_{j\ell}) M_s(X_{j\ell})\,\exp\left\{i\,t\,  \varepsilon_{j\ell}\right\} \left\{ K_h\left(0\right)-f_{h,j}(X_{j\ell})\right\} \,,\\
C_{2,n_j}(t) & = \frac{1}{n_j\sqrt{n_j} } \sum_{\ell\ne r }  W_j(X_{j\ell}) M_s(X_{j\ell})\,\exp\left\{i\,t\,  \varepsilon_{j\ell}\right\} \left\{ K_h\left(X_{j\ell}-X_{jr}\right)-f_{h,j}(X_{j\ell})\right\} \\
& = \frac{1}{2 n_j\sqrt{n_j} } \sum_{\ell\ne r } H(t,X_{j\ell}, \varepsilon_{j\ell}; X_{jr})+H(t,X_{jr}, \varepsilon_{jr}; X_{j\ell})\,,
 \end{align*}
 where 
 $$ H(t,X_{j\ell}, \varepsilon_{j\ell}; X_{jr})= W_j(X_{j\ell}) M_s(X_{j\ell})\,\exp\left\{i\,t\,  \varepsilon_{j\ell}\right\} \left\{ K_h\left(X_{j\ell}-X_{jr}\right)-f_{h,j}(X_{j\ell})\right\}  \,.$$
 The fact that $nh^2\to \infty$ implies that $\sup_{t\in \real} |C_{1,n_j}(t)|=o_{\prob}(1)$.
  Using similar arguments to those considered in \citet{Pardo:etal:2015} for $B_{1,s,n_j}(t)$, we conclude that $\|C_{1,n_j}\|_w=o_{\prob}(1)$. 
  \end{proof}
  
\vskip0.1in
\begin{proof}[Proof of Theorem \ref{teo:theorem1}.] The proof of b) follows as in \citet{Pardo:etal:2015}, so we will only derive a).
Recall that $\wDelta_j(x)=(\wmu_0(x)-\wm_j(x))/{\wsigma_j}$, $\wUps_{0,j}(x)=(\mu_0(x)-\wmu_0(x))/{\wsigma_j}$,  $\wUps_j(x)=(m_j(x)-\wm_j(x))/{\wsigma_j}$ and $\wGamma_j(x)= n^{-1/2}(\Delta_j(x)-\Delta_0(x))/{\wsigma_j}$. Hence, using a Taylor's expansion of order 2, we get that
\begin{align}
\sqrt{n_j}\left(\wvarphi_j(t)-\wvarphi_{0j}(t)\right) 
%&= \frac{1}{\sqrt{n_j}} \sum_{\ell=1}^{n_j} W_j(X_{j\ell})\,\left\{\exp(i\,t\, \weps_{j\ell})-\exp(i\,t\, \weps_{0j\ell})\right\} \nonumber\\
&= \frac{1}{\sqrt{n_j}} \sum_{\ell=1}^{n_j} W_j(X_{j\ell})\,\exp(i\,t\, \weps_{0j\ell})\, \left\{\exp\left(i\,t\, \frac{\wmu_0(X_{j\ell})-\wm_j(X_{j\ell})}{\wsigma_j}\right)-1\right\}
 \nonumber
\\
&=i\,t\, S_{1,n_j}(t)+t^2\, D_{1,n_j}(t) \,,
\label{eq:expresion1}
\end{align}
where
\begin{align}
S_{1,n_j}(t)&= 
%\frac{1}{\sqrt{n_j}} \sum_{\ell=1}^{n_j} W_j(X_{j\ell})\,\exp(i\,t\, \weps_{0j\ell})\,  \frac{\wmu_0(X_{j\ell})-\wm_j(X_{j\ell})}{\wsigma_j}=
 \frac{1}{\sqrt{n_j}} \sum_{\ell=1}^{n_j} W_j(X_{j\ell})\,\exp(i\,t\, \weps_{0j\ell})\,   \wDelta_j(X_{j\ell}) \,,
\label{eq:S1nj}\\
 D_{1,n_j}(t) & =
- \;\frac{1}{\sqrt{n_j}} \sum_{\ell=1}^{n_j} W_j(X_{j\ell})\,\exp(i\,t\, \weps_{0j\ell})\,\left\{  \frac{\wmu_0(X_{j\ell})-\wm_j(X_{j\ell})}{\wsigma_j}\right\}^2 \exp(i\,t\, \xi_{j\ell}) \,,
\nonumber
\end{align}
with $\xi_{j\ell}$ an intermediate point between 0 and $\left\{\wmu_0(X_{j\ell})-\wm_j(X_{j\ell})\right\}/{\wsigma_j}$. Hence, from Lemma \ref{lemma:lema3}, we may conclude that 
\begin{equation}
\sup_{t\in \real} |D_{1,n_j}(t)|=o_{\prob}(1) \,.
\label{eq:tasaR1}
\end{equation}
Recall that, under $H_0$, $ \wGamma_j\equiv 0$, in general under $H_{1,n}$,  from \eqref{eq:wDeltaj}  we may write
\begin{equation}
\weps_{0j\ell}=\frac{\sigma_j}{\wsigma_j}\varepsilon_{j\ell}+ \wUps_{0,j}(X_{j\ell})+ \wGamma_j(X_{j\ell})\,.
\label{eq:weps0jl}
\end{equation}
 This last equality  leads to $S_{1,n_j}(t)=S_{2,n_j}(t) +  i\,t\, D_{2,n_j}(t)$,
where
\begin{align}
 S_{2,n_j}(t) &= \frac{1}{\sqrt{n_j}} \sum_{\ell=1}^{n_j} W_j(X_{j\ell})\,\exp\left\{i\,t\, \frac{\sigma_j}{\wsigma_j}\varepsilon_{j\ell}\right\}\,  \wDelta_j(X_{j\ell}) \,,
\label{eq:S2nj} \\
 D_{2,n_j}(t) & =  \frac{1}{\sqrt{n_j}} \sum_{\ell=1}^{n_j} W_j(X_{j\ell})\, \left[\wUps_{0,j}(X_{j\ell})+ \wGamma_j(X_{j\ell})\right]  \wDelta_j(X_{j\ell}) \exp(i\,t\, \xi_{j\ell}) \,,
\nonumber
\end{align}
and $\xi_{j\ell}$ stands for an intermediate point between 0 and $\wUps_{0,j}(X_{j\ell})+ \wGamma_j(X_{j\ell})$.
Hence, from Lemma \ref{lemma:lema3} we obtain   
\begin{equation}
\sup_{t\in \real} |D_{2,n_j}(t)|=o_{\prob}(1)\,.
\label{eq:tasaR2}
\end{equation}
Let us consider the behaviour of the term $S_{2,n_j}$ under $H_0$. Note that  $ S_{2,n_j} (t)=S_{3,n_j}(t) - D_{3,n_j}(t)$ 
with 
 \begin{align*}
 S_{3,n_j}(t) &= 
\frac{\wsigma_j}{\sigma_j}\frac{1}{\sqrt{n_j}} \sum_{\ell=1}^{n_j} W_j(X_{j\ell})\,\exp\left\{i\,t\, \frac{\sigma_j}{\wsigma_j}\varepsilon_{j\ell}\right\}\,  \wDelta_j(X_{j\ell}) \,,\\
D_{3,n_j}(t) &=  \left(\frac{\wsigma_j-\sigma_j}{ \sigma_j}\right) \frac{1}{\sqrt{n_j}} \sum_{\ell=1}^{n_j} W_j(X_{j\ell})\,\exp\left\{i\,t\, \frac{\sigma_j}{\wsigma_j}\varepsilon_{j\ell}\right\}\,   {\wDelta_j(X_{j\ell})} \,.
  \end{align*}
Using again Lemma \ref{lemma:lema3}, we conclude that
 \begin{equation}
\sup_{t\in \real} |D_{3,n_j}(t)|=o_{\prob}(1)\,.
\label{eq:tasaR3}
\end{equation}
 As in Lemma \ref{lemma:lema4}, denote as $V_s(x)= (n_s/n) (f_s(x)/f(x))$, $M_s(x)=m_s(x)/f(x)$ and $C_s(x)=(n_s/n) M_s(x)$
From $\wDelta_j(x)=(\wmu_0(x)-\wm_j(x))/{\wsigma_j}$ and using that under $H_0$, $m_j(x)=m_0=\mu_0$, for all $j$, we have that  
$\wDelta_j(x) =\left({\wmu_0(x)-\mu_0(x)+m_j(x)-\wm_j(x)}\right){\wsigma_j} = \wUps_j(x)- \wUps_{0,j}(x)\,. $   
Standard arguments together with \ref{ass:A8a}), \eqref{eq:tasamj} and \eqref{eq:tasafj} allow to show that
 \begin{align}
 \sup_{x\in \itS_j} \left|\wmu_0(x)   -\mu_0(x)    -\sum_{s=1}^k V_s(x) \left(\wm_s(x)-m_s(x)\right) \right|=o_{\prob}(n^{-1/2}) \,.
 \label{eq:aprox1}
 \end{align}
  %%%%%%%%%%%%%%%%%%%%%%%%%%%
% HASTA ACA BUSCAR IFTHEN
%%%%%%%%%%%%%%%%%%%%%%%%%%%%
 Define
  \begin{align*}
  S_{3,n_j}^{(1)}(\sigma,t) & = \sum_{s=1}^k \frac{1}{\sigma_j}\frac{1}{\sqrt{n_j}} \sum_{\ell=1}^{n_j} W_j(X_{j\ell})V_s(X_{j\ell})\,\exp\left\{i\,t\, \frac{\sigma_j}{\sigma}\varepsilon_{j\ell}\right\}\,  \left\{ \wm_s(X_{j\ell})-m_s(X_{j\ell})\right\} \,,\\
 S_{3,n_j}^{(2)}(\sigma,t) & = \frac{1}{\sigma_j}\frac{1}{\sqrt{n_j}} \sum_{\ell=1}^{n_j} W_j(X_{j\ell})\,\exp\left\{i\,t\, \frac{\sigma_j}{\sigma}\varepsilon_{j\ell}\right\}\,  \left\{ \wm_j(X_{j\ell})-m_j(X_{j\ell})\right\}\;.
  \end{align*}
Then, \eqref{eq:aprox1} implies that $D_{4,n_j}(t)=S_{3,n_j}(t)- \left(S_{3,n_j}^{(1)}(\wsigma_j,t) - S_{3,n_j}^{(2)}(\wsigma_j,t)  \right) $ is such that $\sup_{t\in \real} |D_{4,n_j}(t)| =o_{\prob}(1)$. Furthermore, Lemma \ref{lemma:lema4}, leads to 
 $S_{3,n_j}(t)= \left(S_{3,n_j}^{(1)}(\sigma_j,t) - S_{3,n_j}^{(2)}(\sigma_j,t)  \right)+D_{4,n_j}(t)+i\,t\,  D_{5,n_j}(t) \,,$
 where  $\sup_{t\in \real} |D_{s,n_j}(t)| =o_{\prob}(1)$, for $s=4,5$.

Therefore, combining \eqref{eq:tasaR1}, \eqref{eq:tasaR2} and using that $\sup_{t\in \real} |D_{s,n_j}(t)| =o_{\prob}(1)$, for $s=1,\ldots,5$, we obtain that $
 \sqrt{n_j}\left(\wvarphi_j(t)-\wvarphi_{0j}(t)\right)  =   i\,t\,\left( S_{3,n_j}^{(1)}(\sigma_j,t)-S_{3,n_j}^{(2)}(\sigma_j,t)\right)+
   i\,t\,\wtD_{1,n_j}(t)+ t^2\wtD_{2,n_j}(t)  $,
where for simplicity we have denoted as   $\wtD_{1,n_j}(t)=D_{4,n_j}(t)-D_{3,n_j}(t) $, $\wtD_{2,n_j}(t)=D_{1,n_j}(t) -D_{2,n_j}(t)-D_{5,n_j}(t)$  with $\sup_{t\in \real} |\wtD_{s,n_j}(t)|=o_{\prob}(1)$, $s=1,2$.

From \eqref{eq:linearmj}, we get that  $ S_{3,n_j}^{(2)} (\sigma_j,t)   =  \{{\wsigma_j}/({\sigma_j\; \nu_{j}})\}\, S_{4,n_j}^{(2)}(\wsigma_j, t)+ D_{6,n_j}(t)$,  where $\sup_{t\in \real} |D_{6,n_j}(t)|=o_{\prob}(1)$, 
\begin{equation}
 S_{4,n_j}^{(2)}(\sigma, t) = \frac{1}{\sqrt{n_j}} \sum_{\ell=1}^{n_j} W_j(X_{j\ell})\,\exp\left\{i\,t\, \varepsilon_{j\ell}\right\}\,   \frac{\wL_j(X_{j\ell},\wsigma_j)}{f_j(X_{j\ell}) } \,,
 \label{eq:S4nj}
 \end{equation}
and $\wL_j(x,\sigma)$ is defined in \eqref{eq:WLj}.  Similarly, recalling that $V_s(x)= (n_s/n) (f_s(x)/f(x))$, we obtain
   \begin{align}
  S_{3,n_j}^{(1)}(\sigma_j,t) 
    & = \sum_{s=1}^k \frac{n_s}{n}\frac{\wsigma_s}{\sigma_j\,\nu_{s}}\frac{1}{\sqrt{n_j}} \sum_{\ell=1}^{n_j} W_j(X_{j\ell}) \,\exp\left\{i\,t\, \varepsilon_{j\ell}\right\}\,  \frac{\wL_s(X_{j\ell},\wsigma_s)}{f(X_{j\ell})}   + D_{7,n_j}(t)
    \nonumber \\
    & = \sum_{s=1}^k \frac{n_s}{n}\frac{\wsigma_s}{\sigma_j\, \nu_{s}} S_{4,n_j}^{(1,s)}(\wsigma_s, t)+ D_{7,n_j}(t) \,,
    \label{eq:S3nj1}
  \end{align} 
 with    $\sup_{t\in \real} |D_{7,n_j}(t)|=o_{\prob}(1)$. We will first
 expand $ S_{4,n_j}^{(1,s)}(\wsigma_s, t)$ as  $S_{4,n_j}^{(1,s)}(\wsigma_s, t) =\wtS_{n_j}^{(1,s)}(\wsigma_s, t)+ \wtS_{n_j}^{(2,s)}(\wsigma_s, t)+ \wtS_{n_j}^{(3,s)}(\wsigma_s, t) $,
where 
  \begin{align*}
 \wtS_{n_j}^{(1,s)}(\sigma , t)&=  \frac{1}{n_s} \, \frac{1}{\sqrt{n_j}} \sum_{r=1}^{n_s}\sum_{\ell=1}^{n_j} W_j(X_{j\ell}) \,\exp\left\{i\,t\, \varepsilon_{j\ell}\right\}\,  \frac{1}{f(X_{j\ell}) }\, K_h\left(X_{j\ell}-X_{sr}\right) \psi_s\left( \frac{\sigma_s}{\sigma} \varepsilon_{sr}\right) \,, \\
\wtS_{n_j}^{(2,s)}(\sigma, t) & =    \frac{1}{n_s} \, \frac{1}{\sqrt{n_j}} \sum_{r=1}^{n_s}\sum_{\ell=1}^{n_j} W_j(X_{j\ell}) \,\exp\left\{i\,t\, \varepsilon_{j\ell}\right\}\,  \frac{1}{f(X_{j\ell}) }\, K_h\left(X_{j\ell}-X_{sr}\right) \psi_s^{\prime}\left( \frac{\sigma_s}{\sigma} \varepsilon_{sr}\right)  \frac{m_0(X_{sr})-m_0(X_{j\ell})}{\sigma} \,, \\
\wtS_{n_j}^{(3,s)}(\wsigma_s, t) & = \frac 12  \frac{1}{n_s} \, \frac{1}{\sqrt{n_j}} \sum_{r=1}^{n_s}\sum_{\ell=1}^{n_j} W_j(X_{j\ell}) \,\exp\left\{i\,t\, \varepsilon_{j\ell}\right\}\,  \frac{1}{f(X_{j\ell}) }\, K_h\left(X_{j\ell}-X_{sr}\right) \psi_s^{\prime\prime}\left(  \xi_{sr, j\ell} \right) \frac{\left(m_0(X_{sr})-m_0(X_{j\ell})\right)^2}{\wsigma_s^2} \,.
   \end{align*}
The term $\wtS_{n_j}^{(3,s)}(\wsigma_s, t) $ can be bounded as $|\wtS_{n_j}^{(3,s)}(\wsigma_s, t) |\le   (1/{\wsigma_s^2} )\iota_{j,f}\,  \| \psi_s^{\prime\prime}\|_{\infty}  U_{j,s}^{1} \,,$
 where $\iota_{j,f}=\inf_{x\in \itS_j} f(x)$  and $U_{j,s}^{1} = \sum_{r=1}^{n_s}\sum_{\ell=1}^{n_j} W_j(X_{j\ell})     \, K_h\left(X_{j\ell}-X_{sr}\right)   {\left(m_0(X_{sr})-m_0(X_{j\ell})\right)^2} / ( n_s  \,  \sqrt{n_j})$.
Using standard $U-$statistics methods on $U_{j,s}^{1}$ and the fact that $n_j h^4\to 0$ (note that $\esp U_{j,s}^{1}=O(\sqrt{n_j h^4})$), we get easily that $ U_{j,s}^{1} \convprob 0$, leading to $\sup_{t\in \real} |\wtS_{n_j}^{(3,s)}(\wsigma_s, t) | =o_{\prob}(1)$.

To obtain that $\sup_{t\in \real} |\wtS_{n_j}^{(2,s)}(\wsigma_s, t) | =o_{\prob}(1)$, note that 
\begin{align*}
\sup_{t} | \wtS_{n_j}^{(2,s)}(\wsigma_s, t)| & \le \frac{1}{\sigma_s}\|\psi_s^{\prime}\|_{\infty} \iota_{j,f} \frac{1}{n_s} \, \frac{1}{\sqrt{n_j}} \sum_{r=1}^{n_s}\sum_{\ell=1}^{n_j}   \, K_h\left(X_{j\ell}-X_{sr}\right)   |m_0(X_{sr})-m_0(X_{j\ell})| \,,
\end{align*}
where the expectation of the right hand side converges to 0,  since $K$ is an even function, $m_s$ is twice continuously differentiable and $n h^4 \to 0$. 
   
 Using that  $\wsigma_s-\sigma_s=o_{\prob}(n^{-1/4})$, $\psi$ is bounded, $\esp \psi_s(t\varepsilon_s)=0$, for any $t>0$ and similar techniques as those considered in \citet{Pardo:etal:2015} when dealing with $\wA_{2j}(t)$, we get that  $\| \wtS_{n_j}^{(1,s)}(\wsigma_s, \cdot)- \wtS_{n_j}^{(1,s)}(\sigma_s, \cdot)\|_w=o_{\prob}(1)$. Therefore, combining the previous results, we conclude that
 $
  S_{4,n_j}^{(1,s)}(\wsigma_s, t)  =   \wtS_{n_j}^{(1,s)}(\sigma_s, t)+   \wtD_{n_j}^{(4,s)}( t) \,,$
   with  $\|\wtD_{n_j}^{(4,s)}( t)\|_w=o_{\prob}(1)$. 
   Therefore, $S_{3,n_j}^{(1)}(\sigma_j,t)= S_{4,n_j}^{(1)}(\sigma_j,t) +D_{8,n_j}(t)$, where $\|D_{8,n_j} \|_w=o_{\prob}(1)$ and
   $$S_{4,n_j}^{(1)}(\sigma_j,t) =\sum_{s=1}^k \pi_s \frac{\sigma_s}{\sigma_j\, \nu_{s}}  \frac{1}{n_s} \, \frac{1}{\sqrt{n_j}} \sum_{r=1}^{n_s}\sum_{\ell=1}^{n_j} W_j(X_{j\ell}) \,\exp\left\{i\,t\, \varepsilon_{j\ell}\right\}\,  \frac{1}{f(X_{j\ell}) }\, K_h\left(X_{j\ell}-X_{sr}\right) \psi_s\left( \varepsilon_{sr}\right) \,.$$
 Similar arguments allow to show that  $ S_{3,n_j}^{(2)}(\sigma_j,t)=  S_{4,n_j}^{(2)}(\sigma_j,t) +D_{9,n_j}(t) $ where $\sup_{t\in \real} |D_{9,n_j}(t)|=o_{\prob}(1)$ and 
$$   S_{4,n_j}^{(2)}(\sigma_j,t)  = \frac{1}{\nu_j } \frac{1}{n_j \sqrt{n_j}} \sum_{1\le s,\ell\le n_j} W_j(X_{j\ell})\,\exp\left\{i\,t\, \varepsilon_{j\ell}\right\}\,   \frac{1}{f_j(X_{j\ell}) }  K_h\left(X_{j\ell}-X_{j s}\right) \psi_j\left( \varepsilon_{js}\right) \,.$$
 Arguing as in  \citet{Pardo:etal:2015}, we may obtain that 
     \begin{align*}
  S_{4,n_j}^{(1)}(\sigma_j,t) 
    &= \frac{1}{\sigma_j} \varphi_j(t)\pi_j^{1/2} \sum_{s=1}^k  \frac{\sigma_s}{\nu_{s}}   \pi_s ^{1/2}   \, \frac{1}{\sqrt{n_s}} \sum_{r=1}^{n_s}  W_j(X_{sr} )   \frac{f_j(X_{sr} )}{f( X_{sr} ) }  \psi_s\left( \varepsilon_{sr}\right)
    +D_{10,n_j}(t) \,, \\
     S_{4,n_j}^{(2)}(\sigma_j,t) &= \frac{1}{\nu_j } \varphi_j(t) \esp\left(W_j(X_{j})\right) \frac{1}{ \sqrt{n_j}} \sum_{1\le s\le n_j}      \psi_j\left( \varepsilon_{js}\right)  +D_{11,n_j}(t)  \,,
  \end{align*} 
which leads to  $\sqrt{n_j}\left(\wvarphi_j(t)-\wvarphi_{0j}(t)\right)= i\,t\,  \varphi_j(t)  Z_{n,j}    +  i\,t\, R_{1,n }(t)+ t^2 R_{2,n}(t) $,  
where $ \{Z_{n,j}\}_{j=1}^k$ are defined in Lemma \ref{lemma:lema1} and $\|R_{s,n }\|_w=o_{\prob}(1)$ for $s=1,2$. The conclusion follows now from Lemma \ref{lemma:lema1}. 
\end{proof}

\vskip0.1in
\begin{proof}[{Proof of Theorem \ref{teo:theorem2}.}] The proof of Theorem \ref{teo:theorem2} follows the same steps as those considered in Theorem \ref{teo:theorem1}. 
Using \eqref{eq:expresion1} and  Lemma \ref{lemma:lema3}, we get that  $
\sqrt{n_j}\left(\wvarphi_j(t)-\wvarphi_{0j}(t)\right)  = i\,t\, S_{1,n_j}(t)+t^2\, D_{1,n_j}(t) $,
where $
S_{1,n_j}(t)$ is defined in \eqref{eq:S1nj} and $\sup_{t\in \real} |D_{1,n_j}(t)|=o_{\prob}(1)$.
 
Recall that from \eqref{eq:weps0jl} $\weps_{0j\ell}=({\sigma_j}/{\wsigma_j})\varepsilon_{j\ell}+ \wUps_{0,j}(X_{j\ell})+ \wGamma_j(X_{j\ell})\,,$ 
 which leads to $S_{1,n_j}(t) = S_{2,n_j}(t) +  i\,t\, D_{2,n_j}(t)$, where  $S_{2,n_j}(t)$ is defined in \eqref{eq:S2nj} and $\sup_{t\in \real} |D_{2,n_j}(t)|=o_{\prob}(1)$ from \eqref{eq:tasaR2}.
 
Let us consider the term $S_{2,n_j}$. From \eqref{eq:wDeltaj} and denoting 
$$D_j(x)=\frac{\wmu_0(x)-\mu_0(x)+m_j(x)-\wm_j(x) - n^{-1/2}(\Delta_j(x)-\Delta_0(x))}{\wsigma_j}= \wUps_j(x)- \wUps_{0,j}(x)-\wGamma_j(x) \,,$$ 
we have that  $ S_{2,n_j} (t)=    \sum_{\ell=1}^{n_j} W_j(X_{j\ell})\,\exp\left\{i\,t\,  \sigma_j \varepsilon_{j\ell}/{\wsigma_j}\right\}\,  D_j(X_{j\ell})/({\sqrt{n_j}}{\wsigma_j})=  S_{3,n_j}(t) - D_{3,n_j}(t) $,
with 
 \begin{align*}
 S_{3,n_j}(t) &= 
\frac{1}{\sigma_j}\frac{1}{\sqrt{n_j}} \sum_{\ell=1}^{n_j} W_j(X_{j\ell})\,\exp\left\{i\,t\, \frac{\sigma_j}{\wsigma_j}\varepsilon_{j\ell}\right\}\,  D_j(X_{j\ell}) \,, \\
D_{3,n_j}(t)&=  \left(\frac{\wsigma_j-\sigma_j}{ \sigma_j}\right) \frac{1}{\sqrt{n_j}} \sum_{\ell=1}^{n_j} W_j(X_{j\ell})\,\exp\left\{i\,t\, \frac{\sigma_j}{\wsigma_j}\varepsilon_{j\ell}\right\}\,  \frac{D_j(X_{j\ell})}{\wsigma_j} \,.
  \end{align*}
As when considering \eqref{eq:tasaR3}, using  Lemma \ref{lemma:lema3}, we conclude  that
 \begin{equation}
\sup_{t\in \real} |D_{3,n_j}(t)|=o_{\prob}(1)\,.
\label{eq:tasaR3bis}
\end{equation}
 As in the proof of Theorem \ref{teo:theorem1}, denote as $V_s(x)= (n_s/n) (f_s(x)/f(x))$, $M_s(x)=m_s(x)/f(x)$ and $C_s(x)=(n_s/n) M_s(x)$.
Again standard arguments   allow to show that
 \begin{align}
 \sup_{x\in \itS_j} \Bigg|\wmu_0(x)   -\mu_0(x)  & -\sum_{s=1}^k V_s(x) \left(\wm_s(x)-m_s(x)\right)-\sum_{s=1}^k  \left(\wefe_s(x)-f_s(x)\right)C_s(x) \nonumber\\
 &\left.-\sum_{s=1}^k V_s(x)\left(f(x)-\wefe(x)\right)\, M_s(x)\right|=o_{\prob}(n^{-1/2})\;.
 \label{eq:aprox1bis}
 \end{align}
%%%%%%%%%%%%%%%%%%%%%%%
% TERMINA EL CANCELO
%%%%%%%%%%%%%%%%%%%%
   Define
  \begin{align*}
  S_{3,n_j}^{(1)}(\sigma,t) & = \sum_{s=1}^k \frac{1}{\sigma_j}\frac{1}{\sqrt{n_j}} \sum_{\ell=1}^{n_j} W_j(X_{j\ell})V_s(X_{j\ell})\,\exp\left\{i\,t\, \frac{\sigma_j}{\sigma}\varepsilon_{j\ell}\right\}\,  \left\{ \wm_s(X_{j\ell})-m_s(X_{j\ell})\right\}  \,, \\
 S_{3,n_j}^{(2)}(\sigma,t) & = \frac{1}{\sigma_j}\frac{1}{\sqrt{n_j}} \sum_{\ell=1}^{n_j} W_j(X_{j\ell})\,\exp\left\{i\,t\, \frac{\sigma_j}{\sigma}\varepsilon_{j\ell}\right\}\,  \left\{ \wm_j(X_{j\ell})-m_j(X_{j\ell})\right\}  \,, \\
  S_{3,n_j}^{(3)}(\sigma,t) & = \frac{1}{\sigma_j}\, \left(\frac{n_j}{n}\right)^{1/2}\,\frac{1}{n_j} \sum_{\ell=1}^{n_j} W_j(X_{j\ell})\,\exp\left\{i\,t\, \frac{\sigma_j}{\sigma}\varepsilon_{j\ell}\right\}  (\Delta_j(X_{j\ell})-\Delta_0(X_{j\ell})) \,, \\
  S_{3,n_j}^{(4)}(\sigma,t) & = \sum_{s=1}^k \frac{1}{\sigma_j}\frac{1}{\sqrt{n_j}} \sum_{\ell=1}^{n_j} W_j(X_{j\ell}) M_s(X_{j\ell})\,\exp\left\{i\,t\, \frac{\sigma_j}{\sigma}\varepsilon_{j\ell}\right\}\,  \left\{ \wefe_s(X_{j\ell})-f_s(X_{j\ell})\right\}  \,, \\
  S_{3,n_j}^{(5)}(\sigma,t) & = \sum_{s=1}^k \frac{1}{\sigma_j}\frac{1}{\sqrt{n_j}} \sum_{\ell=1}^{n_j} W_j(X_{j\ell}) V_s(X_{j\ell}) M_s(X_{j\ell})\,\exp\left\{i\,t\, \frac{\sigma_j}{\sigma}\varepsilon_{j\ell}\right\}\,  \left\{ \wefe(X_{j\ell})-f(X_{j\ell})\right\}\,.
  \end{align*}
Note that $ S_{3,n_j}^{(1)}(\sigma,t)$ and $ S_{3,n_j}^{(1)}(\sigma,t)$ have been already defined in the proof of Theorem \ref{teo:theorem1}. Then, \eqref{eq:aprox1bis} implies that $D_{4,n_j}(t)=S_{3,n_j}(t)- \left(S_{3,n_j}^{(1)}(\wsigma_j,t)+S_{3,n_j}^{(4)}(\wsigma_j,t)-S_{3,n_j}^{(5)}(\wsigma_j,t) - S_{3,n_j}^{(2)}(\wsigma_j,t)- S_{3,n_j}^{(3)}(\wsigma_j,t) \right) $ is such that $\sup_{t\in \real} |D_{4,n_j}(t)| =o_{\prob}(1)$. As in the proof of Theorem \ref{teo:theorem1} using Lemma \ref{lemma:lema4} we get that
 $$S_{3,n_j}(t)= \left(S_{3,n_j}^{(1)}(\sigma_j,t)+S_{3,n_j}^{(4)}(\sigma_j,t)-S_{3,n_j}^{(5)}(\sigma_j,t) - S_{3,n_j}^{(2)}(\sigma_j,t)- S_{3,n_j}^{(3)}(\sigma_j,t) \right)+D_{4,n_j}(t)+i\,t\,  D_{5,n_j}(t) \,,$$
 where  $\sup_{t\in \real} |D_{s,n_j}(t)| =o_{\prob}(1)$, for $s=4,5$.
  
As in \citet{Pardo:etal:2015}, by the strong law of large numbers in Hilbert spaces, we obtain that $\|D_{6,n_j}(t)\|_w=o_{\prob}(1)$ where
 $$D_{6,n_j}(t)=  i\,t\,S_{3,n_j}^{(3)}(\sigma_j,t)-  i\,t\, \varphi_j(t)  \left({\pi_j^{1/2}}/{\sigma_j} \right) \esp \left\{W_j(X_{j} )\, \left[\Delta_j(X_{j})-\Delta_0(X_{j})\right]\right\}   \,.$$ 
 Furthermore,  Lemma \ref{lemma:lema5}, entails that $D_{7,n_j}(t)=S_{3,n_j}^{(4)}(\sigma_j,t)-S_{3,n_j}^{(5)}(\sigma_j,t)$ is such that $\|  D_{7,n_j} \|_{w} =o_{\prob}(1)$.   
Therefore, combining \eqref{eq:tasaR1}, \eqref{eq:tasaR2} and using that $\sup_{t\in \real} |D_{s,n_j}(t)| =o_{\prob}(1)$, for $s=1,\ldots,5$ and $s=7$, we obtain that
 \begin{align*}
 \sqrt{n_j}\left(\wvarphi_j(t)-\wvarphi_{0j}(t)\right) & =   i\,t\,\left( S_{3,n_j}^{(1)}(\sigma_j,t)-S_{3,n_j}^{(2)}(\sigma_j,t)-\frac{\pi_j^{1/2}}{\sigma_j} \varphi_j(t) \esp W_j(X_{j} )\, \left\{\Delta_j(X_{j})-\Delta_0(X_{j})\right\}\right) \nonumber\\
 & =   i\,\wtD_{1,n_j}(t)+ t^2\wtD_{2,n_j}(t) + \wtD_{3,n_j}(t)  \,,
\label{eq:expansionbis}
  \end{align*}
where for simplicity we have denoted as   $\wtD_{1,n_j}(t)=t\,\left(D_{4,n_j}(t)-D_{3,n_j}(t)+D_{7,n_j}(t)\right)$, $\wtD_{2,n_j}(t)=D_{1,n_j}(t) -D_{2,n_j}(t)-D_{5,n_j}(t)$ and $\wtD_{3,n_j}(t)=D_{6,n_j}(t)$ with $\sup_{t\in \real} |\wtD_{2,n_j}(t)|=o_{\prob}(1)$,  while $\|\wtD_{s,n_j}\|_w=o_{\prob}(1)$, $s=1,3$.

As in the proof of Theorem \ref{teo:theorem1},   \eqref{eq:linearmj} leads to
 \begin{align*}
  S_{3,n_j}^{(2)} (\sigma_j,t) & = \frac{\wsigma_j}{\sigma_j}\frac{1}{\sqrt{n_j}} \sum_{\ell=1}^{n_j} W_j(X_{j\ell})\,\exp\left\{i\,t\, \varepsilon_{j\ell}\right\}\,   \frac{ \wL_j(X_{j\ell},\wsigma_j)}{f_j(X_{j\ell})\nu_{j}}\, + D_{8,n_j}(t) = \frac{\wsigma_j}{\sigma_j\; \nu_{j}} S_{4,n_j}^{(2)}(\wsigma_j, t)+ D_{8,n_j}(t)\,,  
   \end{align*}
where $\sup_{t\in \real} |D_{8,n_j}(t)|=o_{\prob}(1)$, $\wL_j(x,\sigma)$ is defined in \eqref{eq:WLj} and $
 S_{4,n_j}^{(2)}(\sigma, t) $ is defined in \eqref{eq:S4nj}.
 Similarly to the expansion considered in \eqref{eq:S3nj1} and recalling that $V_s(x)= (n_s/n) (f_s(x)/f(x))$, we get
 $S_{3,n_j}^{(1)}(\sigma_j,t)    = \sum_{s=1}^k ({n_s}/{n})\{{\wsigma_s}/({\sigma_j\, \nu_{s}})\} S_{4,n_j}^{(1,s)}(\wsigma_s, t)+ D_{9,n_j}(t)  $, 
 with    $\sup_{t\in \real} |D_{9,n_j}(t)|=o_{\prob}(1)$.

 A similar expansion to that considered in the proof of Theorem \ref{teo:theorem1}, leads to 
   $  S_{4,n_j}^{(1,s)}(\wsigma_s, t)   =  \wtS_{n_j}^{(1,s)}(\wsigma_s, t)+ \wtS_{n_j}^{(2,s)}(\wsigma_s, t)+ \wtS_{n_j}^{(3,s)}(\wsigma_s, t) $,
where now
  \begin{align*}
 \wtS_{n_j}^{(1,s)}(\sigma , t)&=  \frac{1}{n_s} \, \frac{1}{\sqrt{n_j}} \sum_{r=1}^{n_s}\sum_{\ell=1}^{n_j} W_j(X_{j\ell}) \,\exp\left\{i\,t\, \varepsilon_{j\ell}\right\}\,  \frac{1}{f(X_{j\ell}) }\, K_h\left(X_{j\ell}-X_{sr}\right) \psi_s\left( \frac{\sigma_s}{\sigma} \varepsilon_{sr}\right)  \,,\\
\wtS_{n_j}^{(2,s)}(\sigma, t) & =    \frac{1}{n_s} \, \frac{1}{\sqrt{n_j}} \sum_{r=1}^{n_s}\sum_{\ell=1}^{n_j} W_j(X_{j\ell}) \,\exp\left\{i\,t\, \varepsilon_{j\ell}\right\}\,  \frac{1}{f(X_{j\ell}) }\, K_h\left(X_{j\ell}-X_{sr}\right) \psi_s^{\prime}\left( \frac{\sigma_s}{\sigma} \varepsilon_{sr}\right)  \frac{m_s(X_{sr})-m_s(X_{j\ell})}{\sigma}  \,, \\
\wtS_{n_j}^{(3,s)}(\wsigma_s, t) & = \frac 12  \frac{1}{n_s} \, \frac{1}{\sqrt{n_j}} \sum_{r=1}^{n_s}\sum_{\ell=1}^{n_j} W_j(X_{j\ell}) \,\exp\left\{i\,t\, \varepsilon_{j\ell}\right\}\,  \frac{1}{f(X_{j\ell}) }\, K_h\left(X_{j\ell}-X_{sr}\right) \psi_s^{\prime\prime}\left(  \xi_{sr, j\ell} \right) \frac{\left(m_s(X_{sr})-m_s(X_{j\ell})\right)^2}{\wsigma_s^2}  \,,
   \end{align*}
Again $\wtS_{n_j}^{(3,s)}(\wsigma_s, t) $ can be bounded as   $|\wtS_{n_j}^{(3,s)}(\wsigma_s, t) |\le   ( 1/{\wsigma_s^2}) \, \iota_{j,f}\,  \| \psi_s^{\prime\prime}\|_{\infty} \left(U_{j,s}^{1}+U_{j,s}^{2}\right) $, 
 where $\iota_{j,f}=\inf_{x\in \itS_j} f(x)$  and 
 \begin{align*}
 U_{j,s}^{1} &=   \frac{1}{n_s} \, \frac{1}{\sqrt{n_j}} \sum_{r=1}^{n_s}\sum_{\ell=1}^{n_j} W_j(X_{j\ell})     \, K_h\left(X_{j\ell}-X_{sr}\right)   {\left(m_0(X_{sr})-m_0(X_{j\ell})\right)^2}  \,, \\
 U_{j,s}^{2} &= \frac{1}{\sqrt{n}}  \frac{1}{n_s} \, \frac{1}{\sqrt{n_j}} \sum_{r=1}^{n_s}\sum_{\ell=1}^{n_j} W_j(X_{j\ell})     \, K_h\left(X_{j\ell}-X_{sr}\right)   {\left(\Delta_s(X_{sr})-\Delta_s(X_{j\ell})\right)^2}   \,.
 \end{align*}
Note that $U_{j,s}^{2} \le C h$, since $\Delta_s$ is Lipschitz and $K$ has bounded support. On the other hand, using standard $U-$statistics methods on $U_{j,s}^{1}$ and the fact that $n_j h^4\to 0$ (note that $\esp U_{j,s}^{1}=O(\sqrt{n_j h^4})$), we get easily that $ U_{j,s}^{2} \convprob 0$, leading to $\sup_{t\in \real} |\wtS_{n_j}^{(3,s)}(\wsigma_s, t) | =o_{\prob}(1)$.

As in the proof of Theorem \ref{teo:theorem1}, we have that
\begin{align*}
\sup_{t} | \wtS_{n_j}^{(2,s)}(\wsigma_s, t)| & \le \frac{1}{\sigma_s}\|\psi^{\prime}\|_{\infty} \iota_{j,f} \frac{1}{n_s} \, \frac{1}{\sqrt{n_j}} \sum_{r=1}^{n_s}\sum_{\ell=1}^{n_j}   \, K_h\left(X_{j\ell}-X_{sr}\right)   |m_s(X_{sr})-m_s(X_{j\ell})|   \,,
\end{align*}
which entails that $\sup_{t} | \wtS_{n_j}^{(2,s)}(\wsigma_s, t)| =o_{\prob}(1)$, since  $m_s$ is twice continuously differentiable and $n h^4 \to 0$. 
   
 Using that  $\wsigma_s-\sigma_s=o_{\prob}(n^{-1/4})$,  $\psi$ is bounded, $\esp \psi(t\varepsilon_s)=0$, for any $t>0$, similar arguments to those considered in the proof of Theorem \ref{teo:theorem1} allow to show that  $\| \wtS_{n_j}^{(1,s)}(\wsigma_s, t)- \wtS_{n_j}^{(1,s)}(\sigma_s, t)\|_w=o_{\prob}(1)$. Therefore, combining the previous results, we conclude that 
 $  S_{4,n_j}^{(1,s)}(\wsigma_s, t)  =   \wtS_{n_j}^{(1,s)}(\sigma_s, t)+   \wtD_{n_j}^{(4,s)}( t) $, 
   with $\sup_{t} |\wtD_{n_j}^{(4,s)}( t)|=o_{\prob}(1)$. 
   Finally, as in  the proof of Theorem \ref{teo:theorem1}, the consistency of $\wsigma_j$ and the fact that $n_s/n\to \pi_s$ lead to 
 $S_{3,n_j}^{(1)}(\sigma_j,t) = S_{4,n_j}^{(1)}(\sigma_j,t) +D_{10,n_j}(t)$, where  $\|D_{10,n_j}\|_w=o_{\prob}(1)$ and
$$ S_{4,n_j}^{(1)}(\sigma_j,t)=\sum_{s=1}^k \pi_s \frac{\sigma_s}{\sigma_j\, \nu_{s}}  \frac{1}{n_s} \, \frac{1}{\sqrt{n_j}} \sum_{r=1}^{n_s}\sum_{\ell=1}^{n_j} W_j(X_{j\ell}) \,\exp\left\{i\,t\, \varepsilon_{j\ell}\right\}\,  \frac{1}{f(X_{j\ell}) }\, K_h\left(X_{j\ell}-X_{sr}\right) \psi_s\left( \varepsilon_{sr}\right)\,. $$ 
Using similar arguments, we obtain that  $S_{3,n_j}^{(2)}(\sigma_j,t)= S_{4,n_j}^{(2)}(\sigma_j,t) +D_{11,n_j}(t)  $, where  $\|D_{11,n_j}\|_w=o_{\prob}(1)$ and 
$$S_{4,n_j}^{(2)}(\sigma_j,t)= \frac{1}{\nu_j } \frac{1}{n_j \sqrt{n_j}} \sum_{1\le s,\ell\le n_j} W_j(X_{j\ell})\,\exp\left\{i\,t\, \varepsilon_{j\ell}\right\}\,   \frac{1}{f_j(X_{j\ell}) }  K_h\left(X_{j\ell}-X_{j s}\right) \psi_j\left( \varepsilon_{js}\right) \,.$$ 
As in  \citet{Pardo:etal:2015} and in  the proof of Theorem \ref{teo:theorem1}, we may obtain that 
     \begin{align*}
  S_{4,n_j}^{(1)}(\sigma_j,t) 
    &= \frac{1}{\sigma_j} \varphi_j(t)\pi_j^{1/2} \sum_{s=1}^k  \frac{\sigma_s}{\nu_{s}}   \pi_s ^{1/2}   \, \frac{1}{\sqrt{n_s}} \sum_{r=1}^{n_s}  W_j(X_{sr} )   \frac{f_j(X_{sr} )}{f( X_{sr} ) }  \psi_s\left( \varepsilon_{sr}\right) +D_{12,n_j}(t)  \,, \\
     S_{4,n_j}^{(2)}(\sigma_j,t) &= \frac{1}{\nu_j } \varphi_j(t) \esp\left(W_j(X_{j})\right) \frac{1}{ \sqrt{n_j}} \sum_{1\le s\le n_j}      \psi_j\left( \varepsilon_{js}\right)  +D_{13,n_j}(t)   \,.
  \end{align*} 
Recalling that $ \{Z_{n,j}\}_{j=1}^k$ are defined in Lemma \ref{lemma:lema1}, we have that
\begin{align*}
 \sqrt{n_j}\left(\wvarphi_j(t)-\wvarphi_{0j}(t)\right) 
 & =   i\,t\,  \varphi_j(t)\left\{ Z_{n,j}   -\pi_j^{1/2}  \esp W_j(X_{j} )\, \left\{\Delta_j(X_{j})-\Delta_0(X_{j})\right\}\right\} +  i\,t\, R_{1,n }(t)+ t^2 R_{2,n}(t) + \wD_{3,n}(t)\\
 & =   i\,t\,  \varphi_j(t)\left\{ Z_{n,j}   -\pi_j^{1/2}  \esp W_j(X_{j} )\, \left\{\Delta_j(X_{j})-\Delta_0(X_{j})\right\}\right\} +  i\,  R_{1,n }^{\star}(t)+  R_{2,n}^{\star}(t) + \wD_{3,n}(t)  \,,
  \end{align*} 
where $\|R_{s,n }^{\star}\|_w=o_{\prob}(1)$ for $s=1,2$, $\|\wD_{3,n}\|_w=o_{\prob}(1)$. The conclusion follows now from   Lemma \ref{lemma:lema1}.
\end{proof}
 
 \vskip0.1in
 
\begin{proof}[Proof of Proposition \ref{prop:tasascale}] 
Denote  $\chi(u)=\rho(u)-b$, then $\esp \chi(\varepsilon_j)=0$ and 
$$ \frac{1}{n_j}\sum_{\ell=1}^{n_j} \chi\left(\frac{ Y_{j\ell} - \wm_j(X_{j\ell})}{\wsigma_j }\right)=0\,.$$
Therefore, using a Taylor's expansion of order one, we get
$$0= \frac{1}{n_j}\sum_{\ell=1}^{n_j} \chi\left(\frac{ Y_{j\ell} - \wm_j(X_{j\ell})}{\sigma_j }\right) -  (\wsigma_j-\sigma_j) \,\frac{1}{\wtsigma_j}\,\frac{1}{n}\sum_{\ell=1}^{n_j} \eta\left(\frac{ Y_{j\ell} - \wm_j(X_{j\ell})}{\wtsigma_j }\right)\,, $$
where $\wtsigma_j$ is an intermediate point between $\wsigma_j$ and $\sigma_j$.  Thus,
\begin{equation}
\label{eq:probar0}
n_j^{\gamma_0} (\wsigma_j-\sigma_j) = \; {\wtsigma_j}\; A_{n_j} ^{-1} n_j^{\gamma_0} \frac{1}{n_j}\sum_{\ell=1}^{n_j} \chi\left(\frac{ Y_{j\ell} - \wm_j(X_{j\ell})}{\sigma_j }\right)  \,,
\end{equation}
where 
$$A_{n_j}=\frac{1}{n_j}\sum_{\ell=1}^{n_j} \eta\left(\frac{ Y_{j\ell} - \wm_j(X_{j\ell})}{\wtsigma_j }\right)\,.$$
We begin by proving that the fact that $\eta$ and $\eta^{\prime} $ are bounded entail that
\begin{equation}
\label{eq:probar1}
  A_{n_j} \convprob \frac{1}{\sigma_j}\esp  \eta\left(\frac{ Y_{j} - m_j(X_{j})}{\sigma_j }\right)=  \esp \eta(\varepsilon_j)=  A_j\ne 0\,.
\end{equation}
Effectively, using that  $Y_{j\ell} =  m_j(X_{j\ell})+\sigma_j \varepsilon_{j\ell} $ we get $A_{n_j}=  A_{n_j,1}-A_{n_j,2} $,
where
\begin{align*}
 A_{n_j,1}  = &  \frac{1}{n_j}\sum_{\ell=1}^{n_j} \eta\left(\frac{\sigma_j \varepsilon_{j\ell}}{\wtsigma_j }\right) \qquad \mbox{and}\qquad
  A_{n_j,2} =   \frac{1}{n_j}\sum_{\ell=1}^{n_j} \eta^{\prime}\left(   \frac{\sigma_j \varepsilon_{j\ell}  -\zeta_{j\ell}}{\wtsigma_j }\right)  \left( \wm_j(X_{j\ell})- m_j(X_{j\ell})\right) \,,
\end{align*}
with $\zeta_{j\ell}= \theta m_j(X_{j\ell})+ (1-\theta) \wm_j(X_{j\ell})$ and $0\le \theta\le 1$. Standard arguments 
\citep[see][for instance]{boente:fraiman:1989}
allow to show that
$$ \frac{1}{n_j}\sum_{\ell=1}^{n_j} \eta\left(\frac{\sigma_j }{\wtsigma_j }\, \varepsilon_{j\ell}\right)-  \frac{1}{n_j}\sum_{\ell=1}^{n_j} \eta\left(  \varepsilon_{j\ell} \right) \convprob 0\,,$$
since $\wtsigma_j\convprob \sigma_j$, which leads to $ A_{n_j,1} \convprob \esp \eta(\varepsilon_j)=A_j$. Thus, to conclude the proof of \eqref{eq:probar1} it will be enough to show that $A_{n_j,2}\convprob 0$.
Using the Cauchy--Schwartz inequality, we get that
\begin{align*}
|A_{n_j,2}| & \le  \left(\frac{1}{n_j}\sum_{\ell=1}^{n_j}\left[\eta^{\prime}\left(   \frac{\sigma_j \varepsilon_{j\ell}  -\zeta_{j\ell}}{\wtsigma_j }\right) \right]^2\right)^{1/2}\, \left(\frac{1}{n_j}\sum_{\ell=1}^{n_j}\left( \wm_j(X_{j\ell})- m_j(X_{j\ell})\right)^2\right)^{1/2}\\
& \le  \left\|\eta^{\prime}\right\|_{\infty} \left(\frac{1}{n_j}\sum_{\ell=1}^{n_j}\left( \wm_j(X_{j\ell})- m_j(X_{j\ell})\right)^2\right)^{1/2}\,,
\end{align*}
Hence, using \ref{ass:C2}, we get that $A_{n_j,2}\convprob 0$, so $A_{n_j}\convprob \esp \eta(\varepsilon_j)$, as desired.

Therefore, using that    $ \wsigma_j\convprob \sigma_j$, to show that $n_j^{\gamma_0}(\wsigma_j-\sigma_j)=O_{\prob}(1)$, from \eqref{eq:probar0} we only have  to prove that
$$V_{n_j}=n_j^{\gamma_0} \frac{1}{n_j}\sum_{\ell=1}^{n_j} \chi\left(\frac{ Y_{j\ell} - \wm_j(X_{j\ell})}{\sigma_j }\right)=O_{\prob}(1)\,.$$
Note that the fact that $Y_{j\ell} =  m_j(X_{j\ell})+\sigma_j \varepsilon_{j\ell} $ and a Taylor's expansion of order two allows to write $V_{n_j}= n_j^{\gamma_0-1/2} V_{n_j,1}+V_{n_j,2}/\sigma_j +V_{n_j,3}/\sigma_j ^2$, where $V_{n_j,1} =  n_j^{\,-\,1/2}\sum_{\ell=1}^{n_j} \chi\left(\varepsilon_{j\ell}\right)$,
\begin{align*}
V_{n_j,2} = & n_j^{\gamma_0}  \frac{1}{n_j }\sum_{\ell=1}^{n_j} \psi\left(\varepsilon_{j\ell}\right) \left\{m_j(X_{j\ell})- \wm_j(X_{j\ell})\right\} \quad \mbox{and}\quad 
V_{n_j,3} =   n_j^{\gamma_0}  \frac{1}{n_j }\sum_{\ell=1}^{n_j} \psi^{\prime}\left(\varepsilon_{j\ell}+\xi_{j\ell}\right) \left\{m_j(X_{j\ell})- \wm_j(X_{j\ell})\right\}^2\;,
\end{align*}
with $\xi_{j\ell}= \theta m_j(X_{j\ell})+ (1-\theta) \wm_j(X_{j\ell})$ and $0\le \theta\le 1$. 

Taking into account that $\esp \rho(\varepsilon_{j})=b$, i.e., $\esp \chi(\varepsilon_{j})=0$, from the Central Limit Theorem we obtain that $V_{n_j,1} =O_{\prob}(1)$, so $n_j^{\gamma_0-1/2} V_{n_j,1}=o_{\prob}(1)$, since $\gamma_0<1/2$. 

Using that $\psi^{\prime}$ is bounded we get that
$$|V_{n_j,3}|\le \|\psi^{\prime}\|_{\infty}  n_j^{\gamma_0} \frac{1}{n_j}\sum_{\ell=1}^{n_j} \left\{\wm_j(X_{j\ell})- m_j(X_{j\ell})\right\}^2\,.$$
If \ref{ass:C2}b) holds, we immediately obtain that $|V_{n_j,3}|=o_{\prob}(1)$. Besides, if  \ref{ass:C2}a) holds, it is enough to use the bound
\begin{equation}
\label{eq:cota}
\frac{1}{n_j}\sum_{\ell=1}^{n_j} \left\{\wm_j(X_{j\ell})- m_j(X_{j\ell})\right\}^2\le   \sup_{x\in [0,1]}|\wm_j(x)- m_j(x)|^2\,,
\end{equation}
 to conclude that $|V_{n_j,3}|=o_{\prob}(1)$.
 
 To bound $V_{n_j,2}$ note that from the Cauchy--Schwartz inequality and the boundedness of $\psi$, we get
 \begin{align*}
\left| V_{n_j,2}\right| \le & n_j^{\gamma_0} \left\{\frac{1}{n_j }\sum_{\ell=1}^{n_j} \psi^2\left(\varepsilon_{j\ell}\right)  \right\}^{1/2}\left\{\frac{1}{n_j }\sum_{\ell=1}^{n_j}   \left\{m_j(X_{j\ell})- \wm_j(X_{j\ell})\right\}^2\right\}^{1/2}\\
\le & \| \psi\|_{\infty} \left\{ n_j^{2\,\gamma_0} \frac{1}{n_j }\sum_{\ell=1}^{n_j}   \left\{m_j(X_{j\ell})- \wm_j(X_{j\ell})\right\}^2\right\}^{1/2}\,.
 \end{align*}
 If \ref{ass:C2}b) holds, we immediately obtain that $|V_{n_j,2}|=O_{\prob}(1)$. If \ref{ass:C2}a) is valid, using again \eqref{eq:cota} we also obtain that  $|V_{n_j,2}|=O_{\prob}(1)$.
\end{proof}

%%%%%%%%%%%%%%%%%%%%%%%%%%%%%
\small
%\nocite{*}

%\nocite{*}
\bibliographystyle{apalike}
 
\bibliography{referencias2}
 
\end{document}